\documentclass[twocolumn,english,aps,prx,floatfix,amssymb,superscriptaddress,longbibliography]{revtex4-1}
\usepackage[latin9]{inputenc}
\usepackage{amsmath}
\usepackage{graphicx}
\usepackage{esint}

\makeatletter
\usepackage[unicode=true,pdfusetitle,
 bookmarks=true,bookmarksnumbered=false,bookmarksopen=false,
 breaklinks=false,pdfborder={0 0 0},backref=false,colorlinks=true,citecolor=blue]
 {hyperref}

\makeatother

\begin{document}
\title{Scattering mechanisms and electrical transport near an Ising nematic
quantum critical point}
\author{Xiaoyu Wang}
\affiliation{James Frank Institute, University of Chicago, Chicago, IL 60637}
\author{Erez Berg}
\affiliation{James Frank Institute, University of Chicago, Chicago, IL 60637}
\affiliation{Department of Condensed Matter Physics, Weizmann Institute of Science,
Rehovot 76100, Israel}
\begin{abstract}
Electrical transport properties near an electronic Ising-nematic quantum
critical point in two dimensions are of both theoretical and experimental
interest. In this work, we derive a kinetic equation valid in a broad
regime near the quantum critical point using the memory matrix approach.
The formalism is applied to study the effect of the critical fluctuations
on the dc resistivity through different scattering mechanisms, including
umklapp, impurity scattering, and electron-hole scattering in a compensated
metal. We show that electrical transport in the quantum critical regime
exhibits a rich behavior that depends sensitively on the scattering
mechanism and the band structure. In the case of a single large Fermi
surface, the resistivity due to umklapp scattering crosses over from
$\rho\sim T^{2}$ at low temperature to sublinear at high temperature.
The crossover temperature scales as $q_{0}^{3}$, where $q_{0}$ is
the minimal wavevector for umklapp scattering. Impurity scattering
leads to $\rho-\rho_{0}\sim T^{\alpha}$ ($\rho_{0}$ being the residual
resistivity), where $\alpha$ is either larger than $2$ if there
is only a single Fermi sheet present, or $4/3$ in the case of multiple
Fermi sheets. Finally, in a perfectly compensated metal with an equal
density of electrons and holes, the low temperature behavior depends
strongly on the structure of ``cold spots'' on the Fermi surface,
where the coupling between the quasiparticles and order parameter
fluctuations vanishes by symmetry. In particular, for a system where
cold spots are present on some (but not all) Fermi sheets, $\rho\sim T^{5/3}$.
At higher temperatures there is a broad crossover regime where $\rho$
either saturates or $\rho\sim T$, depending on microscopic details.
We discuss these results in the context of recent quantum Monte Carlo
simulations of a metallic Ising nematic critical point, and experiments
in certain iron-based superconductors.
\end{abstract}
\maketitle

\section{Introduction}

Many strongly correlated materials, such as high temperature superconductors,
exhibit electrical transport properties that are markedly different
from those expected in a Fermi liquid. Most famously, in many systems,
the electrical resistivity is a linear function of temperature. Interestingly,
it is often in this regime where the highest superconducting transition
temperature is observed. Understanding the role of strong electronic
correlations in such systems is of utmost importance. A natural way
to explain these properties is to consider a metallic system tuned
near a quantum critical point (QCP), i.e., a ``quantum critical metal''.
This is further motivated by the observation that superconductivity
is typically found near another electronically ordered phase \cite{scalapino12}.
The electronic order is suppressed by varying an external parameter
such as electron concentration or pressure, leading to a putative
QCP. The critical fluctuations of the order parameter mediate long-range,
dynamical interactions between electrons. These interactions can lead
to a breakdown of Fermi liquid behavior at the longest length and
timescales.

Of particular interest is the electrical transport properties near
an Ising-nematic QCP in two spatial dimensions \cite{dellanna07,hartnoll14,lederer17,maslov11}.
An Ising nematic phase refers to a rotation-symmetry-breaking electronic
order where the square lattice tetragonal symmetry is spontaneously
broken down to orthorhombic. It has been observed in various condensed
matter systems \cite{fradkin10}, notably in various high $T_{c}$
superconductors, including iron pnictides \cite{Chu:2010aa,fernandes14}
and iron chalcogenides \cite{hosoi2016,coldea18}. Evidence for nematic
fluctuations has been found in the cuprate superconductors, as well
\cite{Kohsaka:2007aa,Daou_2010}. Recent experiments \cite{kasahara10,reiss17,Urata:2016aa,reiss18}
reported non-Fermi liquid transport in the vicinity of the nematic
QCP.

On the theoretical front, several works have studied trasport near
a nematic QCP~\cite{dellanna07,hartnoll14,maslov11,lederer17}, along with many other investigations of transport in different types of quantum critical metals~\cite{varma1989phenomenology,YongBaek1995,hlubina95, Rosch1999,Syzranov2012,Paul2013,Patel2014,Chubukov2014,Chubukov2017}. However,
the general picture remains unclear. Naively, since the singular fluctuations
of the nematic order parameter occur at small wavevectors, one may
expect them not to strongly modify transport properties. Treating
the order parameter fluctuations as an external ``bath'', Ref. \cite{dellanna07}
found a resistivity that varies as $T^{3/2}$ near the QCP. However,
in an electronic nematic transition, the order parameter fluctuations
are a collective mode of the same electrons that carry the current.
This issue was addressed in Ref. \cite{maslov11}, and the resistivity
was found to vary as $T^{2}$ in the clean case, due to the long-wavelength
nature of the nematic fluctuations. Ref. \cite{hartnoll14} treated
the case where the main source of momentum relaxation is long-wavelength
disorder that couples to the order parameter as a random field; a
variety of behaviors was found depending on the dynamical critical
exponent assumed for the QCP. Indications for non-Fermi liquid transport
due to quantum critical nematic fluctuations were found in recent
numerical simulations \cite{lederer17}; however, these result suffer
from the usual uncertainties associated with analytical continuation
of numerical data, and need to be backed by analytical calculations.
Clearly, a uniform theoretical framework to treat transport near an
Ising-nematic QCP is highly desireable.

The situation is complicated by the fact that despite the substantial progress made
~\cite{metzner03,oganesyan2001,jerome06,lawler06,lohneysen07,sslee09,zacharias09,mross10,metlitski10a,Dalidovich2013,holder15,klein18a,klein18b},
the theory of Ising-nematic quantum criticality at asymptotically low energies
is not well understood. However, as we argue below, there is a broad
range of energy scales where the theory can be controlled; this ``Hertz--Millis--Moriya''
regime is described in terms of coherent electrons interacting with
strongly renormalized, overdamped collective fluctuations of the order
parameter \cite{hertz76,millis93,Moriya_1985}. Evidence for the
existence of such a regime has been found in recent quantum Monte
Carlo simulations \cite{schattner16,lederer17,Berg:2018aa}. At lower
temperatures, the theory becomes strongly coupled, with nematic-mediated
superconducting fluctuations and strong non-Fermi liquid behavior
onsetting at the same energy scale.

In this work, we develop a memory matrix approach for transport in
the intermediate coherent electron regime. Transport in this regime
can be described by a kinetic equation, where the effects of the quantum
critical fluctuations are incorporated in the collision integral.
At low temperatures (in the absence of a superconducting phase), the
electrons become incoherent, and our approach breaks down; however,
in the absence of a magnetic field, in this regime the system generically
becomes a high $T_{c}$ superconductor mediated by the nematic fluctuations.
Our theory is expected to hold down to temperatures of the order of
$T_{c}$.

We apply our technique to study situations where different scattering
mechanisms dominate the transport behavior, including umklapp scattering,
impurity scattering, and momentum-conserving electron-hole scattering
in a compensated metal. We find that depending on the relaxation mechanism,
the electrical resistivity exhibits rich features, including several
regimes beyond the ones discussed in Refs. \cite{dellanna07,maslov11}.
Our main results are summarized as follows:
\begin{enumerate}
\item In a clean system with a large, generic Fermi surface, umklapp scattering
has a low-momentum threshold $q_{0}$ determined by the smallest distance
between Fermi surfaces in different Brillouin zones. At low temperatures,
$T\ll T_{0}\sim q_{0}^{3}$, the typical momentum of the quantum critical
fluctuations is smaller than $q_{0}$. As a result, $\rho(T)\sim T^{2}$,
as in a Fermi liquid, despite the proximity to the QCP \cite{maslov11}.
At higher temperatures, umklapp scattering is governed by critical
fluctuations. In this regime, $\rho(T)$ is strongly enhanced, and
the dynamics of the critical fluctuations is qualitatively modified
by the umklapp processes, and deviates from the naive low-temperature
scaling behavior (characterized by a $z=3$ dynamical critical exponent).
As a result, $\rho(T)$ exhibits a smooth crossover from $T^{2}$
at $T\ll T_{0}$ to sublinear at higher temperatures.
\item In the presence of weak impurity scattering in a single electronic
band and in the absence of umklapp scattering, the critical nematic
fluctuations decouple from the transport properties to lowest order
in temperature. This is since for a single convex Fermi surface, electron-electron
scattering near the Fermi surface conserves the odd moments of the
quasiparticle distribution function \cite{maslov11,ledwidth17}.
Therefore, to lowest order, $\rho(T)=\rho_{0}$, and the correction
scales as $T^{\beta}$ with $\beta>2$. However, for a generic multiband
system, we find $\rho(T)-\rho_{0}\propto T^{4/3}$ due to quantum
critical fluctuations.
\item For a clean compensated metal with an equal density of electrons and
holes, the low-tempearture properties are strongly sensitive to the
presence of ``cold spots'' on the Fermi sheets. At these points,
the lowest-order coupling between the quasiparticles and the nematic
fluctuations vanishes by symmetry. For instance, in the case of a
$B_{1g}$ nematic order parameter (of $x^{2}-y^{2}$ symmetry), the
cold spots occur at the intersection between the diagonals $k_{x}=\pm k_{y}$
and the Fermi surfaces. We show that the non-equilibrium distribution
function displays a strongly non-harmonic form, changing abruptly
at the cold spots. In the case where all the Fermi surfaces have cold
spots, $\rho\sim T^{2}$; if only some of the Fermi surfaces have
cold spots and others do not \footnote{This is the case, for instance, if the order parameter has \textbf{$B_{1g}$}
symmetry and some of the Fermi pockets are centered away from the
high symmetry points $\Gamma=(0,0)$, $M=(\pi,\pi)$ in the Brillouin
zone, as in many of the iron-based superconductors.}, then $\rho\sim T^{5/3}$. If none of the Fermi surfaces have cold
spots, then $\rho\sim T^{4/3}$. At intermediate to strong coupling
strengths \footnote{This regime is accessible, within our model, in the limit where the
number of fermion flavors is large.}, there is a broad crossover regime where the resistivity either saturates
or is linear in temperatrure, depending on microscopic details. The
crossover behavior is due to near elastic scattering of electrons
by thermal nematic fluctuations.
\end{enumerate}
The outline of the paper is given as follows. In Section II, we introduce
a model Hamiltonian that realizes a metallic Ising-nematic QCP, and
argue that there is a parametrically broad temperature regime where
electrical transport can be described by kinetic theory. In Section
III, we derive a memory matrix formalism for calculating the linearized
collision integral, and discuss various conservation laws. In Section
IV, we study the temperature dependence of dc electrical resistivity
near the QCP originating from various current-relaxation mechanisms,
including impurity scattering, umklapp scattering, and momentum conserving
electron-hole scattering in a compensated metal. We conclude in Section
V.

\section{Model}

We consider a simple model on a two-dimensional square lattice that
realizes a metallic Ising-nematic QCP. The Hamiltonian is given by
\begin{equation}
\begin{split}H= & \sum_{\mathbf{k},\alpha=1}^{N}\left(\varepsilon_{\mathbf{k}}-\mu\right)c_{\alpha\mathbf{k}}^{\dagger}c_{\alpha\mathbf{k}}-\frac{1}{2}\sum_{\mathbf{q}}U_{\mathbf{q}}Q_{\mathbf{q}}Q_{-\mathbf{q}}\end{split}
\label{eq:H}
\end{equation}
where
\begin{equation}
Q_{\mathbf{q}}=\sum_{\mathbf{k},\alpha}f_{\mathbf{k},\mathbf{k+q}}c_{\alpha\mathbf{k}+\mathbf{q}}^{\dagger}c_{\alpha\mathbf{k}}.
\end{equation}
Here, $c_{\alpha\mathbf{k}}$ annihilates an electron of flavor $\alpha=1,\dots,N$
with momentum $\mathbf{k}$. $\varepsilon_{\mathbf{k}}$ is the electron's
energy dispersion, and $\mu$ is the chemical potential. Physically,
$N=2$ (for the two spin flavors), but we will keep $N$ general --
for some purposes it will be useful to consider the limit of large
$N$. The nematic form factor $f_{\mathbf{k},\mathbf{k+q}}$ satisfies
$f_{\mathbf{k},\mathbf{k+q}}=-f_{\mathcal{R}_{\pi/2}\mathbf{k},\mathcal{R}_{\pi/2}(\mathbf{k+q})}$,
where $\mathcal{R}_{\pi/2}$ is a rotation by $\pi/2$ around the
axis perpendicular to the plane. The interaction is written as $U_{\mathbf{q}}=\lambda^{2}D_{0,\mathbf{q}}/N$,
where $\lambda$ is the coupling constant (the normalization by $N$
enables us to define the large-$N$ limit), and we choose $D_{0,\mathbf{q}}=1/[r_{0}+2(2-\cos q_{x}-\cos q_{y})]$
(the lattice constant is set to unity) \footnote{The expression for the bare propagator $D_{0,\mathbf{q}}^{-1}=r_{0}+2\left(2-\cos q_{x}-\cos q_{y}\right)$
has lattice effects included, i.e., $D_{0,\mathbf{q+G}}^{-1}=D_{0,\mathbf{q}}^{-1}$,
where $\mathbf{G}$ is any reciprocal lattice wavevector. Later on
when we study transport properties of a compensated metal, we will
instead use the continuous expression: $D_{0,\mathbf{q}}^{-1}=r_{0}+\mathbf{q}^{2}$.}. $r_{0}$ is a tuning parameter used to approach the QCP.

Following the standard procedure, we use a Hubbard-Stratonovich transformation
with a real scalar field $\phi$ to decouple the interaction term,
obtaining the Lagrangian
\begin{equation}
L=L_{0}+\frac{1}{2}\sum_{\mathbf{q}}D_{0,\mathbf{q}}^{-1}|\phi_{\mathbf{q}}|^{2}+\frac{\lambda}{\sqrt{N}}\sum_{\mathbf{q}}\phi_{-\mathbf{q}}Q_{\mathbf{q}},\label{eq:L}
\end{equation}
where $L_{0}$ is a free fermion Lagrangian corresponding to the first
term in Eq. (\ref{eq:H}). The field $\phi$ can be thought of as
describing the spatial and temporal fluctuations of the nematic order
parameter; in the nematic phase, $\langle\phi\rangle\ne0$.

The low-energy continuum field theory governing the critical behavior
has been studied extensively \cite{metlitski10a}. Here we briefly
summarize the physical picture in the vicinity of the QCP:
\begin{enumerate}
\item The nematic fluctuations become overdamped as a result of their coupling
to electrons near the Fermi surface. To leading order in $\lambda$,
$\phi$ acquires a self-energy of the form: $\Pi(\mathbf{q},i\nu_{n})\propto f_{\mathbf{q},\mathbf{q}}^{2}\frac{\gamma|\nu_{n}|}{|\mathbf{q}|}$,
where $\nu_{n}=2\pi Tn$ is the bosonic Matsubara frequency and $\gamma\propto\lambda^{2}\varepsilon_{F}^{-2}$
is the Landau-damping coefficient ($\varepsilon_{F}$ is the Fermi
energy). This leads to a quantum critical scaling with a dynamical
critical exponent $z=3$.
\item The feedback of the Landau-damped critical fluctuations on the electrons
near the Fermi surface leads, to one-loop order, to an electronic
self-energy $\Sigma(\mathbf{k},i\omega_{n})\propto f_{\mathbf{k},\mathbf{k}}^{2}\varepsilon_{F}N^{-1}|\gamma\omega_{n}|^{2/3}$,
where $\omega_{n}=2\pi T\left(n+\frac{1}{2}\right)$ is the fermionic
Matsubara frequency. Below an energy scale $\Omega_{\text{NFL}}\propto\lambda^{4}\varepsilon_{F}^{-1}N^{-3}$,
the self-energy becomes dominant over the bare $i\omega_{n}$ term
in the electron propagator, and the electrons become strongly incoherent.
This regime is currently not well understood theoretically. Within
a large--$N$ expansion, terms which are naively of arbitrarily high
order in $1/N$ turn out to be equally important as the leading ones
in this regime. Superconducting fluctuations are also expected to
become strong at the same energy scale, implying a nematic-mediated
superconducting transition temperature $T_{c}\sim\Omega_{\text{NFL}}$.
A dome-shaped superconducting phase with a maximum $T_{c}$ near the
nematic QCP was indeed found in quantum Monte Carlo simulations \cite{lederer17,Berg:2018aa}.
\end{enumerate}
These considerations suggest that in the weak coupling limit $\lambda^{2}\ll\varepsilon_{F}$
or in the large--$N$ limit, there is a broad regime of temperatures
$\Omega_{\text{NFL}}\ll T\ll\varepsilon_{F}$ above the QCP where
the system can be described in terms of Landau-damped nematic fluctuations
coupled to \emph{coherent} electrons. Evidence for the existence of
such a regime, even at moderate values of the coupling constant, has
been found in numerical simulations \cite{schattner16}. In this
regime, the use of a kinetic equation approach for computing transport
properties is justified, with the effects of the scattering off critical
fluctuations incorporated in the collision integral.

\section{Method}

In the previous section, we argued that above an energy scale $\Omega_{\text{NFL}}$,
the normal state transport is governed by a kinetic (Boltzmann) equation.
Deriving the kinetic equation requires special care: while the Lagrangian
in Eq. (\ref{eq:L}) describes electrons coupled to a fluctuating
boson $\phi_{\mathbf{q}}$, it is important to keep track of the fact
that the bosonic degrees of freedom do not act as a ``bath'' for
the electrons; rather, they are collective modes of the same electron
fluid [as is manifest in the original Hamiltonian in Eq. (\ref{eq:H})].
Therefore, in the absence of an external momentum relaxation mechanism
(such as umklapp or impurity scattering), the total electronic momentum
is conserved.

In this section, we derive the kinetic equation based on the memory
matrix method. This method has been applied widely for studying transport phenomena; see, e.g., Refs. \cite{kadanoff63,forster,gotze72,Rosch2000,Shimshoni2003,Rosch2007,Mahajan2013,Lucas2015,hartnoll18}. It has the
advantage that the ``collision term'' in the kinetic equation is
formulated as a correlation function at equilibrium, that can be computed
using standard perturbative techniques.

The dc resistivity can be calculated by taking the zero-frequency
limit of the real part of optical conductivity, i.e., $\rho^{-1}=\lim_{\Omega\rightarrow0}\text{Re}\sigma(\Omega)$.
Within linear response, the optical conductivity is given by the retarded
current-current correlation function: $\sigma(\Omega)=\frac{1}{i\Omega}\left[G_{J_{x}J_{x}}^{R}\left(\Omega\right)-G_{J_{x}J_{x}}^{R}\left(0\right)\right]$,
where $\mathbf{J}$ is the electrical current operator corresponding
to Eq. (\ref{eq:H}), obtained through replacing $c_{\alpha\mathbf{k}}\rightarrow c_{\alpha\mathbf{k}+e\mathbf{A}}$
and taking $\mathbf{J}=\partial H/\partial\mathbf{A}$. To lowest
order in $\lambda$ or in $1/N$, $\mathbf{J}\approx\sum_{\mathbf{k,\alpha}}\mathbf{v}_{\mathbf{k}}c_{\alpha\mathbf{k}}^{\dagger}c_{\alpha\mathbf{k}}$
\footnote{Since the interaction is momentum dependent, it also contributes a
term to the current operator. However, this term is subleading in
$\lambda$ and in $1/N$, and will be neglected.}.

Within the memory matrix method, the dynamics is projected onto a
set of ``slow operators'' that are nearly conserved by the Hamiltonian.
We briefly review the method in Appendix \ref{subsec:General-formalism}.
In our model, the operators $n_{\alpha\mathbf{k}}=c_{\alpha\mathbf{k}}^{\dagger}c_{\alpha\mathbf{k}}$
are nearly-conserved in the limit of either small $\lambda$ or large
$N$ \footnote{The precise meaning of these operators being nearly conserved is elaborated
on in Appendix \ref{subsec:General-formalism}.}. The optical conductivity can then be cast in the form:
\begin{equation}
\sigma(\Omega)=\sum_{\alpha,\beta,\mathbf{k},\mathbf{k'}}\chi_{J_{x},\alpha\mathbf{k}}\left(\frac{1}{M(\Omega)-i\Omega\chi}\right)_{\alpha\mathbf{k},\mathbf{\beta k'}}\chi_{\beta\mathbf{k}',J_{x}}\label{eq:mm_conductivity}
\end{equation}
where $\chi_{J_{x},\alpha\mathbf{k}}\equiv\int_{0}^{\beta}\mathrm{d}\tau\langle J_{x}(\tau)n_{\alpha\mathbf{k}}(0)\rangle$
and $\chi_{\alpha\mathbf{k},\beta\mathbf{k'}}\equiv\int_{0}^{\beta}\mathrm{d}\tau\left[\langle n_{\alpha\mathbf{k}}(\tau)n_{\beta\mathbf{k'}}(0)\rangle-\langle n_{\alpha\mathbf{k}}\rangle\langle n_{\beta\mathbf{k'}}\rangle\right]$
are thermodynamic susceptibilities. The dynamical properties are encoded
in the structure $\left[M(\Omega)-i\Omega\chi\right]^{-1}$, where
$M_{\alpha\mathbf{k},\beta\mathbf{k}'}(\Omega)$ is the ``memory matrix''.
To leading order in $1/N$, the memory matrix is given by:
\[
M_{\alpha\mathbf{k},\beta\mathbf{k'}}(\Omega)=\frac{1}{i\Omega}\left[G_{\dot{n}_{\alpha\mathbf{k}}\dot{n}_{\beta\mathbf{k}'}}^{R}(\Omega)-G_{\dot{n}_{\alpha\mathbf{k}}\dot{n}_{\beta\mathbf{k}'}}^{R}(0)\right]
\]
where $\dot{n}_{\alpha\mathbf{k}}\equiv i[H,n_{\alpha\mathbf{k}}]$,
and the correlation function $G_{\dot{n}_{\alpha\mathbf{k}}\dot{n}_{\beta\mathbf{k}'}}^{R}(\Omega)$
is computed to order $1/N^{0}$. To perform the computations, it is
convenient to write the equivalent Hamiltonian to Eq. (\ref{eq:L})
\footnote{To do this, one introduces a field $\pi(\mathbf{r})$ which is the
conjugate momentum of $\phi(\mathbf{r})$. The Hamilontian then contains
a term $\pi^{2}(\mathbf{r})/2m$, where the ``mass'' $m$ is taken
to $0$ at the end of the calculation.}. $\dot{n}_{\alpha\mathbf{k}}$ is then given by

\begin{equation}
\begin{split}\dot{n}_{\alpha\mathbf{k}} & =\frac{i\lambda}{\sqrt{N}}\sum_{\mathbf{q}}\phi_{\mathbf{q}}\left(f_{\mathbf{k},\mathbf{k}-\mathbf{q}}c_{\alpha\mathbf{k}}^{\dagger}c_{\alpha\mathbf{k}-\mathbf{q}}-f_{\mathbf{k},\mathbf{k}+\mathbf{q}}c_{\alpha\mathbf{k}+\mathbf{q}}^{\dagger}c_{\alpha\mathbf{k}}\right).\end{split}
\end{equation}
A diagrammatic representation of this operator is shown in Figure
\ref{fig:mm_diagrams}(a).

The real part of the memory matrix describes the scattering rate between
different momentum states. Its eigenvalues are non-negative, and describes
the decay rate of various collective modes on the Fermi surface. If
there is a conserved mode, for example the total electron number,
then the corresponding eigenvalue is zero. The memory matrix is related
to the collision integral of the Boltzmann equation; as we demonstrate
in Appendix \ref{sec:boltzmann}, the memory matrix formalism coincides
with the standard Boltzmann kinetic equation away from the critical
point (where the interactions between electrons can be treated as
static), but incorporates more complicated scattering processes important
near the QCP.

Below we first construct the memory matrix in the absence of any current-relaxing
mechanisms, and study its structure and conservation laws. Next we
will study how the memory matrix and the transport properties are
affected by different scattering mechanisms.

\subsection{Feynman diagrams and conservation laws}

\begin{figure}
\includegraphics[width=0.8\linewidth]{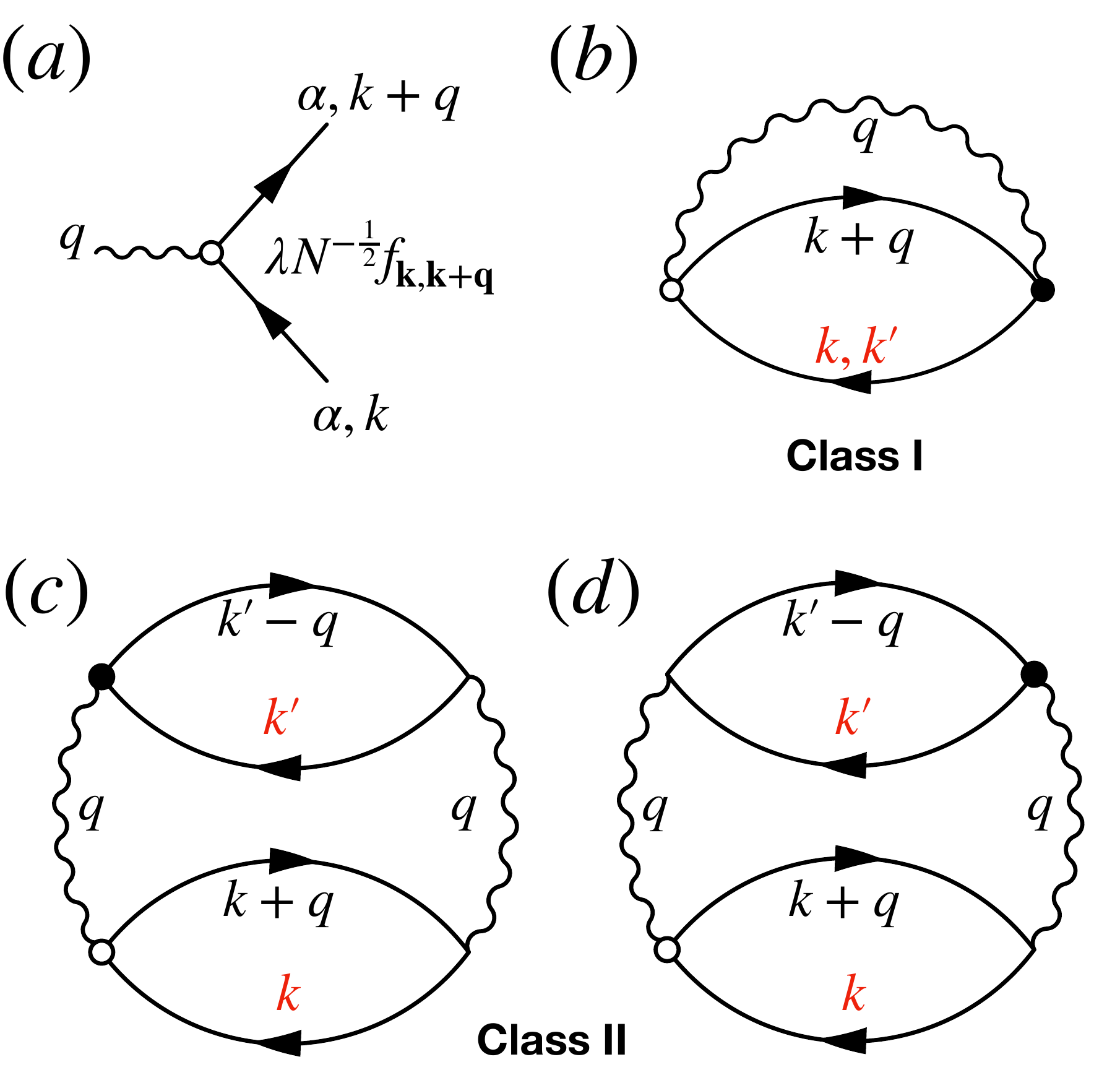}

\caption{\label{fig:mm_diagrams}(a) Diagrammatic representation of $\dot{n}_{\alpha\mathbf{k}}(\Omega)$.
The open/closed circles denote inflow/outflow of an external frequency
$\Omega$. The vertex function changes sign depending on if the $\alpha\mathbf{k}$
state is an initial (final) state of the scattering. (b-d) Two class
of diagrams for the memory matrix $M_{\alpha\mathbf{k},\beta\mathbf{k'}}\left(\Omega\right)$.
The red colored labels correspond to external momenta $\mathbf{k}$
and $\mathbf{k'}$. For convenience we omitted the flavor index on
the fermionic Green's functions, and the summation over internal frequencies
is implicit.}
\end{figure}
At temperatures $T\gg\Omega_{\text{NFL}}$, we compute $M_{\mathbf{\alpha k},\mathbf{\beta k}'}$
in a $1/N$ expansion. Depending on the placement of the two momentum
states $\mathbf{k}$ and $\mathbf{k}'$, there are two classes of
Feynman diagrams that contribute to leading order, as illustrated
in Figure \ref{fig:mm_diagrams}(b-d). In these diagrams, the nematic
propagator includes the one-loop self energy: $D^{-1}(\mathbf{q},i\nu_{n})=D_{0}^{-1}-\Pi(\mathbf{q},i\nu_{n})$,
where
\begin{equation}
\Pi(\mathbf{q},i\nu_{n})=-\lambda^{2}T\sum_{\omega_{m},\mathbf{k}}f_{\mathbf{k},\mathbf{k}+\mathbf{q}}^{2}G_{0}(\mathbf{k},i\omega_{m})G_{0}(\mathbf{k}+\mathbf{q},i\omega_{m}+i\nu_{n}),
\end{equation}
whereas the fermion propagators $G_{0}(\mathbf{k},i\omega_{n})=1/(i\omega_{n}-\varepsilon_{\mathbf{k}})$
are the bare ones. At low frequencies, we write $D^{-1}(\mathbf{q},i\nu_{n})=r_{\mathbf{q}}+\gamma_{\mathbf{q}}|\nu_{n}|$,
where $r_{\mathbf{q}}$ includes the renormalization effects from
$\Pi(\mathbf{q},i\nu_{n}=0)$, and $\gamma_{\mathbf{q}}=\pi\lambda^{2}\sum_{\mathbf{k}}f_{\mathbf{k},\mathbf{k}+\mathbf{q}}^{2}\delta(\varepsilon_{\mathbf{k}})\delta(\varepsilon_{\mathbf{k}+\mathbf{q}})$.

A detailed derivation of the expressions corresponding to these Feynamn
diagrams is presented in Appendix \ref{subsec:mm_QCP}. For class
I {[}Fig. \ref{fig:mm_diagrams}(b){]}, we obtain:

\[
\begin{split}M_{\alpha\mathbf{k},\beta\mathbf{k'}}^{(1)}(i\Omega_{n})= & \delta_{\alpha\beta}\frac{\lambda^{2}T}{N\Omega_{n}}\sum_{\mathbf{q},\nu_{n}}D_{\mathbf{q},\nu_{n}+\Omega_{n}}\\
\times & \sum_{\zeta=\pm1}\left(\delta_{\mathbf{k}'-\mathbf{k},\zeta\mathbf{q}}-\delta_{\mathbf{kk}'}\right)f_{\mathbf{k},\mathbf{k+\zeta q}}^{2}R_{\mathbf{k},\mathbf{k+\zeta q},\zeta\nu_{n}}
\end{split}
\]
where $R_{\mathbf{k},\mathbf{k+q},\nu_{n}}$ is the polarization bubble
summed over the internal fermionic Matsubara frequencies:
\begin{equation}
R_{\mathbf{k},\mathbf{k+q},\nu_{n}}=\frac{n_{F}\left(\varepsilon_{\mathbf{k+q}}\right)-n_{F}\left(\varepsilon_{\mathbf{k}}\right)}{\varepsilon_{\mathbf{k}}-\varepsilon_{\mathbf{k+q}}+i\nu_{n}}.\label{eq:phbubble_uncontracted}
\end{equation}
Here, $n_{F}(\varepsilon)$ is the Fermi function. Similarly, the
expression for class II diagrams is
\begin{equation}
\begin{split}M_{\alpha\mathbf{k},\beta\mathbf{k'}}^{(2)}(i\Omega_{n}) & =-\frac{\lambda^{4}T}{N^{2}\Omega_{n}}\sum_{\mathbf{q},\nu_{n}}D_{\mathbf{q},\nu_{n}}D_{\mathbf{q},\nu_{n}+\Omega_{n}}\\
 & \times\sum_{\zeta\zeta'=\pm1}\zeta\zeta'f_{\mathbf{k},\mathbf{k}+\zeta\mathbf{q}}^{2}f_{\mathbf{k}',\mathbf{k}'+\zeta'\mathbf{q}}^{2}R_{\mathbf{k},\mathbf{k}+\zeta\mathbf{q},\zeta\nu_{n}}\\
 & \times\left[R_{\mathbf{k}',\mathbf{k}'+\zeta'\mathbf{q},\zeta'\nu_{n}}-R_{\mathbf{k}',\mathbf{k}'+\zeta'\mathbf{q},\zeta'\left(\nu_{n}+\Omega_{n}\right)}\right]
\end{split}
\end{equation}
For simplicity, we have omitted the static part $G_{\dot{n}_{\alpha\mathbf{k}}\dot{n}_{\beta\mathbf{k}'}}^{R}(0)$
in the expressions. However, they are always subtracted in later calculations.
The memory matrix are expressed in Matsubara frequencies. To obtain
real-time dynamics, we perform an analytic continuation $i\Omega_{n}\rightarrow\Omega+i\delta$.

At high frequencies, these terms in the memory matrix reproduce the
standard Feynman diagrams describing contributions to optical conductivity,
namely Maki-Thompson (MT), Density of States (DOS) and Aslamazov-Larkin
(AL) diagrams. This is shown in Appendix \ref{sec:Connection-to-diagrams}.
In particular, MT and DOS diagrams combine to give class I diagram,
corresponding to the $\delta_{\mathbf{k'-k},\mathbf{q}}$ and $\delta_{\mathbf{kk'}}$
terms respectively. AL is equivalent to class II diagrams.

In the dc limit, our memory matrix approach is equivalent to the quantum
Boltzmann equation discussed in the Kadanoff-Baym-Keldysh framework
\cite{kadanoff_book,rammer1986,kamenev11}, and that conservation
laws are explicitly built in. We discuss the conservation laws and
their implications for the structure of the memory matrix. One can
easily check that $\sum_{\alpha\mathbf{k}}M_{\alpha\mathbf{k},\beta\mathbf{k'}}=0$,
corresponding to electron number conservation. Note that the two class
of diagrams separately conserve particle number. In Appendix \ref{sec:Momentum-conservation}
we show that in the absence of impurity and umklapp scattering, the
total electronic momentum is conserved: $\sum_{\alpha\mathbf{k}}\mathbf{k}M_{\alpha\mathbf{k},\beta\mathbf{k'}}=0$.
Momentum conservation crucially relies on the fact that the nematic
fluctuations gain their dynamics only as a result of their coupling
to the electrons. As a result, they do not act as a ``sink'' for
the total electronic momentum. It is worth noting that momentum is
conserved only when the two classes of diagrams are combined, but
not for each class separately.

\subsection{Low temperature and dc limit}

\begin{figure}
\includegraphics[width=0.9\linewidth]{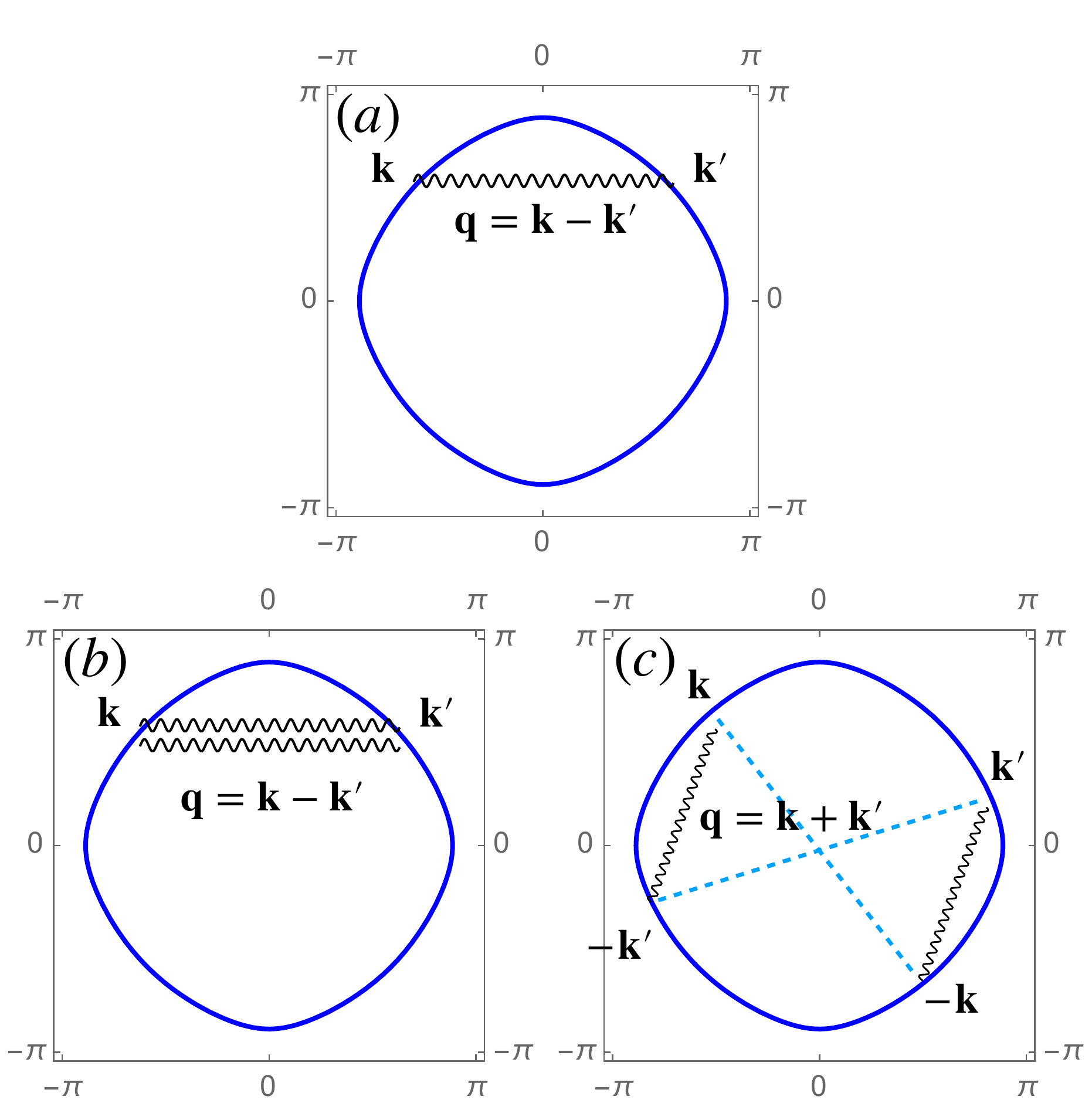}\caption{\label{fig:fs_projection}Three types of scattering processes when
the relevant momentum states are in the vicinity of the Fermi surface.
(a) class I diagram describes direct scattering between $\mathbf{k}$
and $\mathbf{k'}$ mediated by exchanging an nematic boson $\phi_{\mathbf{q}}$
with $\mathbf{q}=\mathbf{k'-k}$. (b) and (c) are two-electron scatterings
mediated by two nematic bosons, described by class I diagram. (b)
describes the momentum exchange scattering, where two electrons with
momentum $\mathbf{k}$ and $\mathbf{k'}$ exchange their momentum.
(c) describes the head-on scattering, where two electrons located
at $\mathbf{k}$ and $\mathbf{-k}$ are scattered into $\mathbf{k'}$
and $-\mathbf{k'}$. }
\end{figure}
At low temperatures compared to the Fermi energy $\varepsilon_{F}$,
the dominant scattering processes occur in the vicinity of the Fermi
surface. For two given momentum states on the Fermi surface $\mathbf{k}$
and $\mathbf{k'}$, there are three types of processes described by
class I and class II diagrams, as depicted in Figure \ref{fig:fs_projection}.
Class I diagram describes the direct scattering between $\mathbf{k}$
and $\mathbf{k'}$ states mediated by a nematic boson with momentum
$\mathbf{q}=\mathbf{k'}-\mathbf{k}$. Class II diagram describes two-electron
scattering which involves two additional momentum states $\mathbf{k+q}$
and $\mathbf{k'\pm q}$. We consider a convex Fermi surface with inversion
symmetry \footnote{Complications due to non-convex Fermi surfaces \cite{maslov11} can
easily be incorporated within the same formalism, but will not be
considered here. }. Due to the one-dimensionality of the Fermi surface, there are only
two allowed scattering processes which satisfy that all four momentum
states are on the Fermi surface, as shown in Figs. \ref{fig:fs_projection}(b)
and (c). Panel (b) describes momentum exchange scattering, with $\mathbf{q=\mathbf{k'-k}}$.
Here two electrons at $\mathbf{k}$ and $\mathbf{k'}$ are scattered
into each other. (c) is a head-on collision, with $\mathbf{q=k+k'}$.
Here two electrons initially at $\mathbf{k}$ and $\mathbf{-k}$ are
scattered into $\mathbf{k'}$ and $-\mathbf{k'}$.

Since the dominant contribution to the memory matrix comes from the
vicinity of the Fermi surface, we can approximate $R$ defined in
Eq. (\ref{eq:phbubble_uncontracted}) as follows:
\begin{equation}
R_{\mathbf{k},\mathbf{k'+q},\nu_{n}}\approx-\pi|\nu_{n}|\delta\left(\varepsilon_{\mathbf{k}}\right)\delta\left(\varepsilon_{\mathbf{k+q}}\right).
\end{equation}
This approximation is justified in Appendix \ref{sec:Low-temperature-and}.
After performing summation over the bosonic Matsubara frequencies,
we can simplify the memory matrix to be:

\begin{equation}
\begin{split}M_{\alpha\mathbf{k},\beta\mathbf{k'}}^{(1)} & =\delta_{\alpha\alpha'}\frac{2\pi\lambda^{2}}{N}\sum_{\mathbf{q}}V_{\mathbf{q}}(T)\left(\delta_{\mathbf{kk}'}-\delta_{\mathbf{k}'-\mathbf{k},\mathbf{q}}\right)\\
 & \times f_{\mathbf{k},\mathbf{k+q}}^{2}\delta\left(\varepsilon_{\mathbf{k}}\right)\delta\left(\varepsilon_{\mathbf{k+q}}\right),
\end{split}
\label{eq:mm_classI}
\end{equation}
and
\begin{equation}
\begin{split}M_{\alpha\mathbf{k},\beta\mathbf{k'}}^{(2)} & =-\frac{2\pi^{2}\lambda^{4}}{N^{2}}\sum_{\mathbf{q}}\frac{V_{\mathbf{q}}(T)}{\gamma_{\mathbf{q}}}f_{\mathbf{k},\mathbf{k}+\mathbf{q}}^{2}\delta\left(\varepsilon_{\mathbf{k}}\right)\delta\left(\varepsilon_{\mathbf{k+q}}\right)\\
 & \times\sum_{\zeta'=\pm1}\zeta'f_{\mathbf{k}',\mathbf{k}'+\zeta'\mathbf{q}}^{2}\delta\left(\varepsilon_{\mathbf{k'}}\right)\delta\left(\varepsilon_{\mathbf{k'+\zeta'q}}\right),
\end{split}
\label{eq:mm_classII}
\end{equation}
where we have defined
\begin{equation}
V_{\mathbf{q}}(T)\equiv\int_{-\infty}^{\infty}\frac{\mathrm{d}\omega}{\pi}\omega\text{Im}D_{\mathbf{q},\omega}\left(-\frac{\partial n_{B}(\omega)}{\partial\omega}\right),\label{eq:Vq}
\end{equation}
The imaginary part of the nematic propagator is $\text{Im}D_{\mathbf{q},\omega}=\gamma_{\mathbf{q}}\omega/(r_{\mathbf{q}}^{2}+\gamma_{\mathbf{q}}^{2}\omega^{2})$,
and $n_{B}(\omega)$ is the Bose-Einstein distribution function.

In the case of a single convex Fermi sheet with no umklapp scattering,
there are additional conservation laws that emerge at low temperature.
In particular, all the odd-parity deformations of the Fermi surface
are quasi-conserved, in the sense that the lifetimes of these modes
are parametrically larger than those of the even-parity modes \cite{gurzhi95,maslov11,ledwidth17}.
This is because the only possible collisions on a one dimensional
convex Fermi surface are either forward scattering or head-on collisions,
and in both cases the odd moments of the distribution function do
not change. These approximate conservation laws are manifest in the
structure of the memory matrix, as demonstrated in Appendix \ref{sec:Harmonic-basis-and}.

\section{Current relaxation mechanisms and dc resistivity}

We now turn to discuss how the quantum critical fluctuations manifest
themselves in the temperature dependence of the dc resistivity, $\rho(T)$,
through different current relaxation mechanisms. In a metal with a
generic electron density, the electrical current cannot relax completely
(and hence $\rho=0$) unless momentum conservation is violated. Having
a non-zero resistivity therefore relies crucially on the momentum
relaxation mechanism, either through impurity or umklapp scattering.
In the special case of a compensated metal with an equal density of
electrons and holes, the resistivity is finite even if the total momentum
is conserved.

We note that in our calculation, the temperature dependence comes
from the scattering rates, encoded in the memory matrix. The thermodynamic
susceptibilities, $\chi_{J_{x},\alpha\mathbf{k}}$ and $\chi_{\alpha\mathbf{k},\beta\mathbf{k'}}$,
defined in Eq. (\ref{eq:mm_conductivity}), can be regarded as temperature
independent. While this is not true for the Ising-nematic (B$_{1g}$)
channel, where the thermodynamic susceptibility diverges at the QCP,
this divergence does not affect the transport properties, since the
current operator (as well as any other odd-partiy deformation of the
Fermi surface) is orthogonal to the nematic order parameter.

We will present our results for $\rho(T)$ in dimensionless units.
The temperature is rescaled by the energy scale set by the Landau
damping term: $\Omega_{L}\equiv\varepsilon_{F}^{2}\lambda^{-2}$.
As we discuss in Appendix \ref{subsec:mm_QCP}, the natural scale
for the resistivity in our problem is $\rho_{L}\equiv\frac{\hbar}{e^{2}}\frac{\lambda^{4}}{\varepsilon_{F}^{2}}\frac{1}{N^{2}}$.
Note that in the large-$N$ or weak coupling limits, where our approach
is valid, $\rho_{L}$ is always much smaller than the quantum of resistance
$\hbar/e^{2}$.

\subsection{Impurity scattering}

We consider quenched disorder, modeled by a random potential: $H_{\text{imp}}=\frac{1}{\sqrt{N}}\sum_{\alpha\mathbf{r}}V_{\text{imp}}(\mathbf{r})c_{\alpha\mathbf{r}}^{\dagger}c_{\alpha\mathbf{r}}$,
where $V_{\text{imp}}(\mathbf{r})$ has the following properties:
$\overline{V_{\text{imp}}(\mathbf{r})}=0$ and $\overline{V_{\text{imp}}(\mathbf{r})V_{\text{imp}}(\mathbf{r'})}=g_{\text{imp}}^{2}\delta\left(\mathbf{r-r'}\right)$,
where $\overline{\cdots}$ denotes disorder averaging. The time-derivative
of the electron occupation number is given by $(\dot{n}_{\alpha\mathbf{k}})_{\text{imp}}=\frac{i}{\sqrt{N}}\sum_{\mathbf{q}}V_{\text{imp},\mathbf{q}}\left(c_{\alpha\mathbf{k}}^{\dagger}c_{\alpha\mathbf{k+q}}-c_{\alpha\mathbf{k-q}}^{\dagger}c_{\alpha\mathbf{k}}\right)$.
As a result, for weak disorder strength, the leading correction to
the memory matrix is given by
\begin{equation}
\begin{split}M_{\alpha\mathbf{k},\beta\mathbf{k'}}^{\text{imp}}(i\Omega_{n}) & =\delta_{\alpha\beta}\frac{1}{N\Omega_{n}}\sum_{\mathbf{q},i\nu_{n}}D_{\mathbf{q},\nu_{n}}^{\text{imp}}\sum_{\zeta=\pm1}f_{\mathbf{k},\mathbf{k+\zeta q}}^{2}\\
 & \times\left(\delta_{\mathbf{k}'-\mathbf{k},\zeta\mathbf{q}}-\delta_{\mathbf{kk}'}\right)R_{\mathbf{k},\mathbf{k+\zeta q},\zeta\nu_{n}+\zeta\Omega_{n},}
\end{split}
\end{equation}
where the correlator of the disorder potential is static and momentum-independent,
i.e., $D_{\mathbf{q},\nu_{n}}^{\text{imp}}=g_{\text{imp}}^{2}\delta_{\nu_{n},0}$.
To leading order in impurity strength, we neglect the cross-terms
involving both impurity and quantum critical scattering \footnote{The cross term is negligible as long as the disorder potential does
not couple linearly to the nematic order parameter. Such coupling
is generated at higher order in disorder, but we assume it to be small.
Transport in the presence of ``random field'' disorder that couples
linearly to the nematic order parameter was considererd in Ref. \cite{hartnoll14}.
At sufficiently low temperature, this type of disorder is likely to
qualitatively modify the properties of the quantum critical point,
as well as its transport properties.}. At low temperatures, $T\ll\varepsilon_{F}$, we project the processes
onto the Fermi surface, and approximate $R(\mathbf{k},\mathbf{k+q},i\Omega_{n})\approx-\pi|\Omega_{n}|\delta\left(\varepsilon_{\mathbf{k}}\right)\delta\left(\varepsilon_{\mathbf{k+q}}\right)$.
As a result, in the dc limit,
\begin{equation}
M_{\alpha\mathbf{k},\beta\mathbf{k'}}^{\text{imp}}=\delta_{\alpha\beta}\frac{2\pi g_{\text{imp}}^{2}}{N}\sum_{\mathbf{q}}\left(\delta_{\mathbf{kk'}}-\delta_{\mathbf{k'-k,q}}\right)\delta\left(\varepsilon_{\mathbf{k}}\right)\delta\left(\varepsilon_{\mathbf{k+q}}\right).
\end{equation}
One can verify that momentum is no longer conserved under impurity
scattering, by observing that
\begin{equation}
\sum_{\alpha\mathbf{k}}\mathbf{k}M_{\alpha\mathbf{k},\beta\mathbf{k'}}^{\text{imp}}=-2\pi g_{\text{imp}}^{2}\sum_{\mathbf{q}}\mathbf{q}\delta\left(\varepsilon_{\mathbf{k}'}\right)\delta\left(\varepsilon_{\mathbf{k'+q}}\right).
\end{equation}
This expression does not vanish for a general value of $\mathbf{k'}$,
due to the asymmetry of momentum states $\mathbf{k'\pm q}$.

We first show that when there is only a single convex electron Fermi
surface, quantum critical scattering does not contribute to the dc
resistivity within our lowest-order approximation in $T/\varepsilon_{F}$.
In this case, $\rho\approx\rho_{0}\propto g^{2}\nu_{F}$ coming from
impurity scattering, where $\nu_{F}$ is the density of states at
the Fermi level. To see this, we work in the basis of Fermi surface
harmonics, $e^{in\theta_{\mathbf{k}}}$, where $\theta_{\mathbf{k}}$
is the angle between a point $\mathbf{k}$ on the Fermi surface and
the $x$ axis. Since our problem is inversion-symmetric on average,
$M_{\alpha n,\beta m}^{\text{imp}}$ is non zero only if $m$ and
$n$ have the same parity. Electron-electron scattering on a single,
convex Fermi surface conserves all the odd-parity modes \cite{gurzhi95,maslov11,ledwidth17}
(See Appendix \ref{sec:Harmonic-basis-and}). Since the electrical
current is parity-odd, it is completely decoupled from the quantum
critical scattering to lowest order in $T/\varepsilon_{F}$. Hence,
to lowest order in $T/\varepsilon_{F}$, $\rho$ is temperature independent.
Scattering processes away from the Fermi surface can give rise to
$\rho(T)\sim T^{2+\alpha},$ with $\alpha=4/3$, as argued in Ref.
\cite{maslov11}.

Next, we study the case of multiple Fermi sheets. Different sheets
generically have different energy dispersions and are separated in
momentum space. The small-momentum nematic fluctuations induces intra-sheet
scattering, while impurity scattering can be either intra- or inter-sheet.
We write the electrical current operator as $\mathbf{J}=\sum_{i\mathbf{k}}\mathbf{v}_{i,\mathbf{k}}c_{i,\mathbf{k}}^{\dagger}c_{i\mathbf{k}}$,
where $i$ is the sheet index. Due to differences in the Fermi velocities
in each Fermi sheet, the current is not conserved by nematic scattering
(unlike total momentum). As a result, the resistivity acquires a temperature-dependent
correction $\sim T^{\alpha}$, whose exponent is determined by the
decay of odd-parity modes on the Fermi surface. Scaling arguments
suggest that nematic contrinbution to the current decay rate scales
as $\bar{q}^{2}\tau_{\text{sp}}^{-1}$, where $\bar{q}\sim T^{1/3}$
is the typical momentum transfer following quantum critical scaling,
and $\tau_{\text{sp}}^{-1}\sim T^{2/3}$ is the single-electron decay
rate. As a result, for a multi-band system subject to impurity scattering,
we expect that $\rho-\rho_{0}\propto T^{4/3}$ .

\begin{figure}
\includegraphics[width=0.96\linewidth]{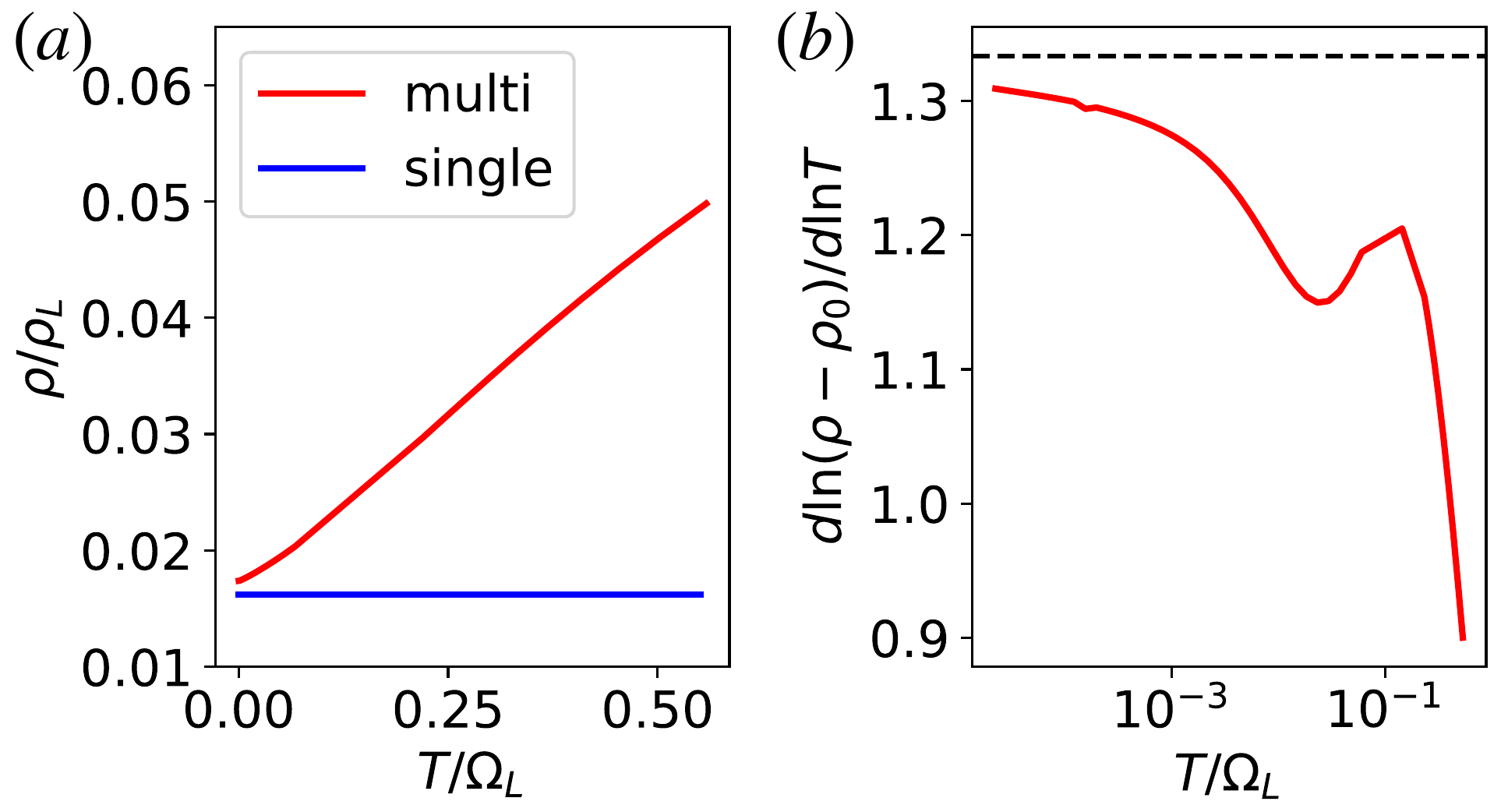}\caption{\label{fig:rho_impurity}(a) Resistivity due to impurity scattering
for a system with a single Fermi surface (blue) and multiple Fermi
sheets (red) cases. In (b) we plot $d\ln\left(\rho-\rho_{0}\right)/d\ln T$
for multi-sheet case to extract the power-law dependence on temperature.
The dashed line marks the value $4/3$. The dispersion used for one-band
case is $\varepsilon_{\mathbf{k}}=-2t\left(\cos k_{x}+\cos k_{y}\right)-4t'\cos k_{x}\cos k_{y}-\mu$
, with $t=1$, $t'=-0.3$ and $\mu=-1$. In the multi-sheet case,
we used three circularly-shaped Fermi pockets of equal size, with
two electron-like and one hole-like dispersions. Impurity scattering
strength is taken to be $g_{\text{imp}}=0.02\varepsilon_{F}$. The
nematic-electronic coupling strength is $\lambda^{2}\approx0.83\varepsilon_{F}$.
The Fermi energy is defined as $\varepsilon_{F}=\langle v_{F}k_{F}\rangle_{\text{FS}}$. }
\end{figure}

In Figure \ref{fig:rho_impurity}, we present a numerical calculation
of $\rho(T)$ for system with either a single Fermi surface or multiple
Fermi sheets, supporting the scaling argument discussed above. More
details of the numerical procedure are given in Appendix \ref{sec:Numerical-construction-of}.
In the single Fermi surface case, the resistivity is completely temperature-independent;
we expect that inclusion of scattering processes away from the Fermi
surface will give rise to a weak temperature dependence. In the multi-Fermi
sheet case, we find a substantial temperature dependence. In panel
(b) we show$\mathrm{d}\ln\left(\rho-\rho_{0}\right)/\mathrm{d}\ln T$
as a function of $T$ for the multi-sheet case. At low temperature,
the exponent approaches $4/3$, as expected from the qualitative analysis
above. However, at intermediate temperatures, the exponent drifts
below $4/3$ and can even exhibit sublinear behavior. The cause of
the drifting exponent will be discussed in Sec. \ref{subsec:Quasi-elastic-thermal-fluctuatio}.

\subsection{Umklapp scattering}

\begin{figure}
\includegraphics[width=0.8\linewidth]{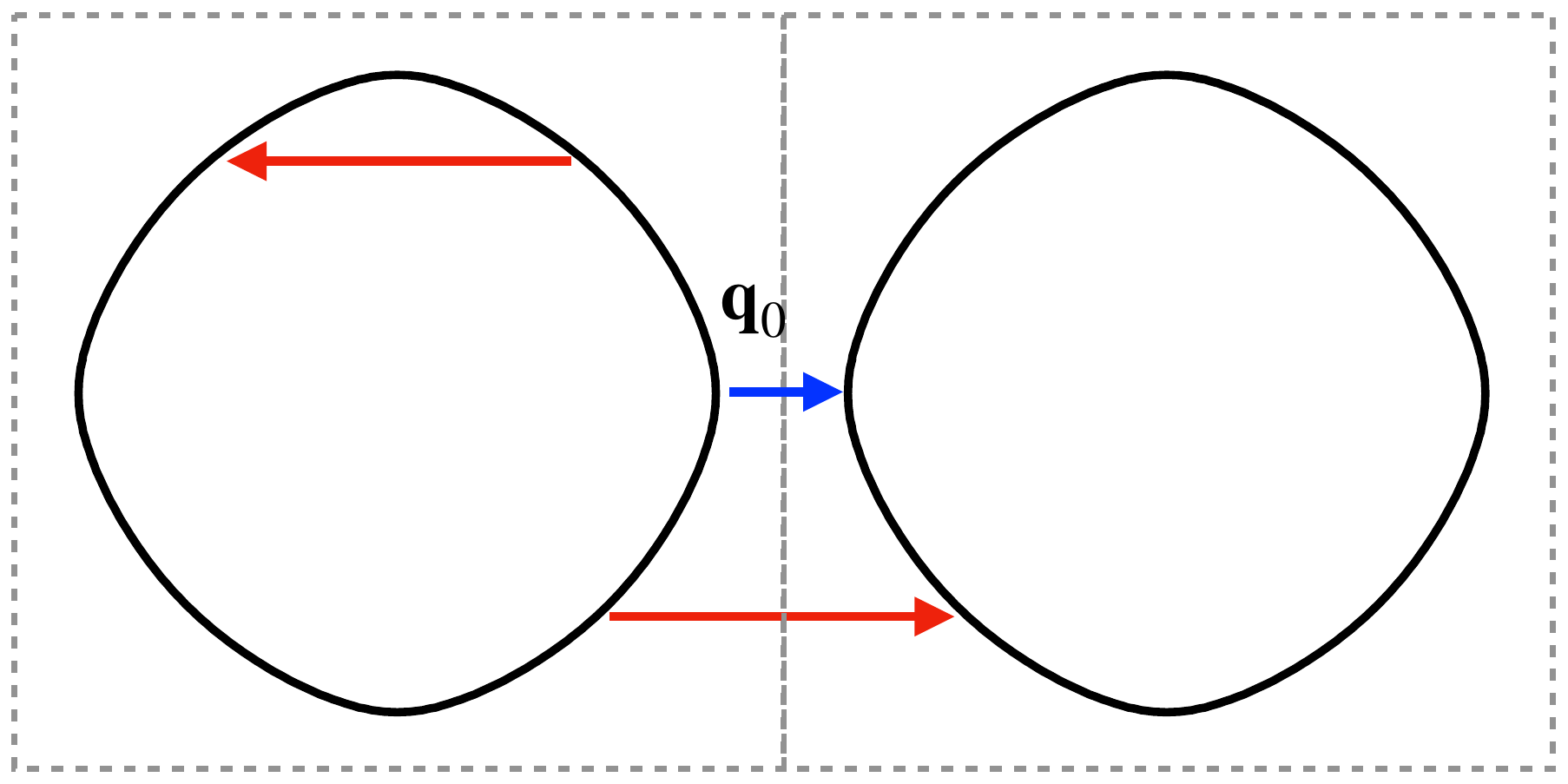}

\caption{\label{fig:umklapp_schem}Two-particle collision involving a Umklapp
process. One of the final states is a different Brillouin zone. Therefore,
the total momentum changes by the crystal momentum $\mathbf{G}$.
$\mathbf{q}_{0}$ is the minimal momentum transfer in an umklapp scattering
event.}
\end{figure}
In the presence of an underlying lattice, the electronic momentum
is only conserved up to a reciprocal lattice vector $\mathbf{G}$.
In an umklapp scattering process, two of the initial states and one
of the final states are on the Fermi surface in the first Brillouin
zone, while the other final state is in a neighboring Brollouin zone.
An umklapp scattering process on the Fermi surface involves a minimum
momentum transfer $\mathbf{q}_{0}$ that depends on the geometry of
the Fermi surface, as illustrated in Figure \ref{fig:umklapp_schem}.
Since the critical nematic fluctuations carry small momentum, they
cannot induce umklapp scattering at asymptotically low temperatures.
Therefore, below a characteristic temperature $T_{0}$ that depends
on $\mathbf{q}_{0}$, we expect the dc transport to be governed by
non-critical fluctuations. For $T\ll T_{0}$, the Fermi liquid behavior
$\rho(T)\sim T^{2}$ behavior should be recovered \cite{maslov11}.
On scaling grounds, we expect $T_{0}\sim|\mathbf{q}_{0}|^{z}$, where
$z$ is the dynamical critical exponent of the transition. At temperatures
$T>T_{0}$ , there are two main factors determining the transport
behavior: (1) As the typical wavevector for critical fluctuations
$|\mathbf{q}|\sim T^{1/z}>|\mathbf{q}_{0}|$, critical fluctuations
contribute to umklapp scattering and directly modify the dc transport
properties. (2) The umklapp processes also modify the spectrum of
the nematic critical fluctuations. In particular, they modify the
coefficient $\gamma_{\mathbf{q}}$ of the Landau damping term in the
bosonic self-energy, which depends on the angle between the Fermi
velocities at the points $\mathbf{k}$ and $\mathbf{k}+\mathbf{q}$
on the Fermi surface. This can lead to a breakdown of the $z=3$ quantum
critical scaling, which relies on the relation $\gamma_{\mathbf{q}}\sim1/|\mathbf{q}|$.

\begin{figure}
\includegraphics[width=1\linewidth]{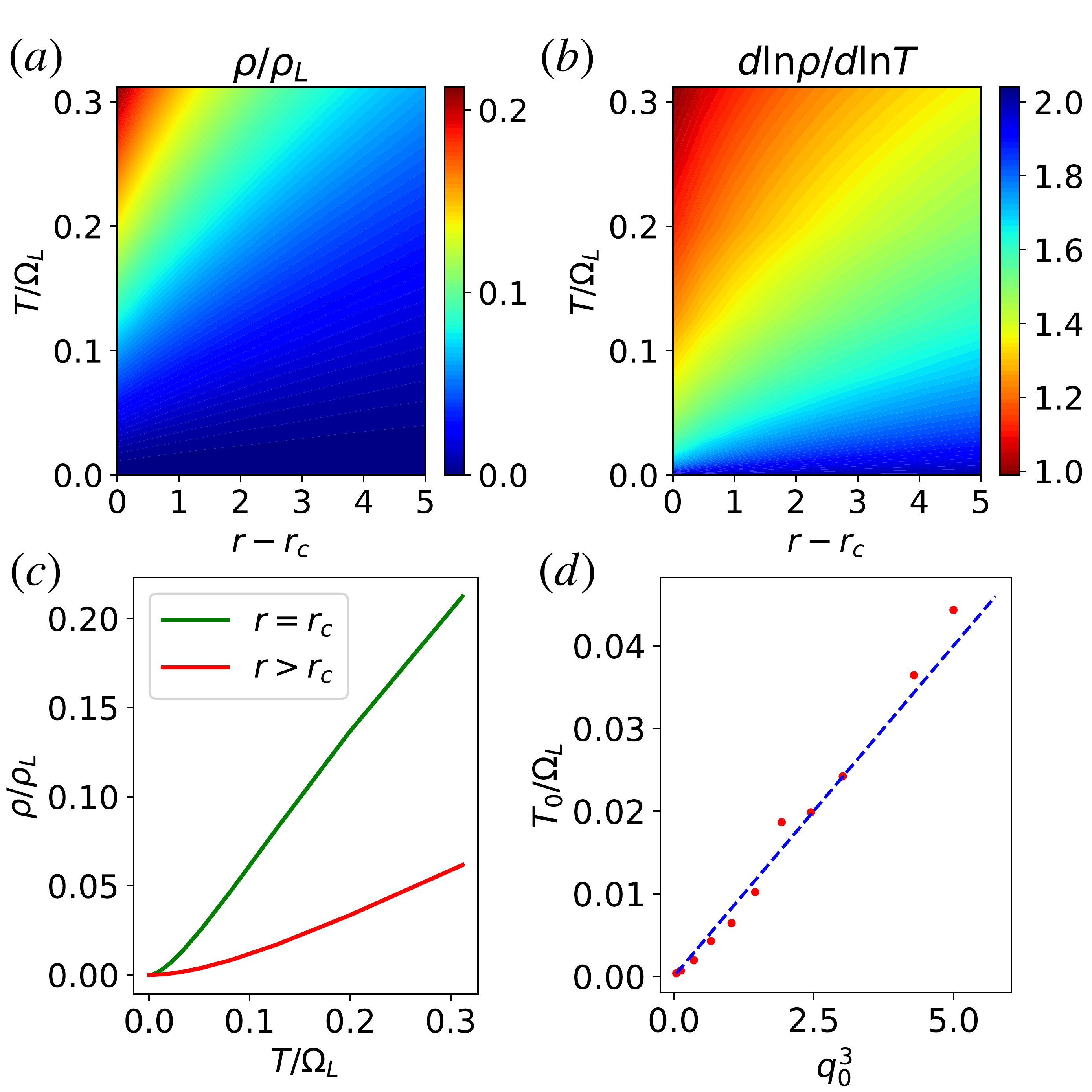}

\caption{\label{fig:rho_umklapp}(a) Resistivity from umklapp scattering at
and away near an Ising-nematic QCP, as a function of $r-r_{c}$ and
$T$. Here, we use the same band dispersion studied for the single-band
case for impurity scattering (see caption of Fig. \ref{fig:rho_impurity}).
The coupling strength $\lambda^{2}=1.88\varepsilon_{F}$. (b) $d\ln\rho/d\ln T$
as a function of $r-r_{c}$ and $T$. (c) Temperature cuts at the
QCP (green) and in the disordered state ($r>r_{c}$, red). (d) Crossover
temperature to Fermi-liquid $T^{2}$ behavior as a function of the
umklapp threshold wavevector $|\mathbf{q}_{0}|^{3}$. $\mathbf{q}_{0}$
was varied by changing the chemical potential $\mu$. We define $T_{0}$
as the temperature where $d\ln\rho/d\ln T=1.8$. }
\end{figure}

We study $\rho(T)$ for a model with a generic, large Fermi surface,
similar to the one shown in Fig. \ref{fig:umklapp_schem}. This is
done by numerically computing the memory matrix from Eqs. (\ref{eq:mm_classI},\ref{eq:mm_classII}),
including the effects of umklapp scattering, and inverting it to obtain
the dc conductivity according to Eq. (\ref{eq:mm_conductivity}). In
Fig. \ref{fig:rho_umklapp}(a) and (c), we present $\rho(T)$ at and
away from the Ising-nematic QCP, with a nematic correlation length
$\xi^{-2}(T)=T+r-r_{c}$. Here, we assume that the at $r=r_{c}$,
there is a ``thermal mass'' assumed to be proportional to $T$. The
model parameters, listed in the caption of Fig. \ref{fig:rho_umklapp},
are chosen to be similar to the ones used in the quantum Monte Carlo
simulations in Ref. \cite{schattner16}. Figure \ref{fig:rho_umklapp}(b)
shows $d\ln\rho/d\ln T$ as a function of $T$ and $r-r_{c}$. As
expected, at sufficiently low temperatures, $\rho\sim T^{2}$. At
higher temperature, $d\ln\rho/d\ln T$ decreases, becoming sublinear
at temperatures of the order of $\Omega_{L}$. Defining $T_{0}$ as
the temperature where $d\ln\rho/d\ln T=1.8$, we find that at the
QCP, $T_{0}\propto|\mathbf{q}_{0}|^{3}$ {[}Fig. \ref{fig:rho_umklapp}(d){]}.
Away from the QCP, the crossover to $T^{2}$ occurs at higher temperatures,
and $T_{0}$ scales linearly with the distance to the QCP $r$.

Although for $r=r_{c}$ and $T>T_{0}$, there is a temperature window
where $\rho(T)$ appears to be approximately linear (Fig. \ref{fig:rho_umklapp}c),
it is important to note that $d\ln\rho/d\ln T$ actually changes continuously
in this regime. It is also worth noting that the dc resistivity obtained
in our calculation reproduces many of the qualitative features observed
in the QMC simulations of Ref. \cite{schattner16}, including the
broad quasi-linear regime near the QCP, and the gradual change of
the slope of $\rho(T)$ as $r$ is tuned away from the QCP.

\subsection{Compensated metal\label{subsec:Compensated-metal}}

The mechanisms for current dissipation discussed so far rely on breaking
the conservation of total electron momentum, either by impurity or
by umklapp scattering. It is well-known that in a compensated metal
with equal number of electron and hole-like charge carriers, electron-electron
interactions alone can lead to a finite electrical resistivity, even
in the absence of momentum relaxation \citep{baber37}. This is because
scattering events between electrons and holes can relax the current,
despite the fact that momentum is conserved. In a compensated metal,
the electrical current and the total electronic momentum are orthogonal
to each other: $\chi_{\mathbf{JP}}=0$, and the current dissipation
is driven by relaxation of other odd-parity modes of the Fermi surface.
For completeness, this result is re-derived in Appendix \ref{sec:Totally-compensated-metal}.

Several of the iron-based superconductors known for exhibiting non-Fermi
liquid transport, such as $\mathrm{FeSe_{1-x}S_{x}}$ \cite{coldea18}
and $\mathrm{BaFe_{2}(As_{1-x}P_{x})_{2}}$ \cite{kasahara10}, are
isovalently doped. These systems have multiple small Fermi surfaces
(pockets) that are of either electron or hole character, with an equal
density of electron and hole carriers. It is reasonable to assume
that such materials are not far from being a compensated metal, and
that normal electron-hole scattering mediated by quantum critical
fluctuations plays an important role in their transport behavior.

In this Section, we study the temperature dependence of dc resistivity
for a compensated metal near a nematic quantum critical point. Motivated
by the case of the iron-based superconductors, we focus on the case
where the electron and hole pockets are much smaller than the Brillouin
zone size, such that umklapp scattering is negligible. Since the nematic
critical fluctuations are centered at small momenta, we assume that
they cannot scatter electrons from one pocket to the other. The Hamiltonian
is given by
\begin{equation}
H=\sum_{i=1}^{n}\left(\sum_{\mathbf{k}}\varepsilon_{i\mathbf{k}}c_{i\mathbf{k}}^{\dagger}c_{i\mathbf{k}}+\lambda\sum_{\mathbf{kq}}\phi_{\mathbf{q}}f_{i,\mathbf{k,k+q}}c_{i\mathbf{k+q}}^{\dagger}c_{i\mathbf{k}}\right),
\end{equation}
where $i=1,\dots,n$ is the pocket index, and the pocket dispersions
$\varepsilon_{i\mathbf{k}}$ are assumed to be such that the total
area enclosed by the electron-like pockets is equal to that of the
hole-like pockets. $f_{i,\mathbf{k},\mathbf{k}+\mathbf{q}}$ is the
Ising-nematic form factor of the $i$th pocket, centered at wavevector
$\mathbf{Q}_{i}$. Importantly, depending on the position of the Fermi
pocket in the Brillouin zone, the form factor may or may not contain
``cold spots'' -- points on the Fermi pocket where the form factor
vanishes. For example, a pocket centered at the $\Gamma$ point {[}$\mathbf{Q}_{i}=(0,0)${]}
has cold spots where the diagonals $k_{x}=\pm k_{y}$ intersect the
Fermi surface. In contrast, a pocket at the $X$ or $Y$ points {[}$\mathbf{Q}_{i}=(\pi,0)$
or $(0,\pi)$, respectively{]} does not have any symmetry-imposed
cold spots. The band structures of most of the iron-based superconductors
contain pockets centered at both the $\Gamma$, $X$, and $Y$ points
(in the one Fe per unit cell scheme). As we shall see below, the presence
of cold spots on the Fermi pockets, and whether they occur on all
or only some of the pockets, changes qualitatively transport behavior
at low temperature.

\begin{figure}
\includegraphics[width=0.96\linewidth]{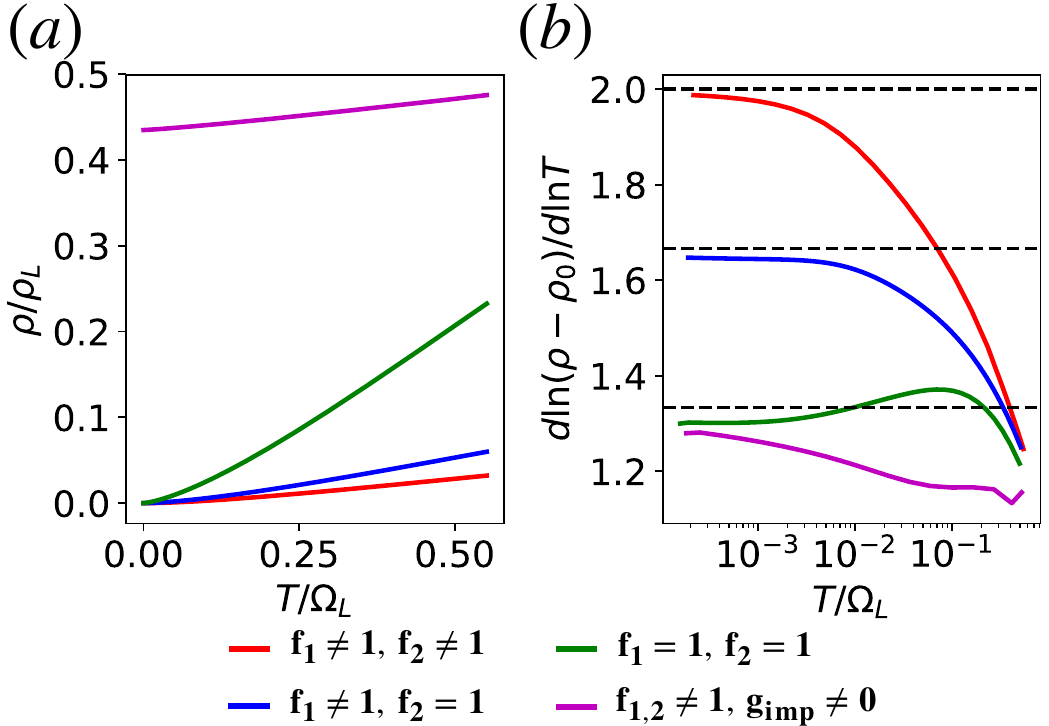}

\caption{\label{fig:rho_compensated} $\rho(T)$ and the log-derivative plot
at the QCP in a perfectly compensated metal with different arrangements
of cold spots on the Fermi pockets. The model consists of two circular
pockets with equal radii, one electron-like and one hole-like. The
nematic form factor is either taken to be $\cos2\theta_{\mathbf{k}}$
when the corresponding pocket has cold spots, or unity for the case
with no cold spots. The coupling constant is $\lambda^{2}\approx0.84\varepsilon_{F}$.
Red curve: both pockets have cold spots along the diagonals. Blue
curve: cold spots on the hole pocket, but not the electron pocket.
Green curve: neither pocket has cold spots. Magenta curve: both pockets
have cold spots, but also subject to impurity scattering of strength
$g_{\text{imp}}\approx0.1\varepsilon_{F}$. The black dashed lines
in (b) correspond to exponents of $4/3$, $5/3$ and $2$. }
\end{figure}

We study $\rho(T)$ in four different scenarios: in a clean system
where (1) all the pockets have cold spots, (2) the hole pockets have
cold spots but the electron pockets do not, (3) no pockets have cold
spots, and (4) a disordered system where all the pockets have cold
spots. For simplicity, we considered a model with two circular pockets
with identical radii, one electron-like and one hole-like. We do not
expect the results to change qualitatively for pockets of a general
shape, as long as the system is perfectly compensated (see Appendix
\ref{sec:form-factor}). The results are summarized in Fig. \ref{fig:rho_compensated}.
In the case when cold spots are absent or if there is weak impurity
scattering (Fig. \ref{fig:rho_compensated} green and magenta lines),
$\rho(T)-\rho_{0}\propto T^{4/3}$, which is the naive quantum critical
scaling exponent discussed previously \cite{lohneysen07,maslov11}.
However, in a clean compensated metal with cold spots on one or both
of the Fermi pockets, the asymptotic low-temperature behavior is different.
When cold spots are present on some (but not all) of the pockets,
$\rho\sim T^{5/3}$ (blue curve), whereas when cold spots are present
on all the pockets, $\rho\sim T^{2}$ (red curve).

To understand the origin of the strong sensitivity of the low-temperature
transport behavior to the presence of cold spots, we examine the non-equibrium
distribution function in the presence of a current. It is convenient
to parametrize the distribution function $\delta n_{i\mathbf{k}}$
in terms of another function $\Phi_{i\mathbf{k}}$ such that $\delta n_{\mathbf{k}}=\left[-\partial_{\varepsilon_{i\mathbf{k}}}n_{F}(\varepsilon_{i\mathbf{k}})\right]\Phi_{i\mathbf{k}}E_{x}$,
where $i$ is the band index, $\mathbf{k}$ is a point on the Fermi
surface and $\mathbf{E}=E_{x}\hat{x}$ is the electric field applied
along the $x$ axis. $\Phi_{i\mathbf{k}}$ can be computed from the
memory matrix according to: $\Phi_{i\mathbf{k}}=\left(M^{-1}\right)_{i\mathbf{k},j\mathbf{k'}}\chi_{j\mathbf{k'},J_{x}}$.
More details on relation between our memory matrix approach and the
non-equilibrium distribution function are given in Appendix \ref{sec:boltzmann}.
Fig. \ref{fig:compensated_angular_distribution}(a) shows the clean
case without hot spots. In this case, $\Phi_{1\theta_{\mathbf{k}}}=-\Phi_{2\theta_{\mathbf{k}}}=\cos\theta_{\mathbf{k}}$.
However, when cold spots are present on at least one Fermi pocket,
the distribution function on both pockets changes dramatically, as
seen in Fig. \ref{fig:compensated_angular_distribution}(b) and (c).
In this case, the distribution function is nearly constant on different
quadrants of the Fermi surface, bounded by the cold spots. In the
presence of disorder, the distribution function becomes more regular
and approaches a cosine at low temperatures, as seen in Fig. \ref{fig:compensated_angular_distribution}(d).

\begin{figure}
\includegraphics[width=0.96\linewidth]{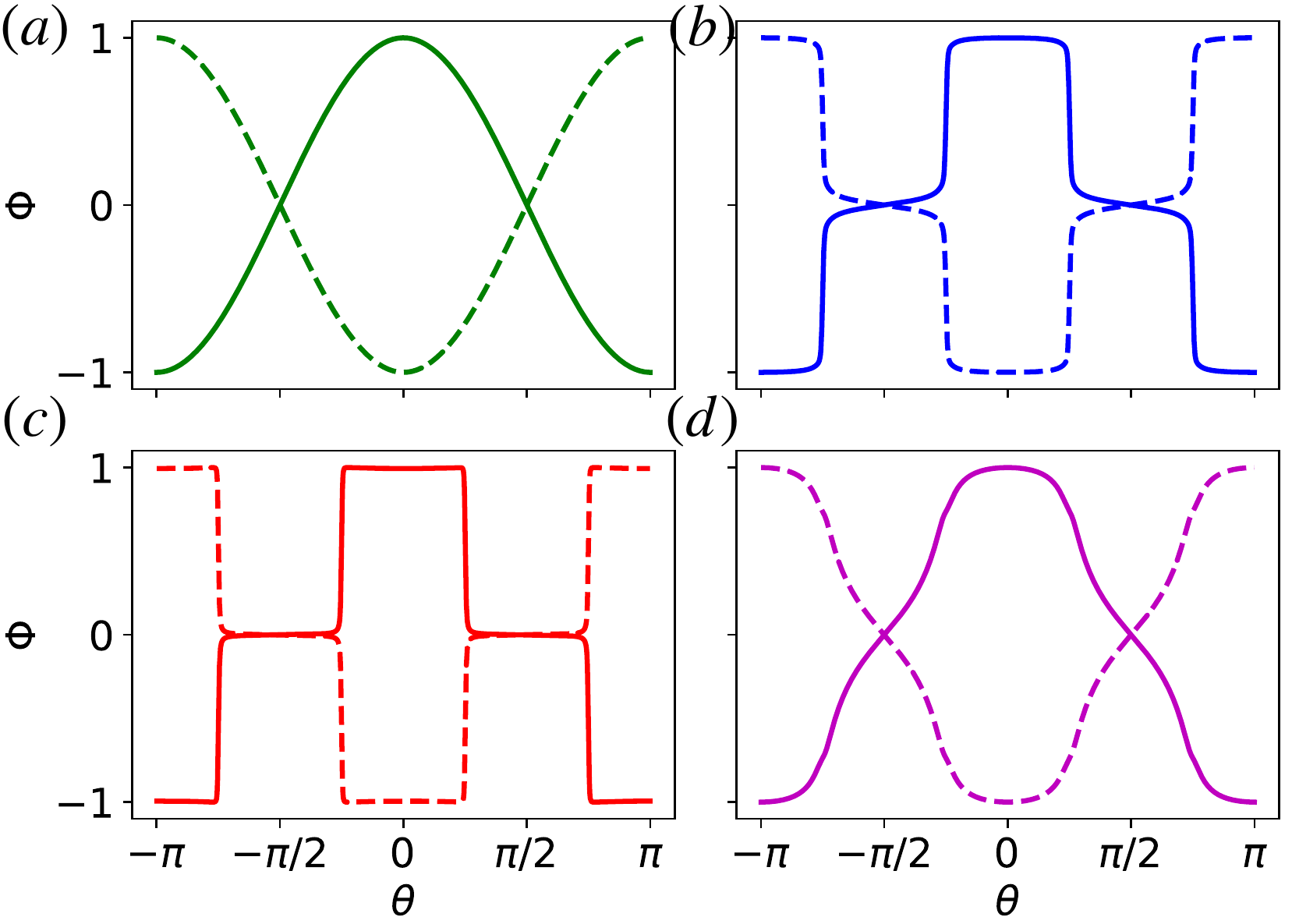}

\caption{\label{fig:compensated_angular_distribution}The function $\Phi_{i,\theta_{\mathbf{k}}}\propto\delta n_{i,\theta_{\mathbf{k}}}$
on Fermi pocket $i=1,2$ in the presence of a current in the $x$
direction, for the same model as in Fig. \ref{fig:rho_compensated}.
Here $\theta_{\mathbf{k}}$ is the angle measured with respect to
the $x$-direction. The solid (dashed) curve is the distribution function
on the electron (hole) pocket. The colors correspond to the same cases
shown in Fig. \ref{fig:rho_compensated}. All four distribution functions
are calculated at the same temperatures $T\approx0.002\varepsilon_{F}$.}
\end{figure}
\begin{figure}
\includegraphics[width=0.5\linewidth]{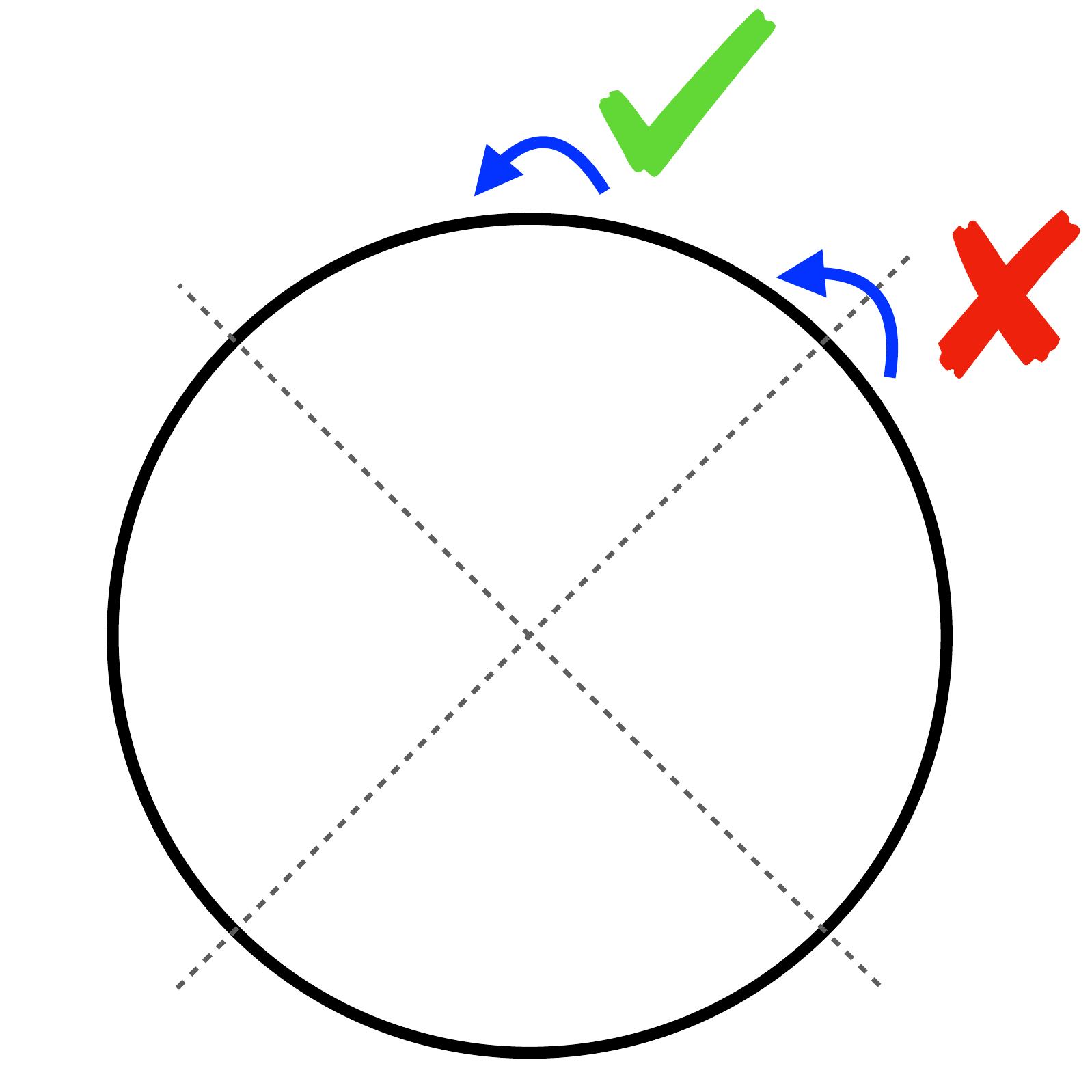}
\caption{\label{fig:compensated_ff_phi_k} Electron scattering on the Fermi
surface due to small wavevector nematic fluctuations. The intersection
between the Fermi surface and the dashed lines are where the nematic
form factor vanishes --- the nematic ``cold spots''. Small wavevector
scattering across the cold spots are strongly suppressed by the nematic
form factor.}
\end{figure}

The shape of the non-equilibrium distribution function in the presence
of cold spots can be understood qualitatively as follows: at low temperatures,
the characteristic momentum transfer $q$ due to quantum critical
scattering, $q\sim k_{F}\left(T/\Omega_{L}\right)^{1/3}$, is small.
The scattering rate from a point $\mathbf{k}$ near one of the cold
spots to a nearby point $\mathbf{k}+\mathbf{q}$ is suppressed by
the nematic form factor. Therefore, the cold spots effectively cut
the Fermi surface into four nearly-disconnected patches. The equilibration
within each patch is much faster than the equilibration between patches
(see Fig. \ref{fig:compensated_ff_phi_k}). In this situation, the
non-equilibrium distribution is nearly constant in each patch, and
the bottleneck for current relaxation becomes the slow inter-patch
scattering. As a result, the resistivity in the presence of cold spots
is reduced than the resistivity with no cold spots. In Appendix \ref{sec:form-factor},
we analyze the transport properties in the presence of cold spots,
using a piecewise-constant distribution function of the form shown
in Fig. \ref{fig:compensated_angular_distribution}(b,c) as a variational
ansatz. We find that the power-law exponents observed in Figure \ref{fig:rho_compensated}
are exactly reproduced.

\subsection{Quasi-elastic thermal fluctuations and intermediate temperature behavior\label{subsec:Quasi-elastic-thermal-fluctuatio}}

Having analyzed the asymptotic low-temperature behavior of the resistivity
due to different scattering mechanisms, we turn to discuss the crossover
behavior at higher temperatures. In particular, depending on microscopic
parameters, there may be a regime where the temperature is comparable
to or larger than the energy scale set by the Landau damping, $\Omega_{L}$,
but still much smaller than the Fermi energy. Since $\Omega_{L}=\varepsilon_{F}^{2}/\lambda^{2}$,
this regime is accessible within our model at strong coupling (or,
equivalently, if the Fermi energy is small). Formally, the calculation
in this regime can still be controlled in the large $N$ limit, as
long as $T\gg\Omega_{\text{NFL}}$. As we shall now show, the resistivity
is determined by quasi-elastic scattering of electrons off thermally
excited nematic fluctuations. Depending on the evolution of the correlation
length with temperature, this may lead to either $\rho\sim T$ or
$\rho\sim\text{const.}$ in the crossover regime.

We begin by examining the temperature dependence of the scattering
cross-section between electrons and nematic fluctuations, $V_{\mathbf{q}}(T)$
{[}see Eqs. (\ref{eq:mm_classI}--\ref{eq:Vq}){]}. We write Eq.
(\ref{eq:Vq}) as $V_{\mathbf{q}}(T)=\frac{1}{\gamma_{\mathbf{q}}}F\left(\frac{T}{\omega_{\mathbf{q}}}\right)$,
where $\omega_{\mathbf{q}}\equiv r_{\mathbf{q}}\gamma_{\mathbf{q}}^{-1}$
(recall that the nematic propagator is written as $D_{\mathbf{q},\nu_{n}}^{-1}=r_{\mathbf{q}}+\gamma_{\mathbf{q}}|\nu_{n}|$)
and
\begin{equation}
F\left(x\right)\equiv\frac{1}{x}\int_{-\infty}^{+\infty}\frac{\mathrm{d}u}{\pi}\frac{u^{2}}{1+u^{2}}\frac{1}{\sinh^{2}\left(\frac{u}{2x}\right)}.\label{eq:F}
\end{equation}
$F\left(x\right)$ has the following asymptotic properties: for $x\ll1$,
$F(x)\approx\pi x^{2}/3$, whereas for $x\gg1$, $F(x)\approx4x$.

We can now estimate the resistivity in the regime $T\gg\Omega_{L}\sim\omega_{\mathbf{q}\sim k_{F}}$,
that corresponds to $x\gg1$ in Eq.~(\ref{eq:F}). In this regime,
the dominant contribution to $V_{\mathbf{q}}(T)$ is from energies
$\omega\sim\omega_{\mathbf{q}}\ll T$, corresponding to quasi-elastic
scattering of electrons off thermally excited nematic fluctuations.
We focus on the case of a compensated metal, where momentum conservation
does not limit the resistivity. At high temperatures, we do not expect
the nematic form factor to have a strong effect on the results, and
will henceforth neglect it. The resistivity can then be estimated
by overlapping the memory matrix described by class I diagram with
the current operator:
\begin{equation}
\rho(T)\propto\sum_{\mathbf{kk'}}\left(\mathbf{v}_{\mathbf{k}}\cdot\mathbf{v}_{\mathbf{k'}}\right)M_{\mathbf{kk'}}^{(1)}\sim\int_{0}^{k_{F}}\mathrm{d}qq^{2}V_{\mathbf{q}}\left(T\right).\label{qualitative_analysis}
\end{equation}
Placing both $\mathbf{k}$ and $\mathbf{k+q}$ on the Fermi surface
(recall that we are still considering $T\ll\varepsilon_{F}$) constrains
the allowed wave-vector $\mathbf{q}$ to a one-dimensional manifold.
The first $q^{2}$ term in the integrand is the well-known transport
factor that suppresses the contribution of small angle scattering.
This factor arises from the term $\sum_{\mathbf{q,kk'}}\left(\mathbf{v}_{\mathbf{k}}\cdot\mathbf{v}_{\mathbf{k'}}\right)\left(\delta_{\mathbf{kk'}}-\delta_{\mathbf{k'-k,q}}\right)\sim\sum_{\mathbf{q,k}}\left(\mathbf{v}_{\mathbf{k}}-\mathbf{v}_{\mathbf{k+q}}\right)^{2}$
that appears when inserting Eq.~(\ref{eq:mm_classI}) in Eq.~(\ref{qualitative_analysis}).

\begin{figure}
\includegraphics[width=0.96\linewidth]{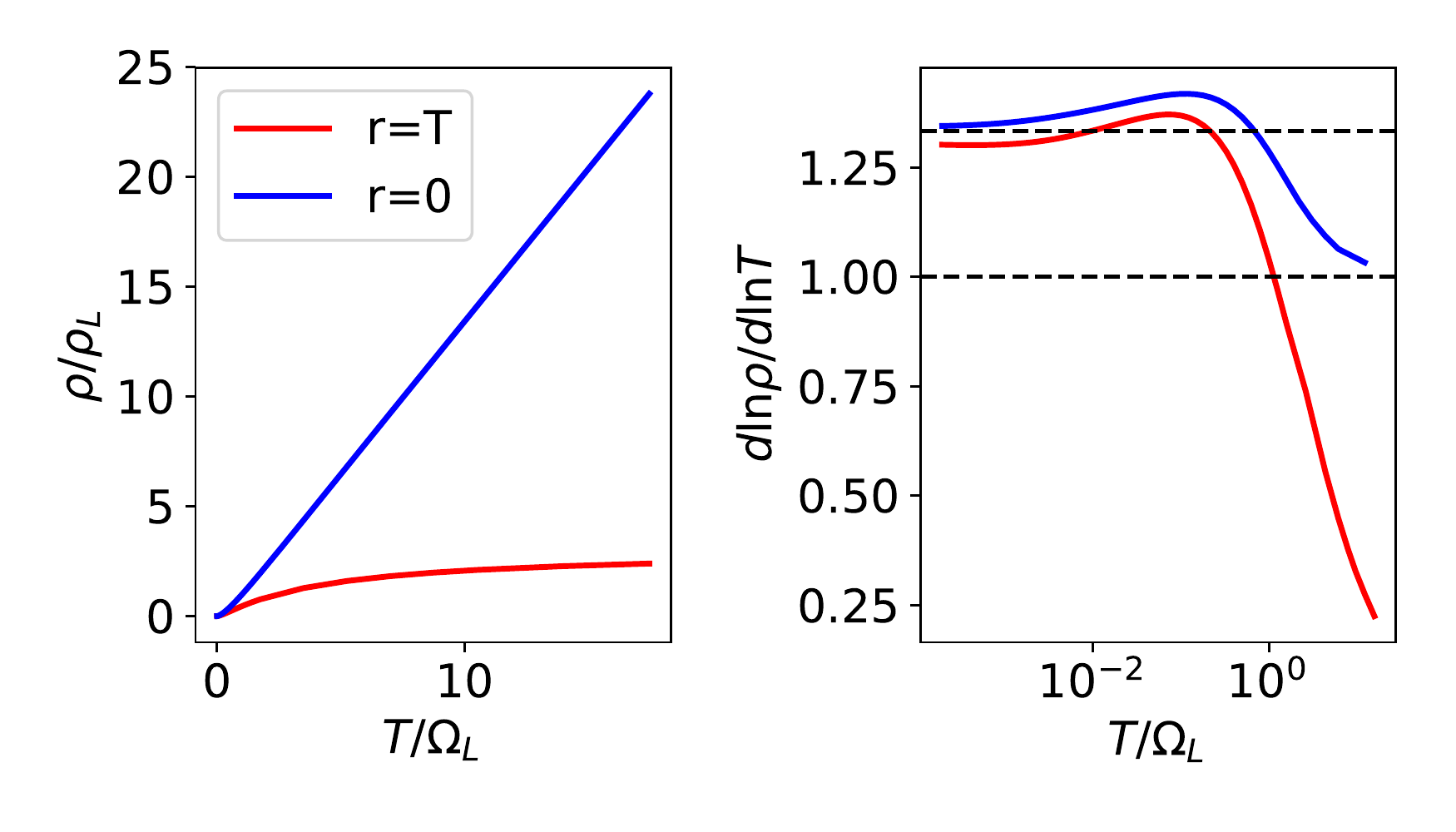}

\caption{\label{fig:compensated_thermal_mass}$\rho(T)$ in the high temperature
regime $T>\Omega_{L}$, calculated for a compensated metal without
cold spots. $r(T)\equiv\xi^{-2}(T)$ is the thermal mass for nematic
fluctuations. The calculation assumes strong coupling $\lambda^{2}\gg\varepsilon_{F}$
and large fermion flavor $N$, so that $\Omega_{L}\ll\varepsilon_{F}$,
and our memory matrix approach remains valid. }
\end{figure}

When $T>\Omega_{L}$, nematic fluctuations with a large wavevector
$q\sim k_{F}$ become thermally excited. From the discussion above
[Eq. (\ref{eq:F})], we find that in this regime $V_{\mathbf{q}}(T)\sim T/r_{\mathbf{q}}$,
and hence
\[
\rho(T)\propto T\int_{0}^{k_{F}}\mathrm{d}q\frac{q^{2}}{\xi^{-2}(T)+q^{2}}
\]
where we have expressed $r_{\mathbf{q}}(T)=\xi^{-2}(T)+\mathbf{q}^{2}$.
If $\xi\approx\xi_{0}$ is only weakly dependent on temperature, or
if $\xi^{-1}\ll k_{F}$, then $\rho(T)\sim T$. In contrast, if $\xi^{-1}(T)\gg k_{F}$,
then the behavior of $\rho(T)$ is determined by the temperature dependence
of $\xi(T)$. For $\xi^{-2}\sim T$ (as was observed in the QMC simulations
of Ref. \cite{lederer17}, and experimentally in Ref. \cite{bohmer16}),
$\rho(T)$ saturates to a constant at high temperature. In Figure
\ref{fig:compensated_thermal_mass}, we illustrate both types of high
temperature behavior for a compensated metal without cold spots, as
discussed in the previous section.

\section{Conclusion}

In summary, we have developed a memory matrix approach to derive a
kinetic equation applicable in a broad temperature regime near an
Ising-nematic QCP. The formalism is applied to study the behavior
of the dc resistivity in the vicinity of the QCP. The resistivity
exhibits a rich behavior that depends on the dominant mechanism for
current dissipation and on the structure of the Fermi surface.

We find several regimes where the resistivity is strongly affected
by nematic critical fluctuations, despite their long--wavelength
nature. As long as the Fermi surface is not very small compared to
the size of the Brillouin zone, there is a broad temperature range
where the resistivity is strongly enhanced near the QCP due to umklapp
scattering; in this regime, the umklapp processes also modify the
spectrum of the critical nematic fluctuations, and $z=3$ dynamical
scaling does not hold. At asymptotically low temperatures, however,
$z=3$ scaling is recovered, and $\rho\sim T^{2}$. In multi-band
systems in the presence of impurities, $\rho-\rho_{0}\propto T^{4/3}$
down to the lowest temperatures, as anticipated from $z=3$ dynamical
scaling. In a compensated metal with an equal density of electrons
and holes, the quantum critical fluctuations can affect the resistivity
even in the absence of impurities and at arbitrarily low temperatures.
In this case we find that the dc resistivity is strongly affected
by the presence of ``cold spots'' on the Fermi sheets, due to the
symmetry of the nematic order parameter. In the clean limit, $\rho\sim T^{\alpha}$,
where $\alpha=2$, $5/3$, or $4/3$, depending on whether there are
cold spots on all the Fermi sheets, on some of the sheets, or on none,
respectively.

It is important to note that we have assumed a purely electronic mechanism for the Ising-nematic QCP, and have neglected the effect of coupling to the lattice. This effect is known to change the properties of the QCP, quenching most of the long-wavelength nematic fluctuations and making the transition more mean-field like~\cite{Zacharias2015,Karahasanovic2016,Paul2017}. As a result, the effect of the critical fluctuations on the resistivity is suppressed, recovering Fermi liquid behavior $\rho-\rho_0\propto T^2$ at the lowest temperatures. The scale at which the crossover to Fermi liquid behavior occurs depends on the strength of the coupling to the lattice.  

Our analysis can be extended straightforwardly to other metallic QCPs
in $d=2$ dimensions that do not involve breaking of translational
symmetry, such as ferromagnetic transitions. In the latter case, there
are generically no cold spots on the Fermi surface.

The diversity of possible behaviors found in our study suggest that
experiments in critical metals should be interpreted with great care.
Nevertheless, it would be interesting to consider our results in the
context of ongoing experiments~\cite{reiss18} in $\mathrm{FeSe_{1-x}S_{x}}$, which
is a compensated system that exhibits an apparent nematic QCP with
no nearby magnetic phase. %\cite{coldea18}.

\vspace{3mm}
\begin{acknowledgments}
The authors would like to thank Andrey Chubukov, Rafael Fernandes,
Sean Hartnoll, Dam T. Son, and Subir Sachdev for helpful discussions and inputs.
XW acknowledges financial support from the James Frank Institute in
University of Chicago. EB was supported by the European Research Council
(ERC) under the European Union Horizon 2020 Research and Innovation
Programme (Grant Agreement No. 817799). Both XW and EB acknowledges
support from the Nordita workshop ``Bounding Transport and Chaos
in Condensed Matter and Holography'', where parts of this work is
carried out.
\end{acknowledgments}

%\pagebreak

\appendix
\widetext
%\begin{center}
%\textbf{\large Supplementary material for: ``Scattering mechanisms and electrical %transport near an Ising nematic quantum critical point''}
%\end{center}
%\setcounter{equation}{0}
%\setcounter{figure}{0}
%\setcounter{section}{0}
%\setcounter{page}{1}
%\renewcommand{\theequation}{S\arabic{equation}}
%\renewcommand{\thefigure}{S\arabic{figure}}

\section{Memory matrix approach to transport properties\label{sec:Memory-matrix-approach}}

\subsection{General formalism\label{subsec:General-formalism}}

In this section, we briefly review the memory matrix formalism and
its application to transport. We then discuss the validity of the
approach in the vicinity of a QCP.

We closely follow the discussion of the memory matrix approach in
Ref. \cite{hartnoll18}. To set up the formalism, we first define
an inner product in the Hilbert space of operators. For two Hermitian
operators $A$, $B$, the inner product is given by

\begin{equation}
(A\vert B)=T\int_{0}^{\beta}d\tau\left[\langle A(\tau)B\rangle-\langle A\rangle\langle B\rangle\right]\equiv T\chi_{AB}.
\end{equation}

Here, $\chi_{AB}$ is a theormodynamic susceptibility relating the
operators $A$ and $B$.

The Liouville ``super operator'' is defined as $\mathcal{L}=-[H,\cdot]$.
The operators satisfy the Heiseberg equation of motion: $\dot{A}=i[H,A]=-i\mathcal{L}A$.
We are interested in calculating a retarded correlation function of
two operators, characterizing the response of the system. This can
be done through the relation \cite{hartnoll18}
\begin{equation}
G_{AB}^{R}(t)\equiv i\Theta(t)\langle[A(t),B(0)]\rangle=-\frac{1}{T}\Theta(t)\partial_{t}C_{AB}(t),
\end{equation}
where $C_{AB}(t)\equiv(A(t)\vert B)$. Fourier transforming both sides,
we obtain
\begin{equation}
\widetilde{C}_{AB}(z)=\frac{T}{iz}\left[\tilde{G}_{AB}^{R}(z)-\tilde{G}_{AB}^{R}(0)\right],\label{eq:CAB}
\end{equation}
where $\widetilde{C}_{AB}(z)\equiv\int_{0}^{\infty}e^{izt}C_{AB}(t)$,
$z$ is a complex number in the upper half plane, and similarly for
$\tilde{G}_{AB}^{R}(z)=\int_{0}^{\infty}e^{izt}G_{AB}^{R}(t)$. For
example, the frequency-dependent conductivity can be written as

\begin{equation}
\sigma(\Omega)\equiv\frac{1}{i\Omega}\left[\tilde{G}_{J_{x}J_{x}}^{R}(\Omega)-\tilde{G}_{J_{x}J_{x}}^{R}(0)\right]=\frac{1}{T}\widetilde{C}_{J_{x}J_{x}}(z=\Omega+i0^{+}).
\end{equation}

The key step in the memory matrix technique is to identify a set of
slow (or nearly-conserved) operators, and project the dynamics onto
these operators. We denote the set of slow operators by $|A_{\alpha})$.
We assumne that the current is a linear combination of $|A_{\alpha})$'s.
Then, to compute $\widetilde{C}_{J_{x}J_{x}}(z)$, it is sufficient
to compute matrix elements of the resolvent (or Green's function)
$\hat{G}(z)=\frac{i}{z-\mathcal{L}}$ in the subspace of $|A_{\alpha})$.
To do this, we define the projection operator
\begin{equation}
\hat{P}=\frac{1}{T}\sum_{\alpha,\beta}\vert A_{\alpha})(\chi^{-1})_{\alpha\beta}(A_{\beta}\vert
\end{equation}
onto the slow subspace, and $\hat{Q}=1-\hat{P}$ is a projection onto
the complementary subspace. The equation for the Green's function
$G$ is

\begin{equation}
\left(\begin{array}{cc}
z\hat{P}-\hat{P}\mathcal{L}\hat{P} & -\hat{P}\mathcal{L}\hat{Q}\\
-\hat{Q}\mathcal{L}\hat{P} & z\hat{Q}-\hat{Q}\mathcal{L}\hat{Q}
\end{array}\right)\left(\begin{array}{cc}
\hat{P}\hat{G}\hat{P} & \hat{P}\hat{G}\hat{Q}\\
\hat{Q}\hat{G}\hat{P} & \hat{Q}\hat{G}\hat{Q}
\end{array}\right)=\left(\begin{array}{cc}
i & 0\\
0 & i
\end{array}\right).
\end{equation}
From this we obtain the two equations

\begin{equation}
\left(z\hat{P}-\hat{P}\mathcal{L}\hat{P}\right)\hat{P}\hat{G}\hat{P}-\hat{P}\mathcal{L}\hat{Q}\left(\hat{Q}\hat{G}\hat{P}\right)=i,
\end{equation}

\begin{equation}
\left(z\hat{Q}-\hat{Q}\mathcal{L}\hat{Q}\right)\hat{Q}\hat{G}\hat{P}-\hat{Q}\mathcal{L}\hat{P}\left(\hat{P}\hat{G}\hat{P}\right)=0.
\end{equation}
Solving for $\hat{P}\hat{G}\hat{P}$, we arrive at

\begin{equation}
\hat{P}\hat{G}\hat{P}=\frac{i}{z\hat{P}-\hat{P}\mathcal{L}\hat{P}-\hat{P}\mathcal{L}\hat{Q}\left(z-\hat{Q}\mathcal{L}\hat{Q}\right)^{-1}\hat{Q}\mathcal{L}\hat{P}}.
\end{equation}

The matrix elements of $\hat{G}$ between two operators in the slow
subspace can then be written as

\begin{equation}
\widetilde{C}_{\alpha\beta}(z)=(A_{\alpha}\vert\frac{i}{z-\mathcal{L}}\vert A_{\beta})=T\chi\frac{1}{-iz\chi+N+M(z)}\chi,
\end{equation}
where we have defined the matrices
\begin{equation}
N_{\alpha\beta}=\frac{i}{T}(A_{\alpha}\vert\mathcal{L}\vert A_{\beta}),
\end{equation}
\begin{equation}
M_{\alpha\beta}(z)=\frac{i}{T}(A_{\alpha}\vert\mathcal{L}\hat{Q}\frac{1}{z-\hat{Q}\mathcal{L}\hat{Q}}\hat{Q}\mathcal{L}\vert A_{\delta}).\label{eq:M}
\end{equation}
So far, the manipulations have all been exact - we have not used the
fact that the operators $\vert A_{\alpha})$ are ``slow''. The slowness
of the operators $A_{\alpha}$ is typically employed when computing
$M_{\alpha\beta}(z)$. We imagine that the Hamiltonian depends on
a parameter $g$, such that to zeroth order in $g$, $\mathcal{L}(g=0)A_{\alpha}=0$.
Then, to leading order in $g$, we can drop the factors of $\hat{Q}$
in the evaluation of the denominator in Eq. (\ref{eq:M}), and the
memory matrix becomes an ordinary dynamical correlation function with
$H(g=0)$.

\subsection{Application to the quantum critical problem\label{subsec:mm_QCP}}

In our quantum critical system, we choose the set of operators to
be the occupations of particles per flavor in momentum space, $\{n_{\alpha\mathbf{k}}\}$.
In order to evaluate the memory matrix, Eq. (\ref{eq:M}), we note
that in the $N\rightarrow\infty$ limit, the operators $\{n_{\alpha\mathbf{k}}\}$
become conserved quantities. This is since
\begin{equation}
\dot{n}_{\alpha\mathbf{k}}=-i\mathcal{L}n_{\alpha\mathbf{k}}=\frac{i\lambda}{\sqrt{N}}\sum_{\mathbf{q}}\phi_{\mathbf{q}}\left(f_{\mathbf{k},\mathbf{k}-\mathbf{q}}c_{\alpha\mathbf{k}}^{\dagger}c_{\alpha\mathbf{k}-\mathbf{q}}-f_{\mathbf{k},\mathbf{k}+\mathbf{q}}c_{\alpha\mathbf{k}+\mathbf{q}}^{\dagger}c_{\alpha\mathbf{k}}\right).
\end{equation}
We use a normalized set of operators, $|\widetilde{n_{\alpha\mathbf{k}}})=(n_{\alpha\mathbf{k}}|n_{\alpha\mathbf{k}})^{-1/2}|n_{\alpha\mathbf{k}})$.
In the presence of time reversal and inversion symmetries, $(\widetilde{n_{\alpha\mathbf{k}}}|\mathcal{L}|\widetilde{n_{\alpha'\mathbf{k}'}})=0$.
Consider the matrix element of $\mathcal{L}$ between $|\widetilde{n_{\alpha\mathbf{k}}})$
and the normalized operator $|\widetilde{\mathcal{L}n_{\alpha'\mathbf{k}'}})\equiv(\mathcal{L}n_{\alpha'\mathbf{k}'}|\mathcal{L}n_{\alpha'\mathbf{k}'})^{-1/2}|\mathcal{L}n_{\alpha'\mathbf{k}'})$:
\begin{equation}
(\widetilde{n_{\alpha\mathbf{k}}}\vert\mathcal{L}|\widetilde{\mathcal{L}n_{\alpha'\mathbf{k}'}})=-\frac{(\mathcal{L}n_{\alpha\mathbf{k}}\vert\mathcal{L}n_{\alpha'\mathbf{k}'})}{(n_{\alpha\mathbf{k}}\vert n_{\alpha\mathbf{k}})^{1/2}(\mathcal{L}n_{\alpha'\mathbf{k}'}|\mathcal{L}n_{\alpha'\mathbf{k}'})^{1/2}}.
\end{equation}
We can easily check that $(\widetilde{n_{\alpha\mathbf{k}}}\vert\mathcal{L}|\widetilde{\mathcal{L}n_{\alpha'\mathbf{k}'}})=\mathcal{O}(N^{-1/2})$
\,\footnote{To appreciate the significance of this point, it is useful to note
that in contrast, the matrix element between the operator $c_{\alpha\mathbf{k}}^{\dagger}c_{\alpha\mathbf{k}+\mathbf{q}}$
and $\mathcal{L}c_{\alpha\mathbf{k}}^{\dagger}c_{\alpha\mathbf{k}+\mathbf{q}}$
is $\mathcal{O}(1)$. This is because $c_{\alpha\mathbf{k}}^{\dagger}c_{\alpha\mathbf{k}+\mathbf{q}}$
does not commute with the kinetic energy, and hence $\mathcal{L}c_{\alpha\mathbf{k}}^{\dagger}c_{\alpha\mathbf{k}+\mathbf{q}}$
is not suppressed by a factor of $1/\sqrt{N}$. }.

This implies that, to leading order in $1/N$, the memory matrix can
be written as

\begin{equation}
M_{\alpha\mathbf{k},\alpha'\mathbf{k}'}=\frac{i}{T}(\dot{n}_{\alpha\mathbf{k}}\vert\frac{1}{z-\mathcal{L}}\vert\dot{n}_{\alpha'\mathbf{k}'}),
\end{equation}
where the correlation function is evaluated in the $N\rightarrow\infty$
limit. In this limit, $\mathcal{L}$ does not connect the slow operators
$n_{\alpha\mathbf{k}}$ to other operators, and the projector $\hat{Q}$
in the denominator of Eq. (\ref{eq:M}) is automatically accounted
for.

To compute the memory matrix, we use Eq. (\ref{eq:CAB}). The retarded
response function can be computed from imaginary-time-ordered correlation
function via analytic continuuation. In Matsubara frequency:

\begin{equation}
\mathcal{G}_{\alpha\mathbf{k},\alpha'\mathbf{k}'}(i\Omega_{n})=-\int_{0}^{\beta}\mathrm{d}\tau e^{i\Omega_{n}\tau}\langle\dot{n}_{\alpha\mathbf{k}}(\tau)\dot{n}_{\alpha'\mathbf{k}'}(0)\rangle.
\end{equation}

Here $\dot{n}_{\alpha\mathbf{k}}\equiv-\left[H,n_{\alpha\mathbf{k}}\right]$
denotes imaginary-time derivative. Let us first observe the general
structure of the response function by treating $\lambda$ as a small
perturbation. To order $\mathcal{O}(\lambda^{2})$:
\begin{equation}
\begin{split}\mathcal{G}_{\alpha\mathbf{k},\alpha'\mathbf{k}'}^{(1)}(i\Omega_{n}) & =\delta_{\alpha\alpha'}\frac{\lambda^{2}}{N}\int_{\tau}e^{i\Omega_{n}\tau}\sum_{\mathbf{q}}D_{0}(\mathbf{q},\tau)\left\{ \left(\delta_{\mathbf{k}'-\mathbf{k},\mathbf{q}}-\delta_{\mathbf{kk}'}\right)f_{\mathbf{k},\mathbf{k}+\mathbf{q}}^{2}G_{0}(\mathbf{k},\tau)G_{0}(\mathbf{k}+\mathbf{q},-\tau)+\left(\tau\leftrightarrow-\tau,\mathbf{q}\leftrightarrow-\mathbf{q}\right)\right\} \end{split}
.
\end{equation}
After Fourier transformation:
\begin{equation}
\mathcal{G}_{\alpha\mathbf{k},\alpha'\mathbf{k}'}^{(1)}(i\Omega_{n})=-\delta_{\alpha\alpha'}\frac{\lambda^{2}T}{N}\sum_{\mathbf{q}\nu_{n}}D_{0}(\mathbf{q},i\nu_{n}+i\Omega_{n})\sum_{\zeta=\pm1}\left(\delta_{\mathbf{k}'-\mathbf{k},\mathbf{q}}-\delta_{\mathbf{kk}'}\right)f_{\mathbf{k},\mathbf{k+\zeta q}}^{2}R\left(\mathbf{k},\mathbf{k+\zeta q},i\zeta\nu_{n}\right).
\end{equation}
where we have defined
\begin{equation}
R(\mathbf{k},\mathbf{k}+\mathbf{q},i\nu_{n})\equiv-T\sum_{\omega_{k}}G_{0}(\mathbf{k},i\omega_{k})G_{0}(\mathbf{k}+\mathbf{q},i\omega_{k}+i\nu_{n}).
\end{equation}
To the next order $\mathcal{O}(\lambda^{4})$, to see that
\begin{equation}
\begin{split}\mathcal{G}_{\alpha\mathbf{k},\alpha'\mathbf{k'}}^{(2)}(i\Omega_{n}) & =-\frac{\lambda^{4}}{N^{2}T}\int_{\tau}e^{i\Omega_{n}\tau}\langle\phi_{\mathbf{q}}\phi_{\mathbf{q}_{1}}\phi_{\mathbf{q}_{2}}\phi_{\mathbf{q}^{\prime}}\rangle_{0}\langle\left(f_{\mathbf{k},\mathbf{k}-\mathbf{q}}c_{\mathbf{k}}^{\dagger}c_{\mathbf{k}-\mathbf{q}}-f_{\mathbf{k},\mathbf{k}+\mathbf{q}}c_{\mathbf{k}+\mathbf{q}}^{\dagger}c_{\mathbf{k}}\right)\left(f_{\mathbf{p}_{1},\mathbf{p}_{1}+\mathbf{q}_{1}}c_{\mathbf{p}_{1}+\mathbf{q}_{1}}^{\dagger}c_{\mathbf{p}_{1}}\right)\rangle_{0}\\
 & \times\langle\left(f_{\mathbf{p}_{2},\mathbf{p}_{2}+\mathbf{q}_{2}}c_{\mathbf{p}_{2}+\mathbf{q}_{2}}^{\dagger}c_{\mathbf{p}_{2}}\right)\left(f_{\mathbf{k}',\mathbf{k}'-\mathbf{q}'}c_{\mathbf{k}'}^{\dagger}c_{\mathbf{k}'-\mathbf{q}'}-f_{\mathbf{k}',\mathbf{k}'+\mathbf{q}'}c_{\mathbf{k}'+\mathbf{q}'}^{\dagger}c_{\mathbf{k}'}\right)\rangle_{0}.
\end{split}
\end{equation}
Here for convenience we have omitted the summation over repeated indices.
The connected random phase diagram has the structure shown in Fig.
3 (b,c) in the main text. The idea is to contract the electron momenta
$\left\{ \mathbf{k},\mathbf{p}_{1}\right\} $ and $\left\{ \mathbf{k}^{\prime},\mathbf{p}_{2}\right\} $,
and bosonic momenta as $\langle\phi_{\mathbf{q}}\phi_{\mathbf{q}^{\prime}}\rangle\langle\phi_{\mathbf{q}_{1}}\phi_{\mathbf{q}_{2}}\rangle$
(b) or $\langle\phi_{\mathbf{q}}\phi_{\mathbf{q}_{2}}\rangle\langle\phi_{\mathbf{q}'}\phi_{\mathbf{q}_{1}}\rangle$.
In either cases, we obtain:
\[
\mathbf{q}=-\mathbf{q}_{1};\mathbf{q}'=-\mathbf{q}_{2};\mathbf{q}'=\pm\mathbf{q}.
\]
After Fourier transformation:
\begin{equation}
\begin{split}\mathcal{G}_{\alpha\mathbf{k},\alpha'\mathbf{k'}}^{(2,b)}(i\Omega_{n}) & =\frac{\lambda^{4}T}{N^{2}}\sum_{\mathbf{q},i\nu_{n}}\sum_{i\omega_{l},i\omega_{s}}D_{0}(\mathbf{q},i\nu_{n})D_{0}(\mathbf{q},i\nu_{n}+i\Omega_{n})\\
 & \times\sum_{\zeta=\pm1}\zeta f_{\mathbf{k},\mathbf{k}+\zeta\mathbf{q}}^{2}R\left(\mathbf{k},\mathbf{k+\zeta q},i\zeta\nu_{n}\right)\\
 & \times\sum_{\zeta'=\pm1}\zeta'f_{\mathbf{k}',\mathbf{k}'+\zeta'\mathbf{q}}^{2}R\left(\mathbf{k}',\mathbf{k'+\zeta'q},i\zeta'\nu_{n}\right),
\end{split}
\end{equation}
and
\begin{equation}
\begin{split}\mathcal{G}_{\alpha\mathbf{k},\alpha'\mathbf{k'}}^{(2,c)}(i\Omega_{n}) & =-\frac{\lambda^{4}T}{N^{2}}\sum_{q,i\nu}\sum_{i\omega_{l},i\omega_{s}}D_{0}(\mathbf{q},i\nu_{n})D_{0}(\mathbf{q},i\nu_{n}+i\Omega_{n})\\
 & \times\sum_{\zeta=\pm1}\zeta f_{\mathbf{k},\mathbf{k}+\zeta\mathbf{q}}^{2}R\left(\mathbf{k},\mathbf{k+\zeta q},i\zeta\nu_{n}\right)\\
 & \times\sum_{\zeta'=\pm1}\zeta'f_{\mathbf{k}',\mathbf{k}'+\zeta'\mathbf{q}}^{2}R\left(\mathbf{k}',\mathbf{k'+\zeta'q},i\zeta'\nu_{n}+i\zeta'\Omega_{n}\right)
\end{split}
\end{equation}

Within random phase approximation corresponding to leading order in
$1/N$, we use the dressed nematic propagator to be $D^{-1}=D_{0}^{-1}-\Pi$,
but keep the fermion Green's function and the coupling vertex unrenormalized.

We obtain the following expression for the memory matrix within the
random phase approximation:
\begin{equation}
\begin{split}M_{\alpha\mathbf{k},\alpha'\mathbf{k'}}^{(1)}(i\Omega_{n}) & =\delta_{\alpha\alpha'}\frac{\lambda^{2}T}{N\Omega_{n}}\sum_{\mathbf{q},\nu_{n}}\left\{ D_{\mathbf{q},\nu_{n}+\Omega_{n}}\sum_{\zeta=\pm1}\left(\delta_{\mathbf{k}'-\mathbf{k},\zeta\mathbf{q}}-\delta_{\mathbf{kk}'}\right)f_{\mathbf{k},\mathbf{k+\zeta q}}^{2}R\left(\mathbf{k},\mathbf{k+\zeta q},i\zeta\nu_{n}\right)\right\} \\
M_{\alpha\mathbf{k},\alpha'\mathbf{k'}}^{(2)}(i\Omega_{n}) & =-\frac{\lambda^{4}T}{N^{2}\Omega_{n}}\sum_{\mathbf{q},\nu_{n}}D_{\mathbf{q},\nu_{n}}D_{\mathbf{q},\nu_{n}+\Omega_{n}}\sum_{\zeta\zeta'=\pm1}\zeta\zeta'f_{\mathbf{k},\mathbf{k}+\zeta\mathbf{q}}^{2}f_{\mathbf{k}',\mathbf{k}'+\zeta'\mathbf{q}}^{2}R(\mathbf{k},\mathbf{k}+\zeta\mathbf{q},i\zeta\nu_{n})\\
 & \times\left[R(\mathbf{k}',\mathbf{k}'+\zeta'\mathbf{q},i\zeta'\nu_{n})-R(\mathbf{k}',\mathbf{k}'+\zeta'\mathbf{q},i\zeta'\nu_{n}+i\zeta'\Omega_{n})\right].
\end{split}
\label{eq:mm_expressions}
\end{equation}
Here for simplicity of writing we have omitted the $\Omega_{n}=0$
term inside the brackets. However, in following calculations the static
contribution is always subtracted. We refer to the two contributions
to the memory matrix in Eq. (\ref{eq:mm_expressions}) as class-I
and class-II diagrams, respectively.

Naively, the class II diagrams are $\mathcal{O}(N^{-2})$ while the
class I diagram is $\mathcal{O}(N^{-1})$, both contributions need
to be kept when studying transport properties. This is because in
class I, the flavor indices $\alpha,$ $\alpha'$ are constrained
to be the same, while in class II they are not.

In the $\left\{ n_{\alpha\mathbf{k}}\right\} $ basis, the optical
conductivity is expressed as:
\begin{equation}
\sigma(\Omega)=\sum_{\alpha,\beta,\mathbf{k},\mathbf{k'}}\chi_{J_{x},\alpha\mathbf{k}}\left(\frac{1}{M(\Omega)-i\Omega\chi}\right)_{\alpha\mathbf{k},\mathbf{\beta k'}}\chi_{\beta\mathbf{k}',J_{x}}\label{eq:mm_conductivity-1}
\end{equation}
where $\chi_{J_{x},\alpha\mathbf{k}}\equiv\int_{0}^{\beta}\mathrm{d}\tau\langle J_{x}(\tau)n_{\alpha\mathbf{k}}(0)\rangle$
and $\chi_{\alpha\mathbf{k},\beta\mathbf{k'}}\equiv\int_{0}^{\beta}\mathrm{d}\tau\left[\langle n_{\alpha\mathbf{k}}(\tau)n_{\beta\mathbf{k'}}(0)\rangle-\langle n_{\alpha\mathbf{k}}\rangle\langle n_{\beta\mathbf{k'}}\rangle\right]$
are thermodynamic susceptibilities.

A straightforward analysis from Eq. (\ref{eq:mm_conductivity-1})
shows that $\sigma(\Omega)\sim\mathcal{O}\left(N^{2}\right)$, where
one factor of $N$ comes from the number of conducting channels, and
the other factor comes from the fact that the eigenvalues of the memory
matrix relevant to transport scale as $\mathcal{O}\left(N^{-1}\right)$.
As a result, the dc resistivity is $\rho\sim\mathcal{O}\left(N^{-2}\right)$
in the large $N$ limit, where our computation is formally justified.

\section{Momentum conservation\label{sec:Momentum-conservation}}

We now demonstrate that in absence of umklapp scattering and impurity
scattering, the total electronic momentum is exactly conserved within
our formalism. To derive momentum conservation, it is sufficient to
show that $\sum_{\alpha'\mathbf{k}'}\mathbf{k}M_{\alpha\mathbf{k},\alpha'\mathbf{k}'}=0$.
We first look at the class I diagram contribution:

\begin{equation}
\begin{split}\sum_{\alpha'\mathbf{k}'}\mathbf{k}'M_{\alpha\mathbf{k},\alpha'\mathbf{k}'}^{(1)} & =\frac{\lambda^{2}T}{N\Omega_{n}}\sum_{\mathbf{k'q}\nu_{n}}\mathbf{k}'D_{\mathbf{q},\nu_{n}+\Omega_{n}}\sum_{\zeta=\pm1}\left(\delta_{\mathbf{k'-k,\zeta q}}-\delta_{\mathbf{kk'}}\right)f_{\mathbf{k},\mathbf{k+\zeta q}}^{2}R\left(\mathbf{k},\mathbf{k+\zeta q},i\zeta\nu_{n}\right)\\
 & =\frac{\lambda^{2}T}{N\Omega_{n}}\sum_{\mathbf{q}\nu_{n}}D_{\mathbf{q},\nu_{n}+\Omega_{n}}\sum_{\zeta=\pm1}\zeta\mathbf{q}f_{\mathbf{k},\mathbf{k+\zeta q}}^{2}R\left(\mathbf{k},\mathbf{k+\zeta q},i\zeta\nu_{n}\right)
\end{split}
\end{equation}
Similarly the class II diagram contribution is:
\begin{equation}
\begin{split}\sum_{\alpha'\mathbf{k}'}\mathbf{k}'M_{\alpha\mathbf{k},\alpha'\mathbf{k'}}^{(2)}(i\Omega_{n}) & =-\frac{\lambda^{4}T}{N\Omega_{n}}\sum_{\mathbf{k'q}\nu_{n}}\mathbf{k}'D_{\mathbf{q},\nu_{n}}D_{\mathbf{q},\nu_{n}+\Omega_{n}}\sum_{\zeta\zeta'=\pm1}\zeta\zeta'f_{\mathbf{k},\mathbf{k}+\zeta\mathbf{q}}^{2}f_{\mathbf{k}',\mathbf{k}'+\zeta'\mathbf{q}}^{2}R(\mathbf{k},\mathbf{k}+\zeta\mathbf{q},i\zeta\nu_{n})\\
 & \times\left[R(\mathbf{k}',\mathbf{k}'+\zeta'\mathbf{q},i\zeta'\nu_{n})-R(\mathbf{k}',\mathbf{k}'+\zeta'\mathbf{q},i\zeta'\nu_{n}+i\zeta'\Omega_{n})\right].
\end{split}
\end{equation}
By change of variables $\mathbf{k}'\rightarrow\mathbf{k}'-\zeta'\mathbf{q}$,
and $\zeta'\rightarrow-\zeta'$, we get

\begin{equation}
\begin{split}\sum_{\alpha'\mathbf{k}'}\mathbf{k}'M_{\alpha\mathbf{k},\alpha'\mathbf{k'}}^{(2)}(i\Omega_{n}) & =-\frac{\lambda^{2}T}{N\Omega_{n}}\sum_{\mathbf{q}\nu_{n}}D_{\mathbf{q},\nu_{n}}D_{\mathbf{q},\nu_{n}+\Omega_{n}}\left(\Pi_{\mathbf{q},\nu_{n}+\Omega_{n}}-\Pi_{\mathbf{q},\nu_{n}}\right)\\
 & \times\sum_{\zeta=\pm1}\zeta\mathbf{q}f_{\mathbf{k},\mathbf{k}+\zeta\mathbf{q}}^{2}R(\mathbf{k},\mathbf{k}+\zeta\mathbf{q},i\zeta\nu_{n}),
\end{split}
\end{equation}
where
\begin{equation}
\Pi_{\mathbf{q},\nu_{n}}=\lambda^{2}\sum_{\mathbf{k}'}f_{\mathbf{k',k'+q}}^{2}R\left(\mathbf{k'},\mathbf{k'+q},i\nu_{n}\right)
\end{equation}
is the polarization bubble. Making use of Dyson's equation $D^{-1}=D_{0}^{-1}-\Pi$,
we obtain:
\begin{equation}
\begin{split}\sum_{\alpha'\mathbf{k}'}\mathbf{k}M_{\alpha\mathbf{k},\alpha'\mathbf{k'}}^{(2)}(i\Omega_{n}) & =-\frac{\lambda^{2}T}{N\Omega_{n}}\sum_{\mathbf{q}\nu_{n}}\left(D_{\mathbf{q},\nu_{n}+\Omega_{n}}-D_{\mathbf{q},\nu_{n}}\right)\sum_{\zeta=\pm1}\zeta\mathbf{q}f_{\mathbf{k},\mathbf{k}+\zeta\mathbf{q}}^{2}R(\mathbf{k},\mathbf{k}+\zeta\mathbf{q},i\zeta\nu_{n})\\
 & -\frac{\lambda^{2}T}{N\Omega_{n}}\sum_{\mathbf{q}\nu_{n}}D_{\mathbf{q},\nu_{n}}D_{\mathbf{q},\nu_{n}+\Omega_{n}}\left(D_{0,\mathbf{q},\nu_{n}+\Omega_{n}}^{-1}-D_{0,\mathbf{q},\nu_{n}}^{-1}\right)\sum_{\zeta=\pm1}\zeta\mathbf{q}f_{\mathbf{k},\mathbf{k}+\zeta\mathbf{q}}^{2}R(\mathbf{k},\mathbf{k}+\zeta\mathbf{q},i\zeta\nu_{n}).
\end{split}
\end{equation}

Summing up the two classes of diagrams:
\begin{equation}
\begin{split}\sum_{\alpha'\mathbf{k}'}\mathbf{k}'M_{\alpha\mathbf{k},\alpha'\mathbf{k'}}^{(1+2)}(i\Omega_{n}) & =\frac{\lambda^{2}T}{N\Omega_{n}}\sum_{\mathbf{q}\nu_{n}}D_{\mathbf{q},\nu_{n}}\sum_{\zeta=\pm1}\zeta\mathbf{q}f_{\mathbf{k},\mathbf{k}+\zeta\mathbf{q}}^{2}R(\mathbf{k},\mathbf{k}+\zeta\mathbf{q},i\zeta\nu_{n})\\
 & -\frac{\lambda^{2}T}{N\Omega_{n}}\sum_{\mathbf{q}\nu_{n}}D_{\mathbf{q},\nu_{n}}D_{\mathbf{q},\nu_{n}+\Omega_{n}}\left(D_{0,\mathbf{q},\nu_{n}+\Omega_{n}}^{-1}-D_{0,\mathbf{q},\nu_{n}}^{-1}\right)\sum_{\zeta=\pm1}\zeta\mathbf{q}f_{\mathbf{k},\mathbf{k}+\zeta\mathbf{q}}^{2}R(\mathbf{k},\mathbf{k}+\zeta\mathbf{q},i\zeta\nu_{n}).
\end{split}
\end{equation}
The first term is the static contribution $\mathcal{G}_{\mathbf{\alpha k,\alpha'k'}}(\Omega_{n}=0)$
, and should be subtracted at the end of the computation. If the
dynamics of nematic fluctuations is generated by coupling to the electrons,
as discussed previously, then $D_{0}$ is frequency independent. In
that case, the second term vanishes as well, and momentum is conserved.
Whether nematic fluctuations can act as a ``momentum sink'' is fundamentally
dependent on whether their propagator has its own independent dynamics.

\section{Low temperature and dc limit\label{sec:Low-temperature-and}}

We derive the expressions of the memory matrix in the dc limit and
for temperature $T$ much smaller than the Fermi energy $\varepsilon_{F}$.
It is convenient to invoke spectral representation for the nematic
propagator. Define $D_{\mathbf{q},\nu_{n}}^{-1}\equiv r_{\mathbf{q}}+\gamma_{\mathbf{q}}|\nu_{n}|$,
we have
\begin{equation}
D_{\mathbf{q},\nu_{n}}=\int\frac{\mathrm{d}\omega}{\pi}\frac{\text{Im}D_{\mathbf{q},\omega}}{\omega-i\nu_{n}},
\end{equation}
where the spectral function is
\begin{equation}
\text{Im}D_{\mathbf{q},\omega}=\frac{\gamma_{\mathbf{q}}\omega}{r_{\mathbf{q}}^{2}+\gamma_{\mathbf{q}}^{2}\omega^{2}}.
\end{equation}

For a given momentum transfer, the spectral function is peaked at
$\omega_{\mathbf{q}}\equiv\frac{r_{\mathbf{q}}}{\gamma_{\mathbf{q}}}\sim\frac{\varepsilon_{F}^{2}}{\lambda^{2}}\left(\frac{|\mathbf{q}|}{k_{F}}\right)^{3}$.
This can be interpreted as the dispersion relation of Landau-damped
nematic fluctuations. At small wave-vectors, $\omega_{\mathbf{q}}\ll\varepsilon_{F}$.

First, consider the class I diagram. The sum over bosonic Matsubara
frequency $\nu_{n}$ can be explicitly performed using spectral representation,
giving rise to:

\begin{equation}
\begin{split}M_{\alpha\mathbf{k},\alpha'\mathbf{k'}}^{(1)}(i\Omega_{n}) & =\delta_{\alpha\alpha'}\frac{\lambda^{2}}{N\Omega_{n}}\sum_{\mathbf{q}}\left(\delta_{\mathbf{k}'-\mathbf{k},\mathbf{q}}-\delta_{\mathbf{kk}'}\right)f_{\mathbf{k},\mathbf{k+q}}^{2}\int\frac{\mathrm{d}\omega}{\pi}\text{Im}D_{\mathbf{q},\omega}\left(n_{F,\mathbf{k+q}}-n_{F,\mathbf{k}}\right)\\
 & \times\left\{ \frac{n_{B}\left(\omega\right)-n_{B}\left(\varepsilon_{\mathbf{k+q}}-\varepsilon_{\mathbf{k}}\right)}{\omega-i\Omega_{n}-\left(\varepsilon_{\mathbf{k+q}}-\varepsilon_{\mathbf{k}}\right)}+h.c.-\text{static part}\right\}
\end{split}
\end{equation}
where $n_{F,\mathbf{k}}$ is short for the Fermi-Dirac distribution
function $n_{F}\left(\varepsilon_{\mathbf{k}}\right)$, and $n_{B}(\omega)$
is the Bose-Einstein distribution function. Following an analytic
continuation $i\Omega_{n}\rightarrow\Omega+i\delta$, we obtain:
\begin{equation}
\begin{split}M_{\alpha\mathbf{k},\alpha'\mathbf{k'}}^{(1)}(\Omega+i\delta) & =-\frac{1}{i\Omega}\delta_{\alpha\alpha'}\frac{\lambda^{2}}{N}\sum_{\mathbf{q}}\left(\delta_{\mathbf{k}'-\mathbf{k},\mathbf{q}}-\delta_{\mathbf{kk}'}\right)f_{\mathbf{k},\mathbf{k+q}}^{2}\int\frac{\mathrm{d}\omega}{\pi}\text{Im}D_{\mathbf{q},\omega}\left(n_{F,\mathbf{k+q}}-n_{F,\mathbf{k}}\right)\\
 & \times\left\{ \frac{n_{B}\left(\omega\right)-n_{B}\left(\varepsilon_{\mathbf{k+q}}-\varepsilon_{\mathbf{k}}\right)}{\omega-\left(\varepsilon_{\mathbf{k+q}}-\varepsilon_{\mathbf{k}}\right)-\Omega-i\delta}+\frac{n_{B}\left(\omega\right)-n_{B}\left(\varepsilon_{\mathbf{k+q}}-\varepsilon_{\mathbf{k}}\right)}{\omega-\left(\varepsilon_{\mathbf{k+q}}-\varepsilon_{\mathbf{k}}\right)+\Omega+i\delta}-\text{static part}\right\}
\end{split}
\end{equation}
Making use of $\frac{1}{\omega-i\delta}=\mathcal{P}\frac{1}{\omega}+i\pi\delta\left(\omega\right)$,
and taking the dc limit $\Omega\rightarrow0$, we observe that the
principle part on the second line cancels the static part, and only
the imaginary part contributes to the real frequency memory matrix.
We obtain:
\begin{equation}
\begin{split}M_{\alpha\mathbf{k},\alpha'\mathbf{k'}}^{(1)} & \rightarrow\delta_{\alpha\alpha'}\frac{2\pi\lambda^{2}}{N}\sum_{\mathbf{q}}\left(\delta_{\mathbf{k}'-\mathbf{k},\mathbf{q}}-\delta_{\mathbf{kk}'}\right)f_{\mathbf{k},\mathbf{k+q}}^{2}\left(n_{F,\mathbf{k+q}}-n_{F,\mathbf{k}}\right)\\
 & \times\int\frac{\mathrm{d}\omega}{\pi}\text{Im}D_{\mathbf{q},\omega}\left(-\frac{\partial n_{B}}{\partial\omega}\right)\delta\left(\omega-\varepsilon_{\mathbf{k+q}}+\varepsilon_{\mathbf{k}}\right).
\end{split}
\end{equation}
This is equivalent to the linearized collision integral studied in
many previous works, e.g. Ref. \cite{hlubina95}, where the class
II diagrams were not considered. From the second line, we see that
the leading contribution to the frequency integration comes from low
frequencies, $\omega\sim\text{min}\left(\omega_{\mathbf{q}},T\right)\ll\varepsilon_{F}$.
As a result, the above expression can be further simplified to be:
\begin{equation}
\begin{split}M_{\alpha\mathbf{k},\alpha'\mathbf{k'}}^{(1)} & \approx\delta_{\alpha\alpha'}\frac{2\pi\lambda^{2}}{N}\sum_{\mathbf{q}}\left(\delta_{\mathbf{kk}'}-\delta_{\mathbf{k}'-\mathbf{k},\mathbf{q}}\right)f_{\mathbf{k},\mathbf{k+q}}^{2}V_{\mathbf{q}}\delta\left(\varepsilon_{\mathbf{k}}\right)\delta\left(\varepsilon_{\mathbf{k+q}}\right)\end{split}
,\label{eq:mm_classI-1}
\end{equation}
where we have used $n_{F,\mathbf{k+q}}-n_{F,\mathbf{k}}\approx-\omega\delta\left(\varepsilon_{\mathbf{k}}\right)$,
and defined
\begin{equation}
V_{\mathbf{q}}\equiv\int\frac{\mathrm{d}\omega}{\pi}\omega\text{Im}D_{\mathbf{q},\omega}\left(-\frac{\partial n_{B}}{\partial\omega}\right).
\end{equation}
We see that the dominant processes contributing to class I diagram
comes from the Fermi surface.

Next we derive the expression for class II diagrams in the dc limit.
There are two terms shown in Eq. (\ref{eq:mm_expressions}), corresponding
to Figs. 1(c) and (d) of the main text. We discuss the (c) contribution
first. We can rewrite the memory matrix in the following way:
\begin{equation}
\begin{split}M_{\alpha\mathbf{k},\alpha'\mathbf{k'}}^{(2,c)}(i\Omega_{n}) & =-\frac{\lambda^{4}}{N^{2}}\sum_{\mathbf{q};\zeta\zeta'=\pm1}\zeta\zeta'f_{\mathbf{k},\mathbf{k}+\zeta\mathbf{q}}^{2}f_{\mathbf{k}',\mathbf{k}'+\zeta'\mathbf{q}}^{2}\left[\frac{T}{\Omega_{n}}\sum_{\nu_{n}}f\left(i\nu_{n}\right)g\left(i\nu_{n}+i\Omega_{n}\right)\right],\\
f\left(i\nu_{n}\right) & =D_{\mathbf{q},\nu_{n}}R(\mathbf{k},\mathbf{k}+\zeta\mathbf{q},i\zeta\nu_{n})R(\mathbf{k}',\mathbf{k}'+\zeta'\mathbf{q},i\zeta'\nu_{n}),\\
g\left(i\nu_{n}+i\Omega_{n}\right) & =D_{\mathbf{q},\nu_{n}+\Omega_{n}}.
\end{split}
\end{equation}
Here we defined two analytic functions $f$ and $g$. Both $f$ and
$g$ are analytic everywhere except on the real axis. Hence, we can
use a spectral representation for both functions, to obtain:
\begin{equation}
\begin{split}M_{\alpha\mathbf{k},\alpha'\mathbf{k'}}^{(2,c)}(i\Omega_{n}) & =-\frac{\lambda^{4}}{N^{2}}\sum_{\mathbf{q};\zeta\zeta'=\pm1}\zeta\zeta'f_{\mathbf{k},\mathbf{k}+\zeta\mathbf{q}}^{2}f_{\mathbf{k}',\mathbf{k}'+\zeta'\mathbf{q}}^{2}\\
 & \times\frac{1}{\Omega_{n}}\int\frac{\mathrm{d}\omega_{1}\mathrm{d}\omega_{2}}{\pi^{2}}\int\frac{\mathrm{d}z}{2\pi i}n_{B}\left(z\right)\frac{\text{Im}f\left(\omega_{1}\right)}{z-\omega_{1}}\frac{\text{Im}g\left(\omega_{2}\right)}{z-\omega_{2}+i\Omega_{n}}
\end{split}
\end{equation}
Carrying out the Matsubara sum:
\begin{equation}
\begin{split}M_{\alpha\mathbf{k},\alpha'\mathbf{k'}}^{(2,c)}(i\Omega_{n}) & =-\frac{\lambda^{4}}{N^{2}}\sum_{\mathbf{q};\zeta\zeta'=\pm1}\zeta\zeta'f_{\mathbf{k},\mathbf{k}+\zeta\mathbf{q}}^{2}f_{\mathbf{k}',\mathbf{k}'+\zeta'\mathbf{q}}^{2}\\
 & \times\frac{1}{\Omega_{n}}\int\frac{\mathrm{d}\omega_{1}\mathrm{d}\omega_{2}}{\pi^{2}}\text{Im}f\left(\omega_{1}\right)\text{Im}g\left(\omega_{2}\right)\frac{n_{B}\left(\omega_{2}\right)-n_{B}\left(\omega_{1}\right)}{\omega_{1}-\omega_{2}+i\Omega_{n}}
\end{split}
\end{equation}
Following an analytic continuation, $i\Omega_{n}\rightarrow\Omega+i\delta$,
and taking the zero-frequency limit, we obtain:
\begin{equation}
\begin{split}M_{\alpha\mathbf{k},\alpha'\mathbf{k'}}^{(2,c)}(\Omega\rightarrow0) & =\frac{\lambda^{4}}{N^{2}}\sum_{\mathbf{q};\zeta\zeta'=\pm1}\zeta\zeta'f_{\mathbf{k},\mathbf{k}+\zeta\mathbf{q}}^{2}f_{\mathbf{k}',\mathbf{k}'+\zeta'\mathbf{q}}^{2}\times\int\frac{\mathrm{d}\omega}{\pi}\text{Im}f\left(\omega\right)\text{Im}g\left(\omega\right)\left(-\frac{\partial n_{B}}{\partial\omega}\right).\end{split}
\end{equation}

The imaginary part of the $f$ and $g$ functions are, respectively:
\[
\begin{split}\text{Im}f\left(\omega\right) & =\text{Im}\left(DR_{1}R_{2}\right)\\
 & =\text{Im}D\left(\text{Re}R_{1}\text{Re}R_{2}-\text{Im}R_{1}\text{Im}R_{2}\right)+\text{Re}D\left(\text{Re}R_{1}\text{Im}R_{2}+\text{Im}R_{1}\text{Re}R_{2}\right)
\end{split}
\]
and
\[
\text{Im}g\left(\omega\right)=\text{Im}D
\]
Here we have used a short-hand notation: $R_{1}=R(\mathbf{k},\mathbf{k}+\zeta\mathbf{q},i\zeta\nu_{n})$
and $R_{2}=R(\mathbf{k}',\mathbf{k}'+\zeta'\mathbf{q},i\zeta'\nu_{n})$.

Following a similar analysis, the second term in the bracket in the
expression for $M^{(2)}$ {[}Eq. (\ref{eq:mm_expressions}){]} can
also be carried out, except in this case,
\begin{equation}
\begin{split}\tilde{f}\left(i\nu_{n}\right) & =D_{\mathbf{q},\nu_{n}}R(\mathbf{k},\mathbf{k}+\zeta\mathbf{q},i\zeta\nu_{n}),\\
\tilde{g}\left(i\nu_{n}+i\Omega_{n}\right) & =D_{\mathbf{q},\nu_{n}+\Omega_{n}}R(\mathbf{k}',\mathbf{k}'+\zeta'\mathbf{q},i\zeta'\nu_{n}+i\zeta'\Omega_{n}).
\end{split}
\end{equation}
As a result,
\begin{equation}
\begin{split}\text{Im}\tilde{f}\left(\omega\right) & =\text{Im}D\text{Re}R_{1}+\text{Re}D\text{Im}R_{1},\\
\text{Im}\tilde{g}\left(\omega\right) & =\text{Im}D\text{Re}R_{2}+\text{Re}D\text{Im}R_{2}.
\end{split}
\end{equation}
Combining the two contributions, we get the following expression for
class II diagram in dc limit:
\begin{equation}
\begin{split}M_{\alpha\mathbf{k},\alpha'\mathbf{k'}}^{(2)} & =-\frac{\lambda^{4}}{N^{2}}\sum_{\mathbf{q};\zeta\zeta'=\pm1}\zeta\zeta'f_{\mathbf{k},\mathbf{k}+\zeta\mathbf{q}}^{2}f_{\mathbf{k}',\mathbf{k}'+\zeta'\mathbf{q}}^{2}\\
 & \times\int\frac{\mathrm{d}\omega}{\pi}|D_{\mathbf{q},\omega}|^{2}\text{Im}R\left(\mathbf{k},\mathbf{k+\zeta q},\zeta\omega\right)\text{Im}R\left(\mathbf{k}',\mathbf{k'+\zeta'q},\zeta'\omega\right)\left(-\frac{\partial n_{B}}{\partial\omega}\right)
\end{split}
,
\end{equation}
where
\begin{equation}
\text{Im}R\left(\mathbf{k},\mathbf{k+\zeta q},\zeta\omega\right)=\text{Im}\frac{n_{F,\mathbf{p}}-n_{F,\mathbf{k}}}{\varepsilon_{\mathbf{k}}-\varepsilon_{\mathbf{k+\zeta q}}+\zeta\left(\omega+i\delta\right)}=-\pi\zeta\left(n_{F,\mathbf{k+\zeta q}}-n_{F,\mathbf{k}}\right)\delta\left(\varepsilon_{\mathbf{k}}-\varepsilon_{\mathbf{k+\zeta q}}+\zeta\omega\right),
\end{equation}
and
\begin{equation}
\text{Im}R\left(\mathbf{k}',\mathbf{k'+\zeta'q},\zeta'\omega\right)=-\pi\zeta'\left(n_{F,\mathbf{k'+\zeta'q}}-n_{F,\mathbf{k'}}\right)\delta\left(\varepsilon_{\mathbf{k}'}-\varepsilon_{\mathbf{k'+\zeta'q}}+\zeta'\omega\right).
\end{equation}

We see again that the frequency integration is constrained to be at
small frequencies $\omega<\text{min}\left(T,\omega_{\mathbf{q}}\right)$.
Approximating $-\pi\zeta\left(n_{F,\mathbf{k+\zeta q}}-n_{F,\mathbf{k}}\right)\approx\pi\omega\delta\left(\varepsilon_{\mathbf{k}}\right)$
and $-\pi\zeta'\left(n_{F,\mathbf{k'+\zeta'q}}-n_{F,\mathbf{k'}}\right)\approx\pi\omega\delta\left(\varepsilon_{\mathbf{k'}}\right)$,
we see that:
\begin{equation}
\begin{split}M_{\alpha\mathbf{k},\alpha'\mathbf{k'}}^{(2)} & =-\frac{\pi\lambda^{4}}{N^{2}}\sum_{\mathbf{q};\zeta\zeta'=\pm1}\zeta\zeta'f_{\mathbf{k},\mathbf{k}+\zeta\mathbf{q}}^{2}f_{\mathbf{k}',\mathbf{k}'+\zeta'\mathbf{q}}^{2}\delta\left(\varepsilon_{\mathbf{k}}\right)\delta\left(\varepsilon_{\mathbf{k'}}\right)\\
 & \times\int\mathrm{d}\omega\omega^{2}|D_{\mathbf{q},\omega}|^{2}\left(-\frac{\partial n_{B}}{\partial\omega}\right)\delta\left(\varepsilon_{\mathbf{k+\zeta q}}-\zeta\omega\right)\delta\left(\varepsilon_{\mathbf{k'+\zeta'q}}-\zeta'\omega\right)
\end{split}
\end{equation}

Since the frequency is small compared to the Fermi energy, we can
further approximate the two $\delta$-function constraints as $\delta\left(\varepsilon_{\mathbf{\mathbf{k+\zeta q}}}\right)\delta\left(\varepsilon_{\mathbf{k'+\zeta'q}}\right)$,
placing all four momentum states on the Fermi surface. We then make
use of the identity:
\begin{equation}
\int\mathrm{d}\omega\omega^{2}|D_{\mathbf{q},\omega}|^{2}\left(-\frac{\partial n_{B}}{\partial\omega}\right)=\frac{1}{\gamma_{\mathbf{q}}}\int\mathrm{d}\omega\omega\text{Im}D{}_{\mathbf{q},\omega}\left(-\frac{\partial n_{B}}{\partial\omega}\right)=\frac{\pi V_{\mathbf{q}}(T)}{\gamma_{\mathbf{q}}},
\end{equation}
where $V_{\mathbf{q}}(T)$ is defined in Eq. (\ref{eq:Vq}). The final
expression for class II contribution to the memory matrix is:
\begin{equation}
\begin{split}M_{\alpha\mathbf{k},\alpha'\mathbf{k'}}^{(2)} & \approx-\frac{2\pi^{2}\lambda^{4}}{N^{2}}\sum_{\mathbf{q};\zeta'=\pm1}\zeta'\frac{V_{\mathbf{q}}}{\gamma_{\mathbf{q}}}f_{\mathbf{k},\mathbf{k}+\mathbf{q}}^{2}f_{\mathbf{k}',\mathbf{k}'+\zeta'\mathbf{q}}^{2}\delta\left(\varepsilon_{\mathbf{k}}\right)\delta\left(\varepsilon_{\mathbf{k+q}}\right)\delta\left(\varepsilon_{\mathbf{k'}}\right)\delta\left(\varepsilon_{\mathbf{k'+\zeta'q}}\right)\end{split}
\label{eq:mm_classII-1}
\end{equation}
Here we have made use of the invariance under $\left(\zeta,\zeta'\right)\rightarrow\left(-\zeta,-\zeta'\right)$
to eliminate $\zeta$ in the expression.

\section{Harmonic basis and additional conservation laws\label{sec:Harmonic-basis-and}}

In this section we show that in a two-dimensional system with a single,
convex Fermi surface and with no umklapp scattering, there are additional
approximately conserved modes that emerge at low temperatures and
frequencies due to projection of scattering processes onto the Fermi
surface.

We start from Eqs. (\ref{eq:mm_classI}) and (\ref{eq:mm_classII}) for the
memory matrix in the dc limit. It is convenient to split $M_{\alpha\mathbf{k},\alpha'\mathbf{k'}}^{(2)}$
as follows:
\begin{equation}
\begin{split}M_{\alpha\mathbf{k},\alpha'\mathbf{k'}}^{(2,+)} & =-\frac{2\pi^{2}\lambda^{4}}{N^{2}}\sum_{\mathbf{q}}\frac{V_{\mathbf{q}}}{\gamma_{\mathbf{q}}}f_{\mathbf{k},\mathbf{k}+\mathbf{q}}^{2}f_{\mathbf{k}',\mathbf{k}'+\mathbf{q}}^{2}\delta\left(\varepsilon_{\mathbf{k}}\right)\delta\left(\varepsilon_{\mathbf{k+q}}\right)\delta\left(\varepsilon_{\mathbf{k'}}\right)\delta\left(\varepsilon_{\mathbf{k'+q}}\right),\\
M_{\alpha\mathbf{k},\alpha'\mathbf{k'}}^{(2,-)} & =\frac{2\pi^{2}\lambda^{4}}{N^{2}}\sum_{\mathbf{q}}\frac{V_{\mathbf{q}}}{\gamma_{\mathbf{q}}}f_{\mathbf{k},\mathbf{k}+\mathbf{q}}^{2}f_{\mathbf{k}',\mathbf{k}'-\mathbf{q}}^{2}\delta\left(\varepsilon_{\mathbf{k}}\right)\delta\left(\varepsilon_{\mathbf{k+q}}\right)\delta\left(\varepsilon_{\mathbf{k'}}\right)\delta\left(\varepsilon_{\mathbf{k'-q}}\right).
\end{split}
\end{equation}
The two class II terms are related to each other via
\begin{equation}
M_{\alpha\mathbf{k},\alpha'\mathbf{k'}}^{(2,+)}=-M_{\alpha\mathbf{k},\alpha'-\mathbf{k'}}^{(2,-)}.
\end{equation}

The Landau damping parameter $\gamma_{\mathbf{q}}$ appearing in $M_{\alpha\mathbf{k},\alpha'\mathbf{k'}}^{(2,\pm)}$
is self-consistently calculated from the electron polarization bubble:
\begin{equation}
\gamma_{\mathbf{q}}=\pi\lambda^{2}\sum_{\mathbf{k}}f_{\mathbf{k},\mathbf{k+q}}^{2}\delta\left(\varepsilon_{\mathbf{k}}\right)\delta\left(\varepsilon_{\mathbf{k+q}}\right).
\end{equation}
Note that $\gamma_{\mathbf{q}}\sim\mathcal{O}\left(N^{0}\right)$,
since nematic boson can decay into all flavors of electrons.

We define memory matrix in the harmonic basis:
\begin{equation}
M_{\alpha n,\alpha'm}\equiv\sum_{\mathbf{kk'}}h_{n\mathbf{k}}^{*}h_{m\mathbf{k'}}M_{\alpha\mathbf{k},\alpha'\mathbf{k'}},
\end{equation}
where $h_{n\mathbf{k}}=\exp\left(in\theta_{\mathbf{k}}\right)$, and
$h_{n,-\mathbf{k}}=\left(-1\right)^{n}h_{n\mathbf{k}}$. By introducing
a short-hand notation $d_{n\mathbf{kk'}}\equiv h_{n\mathbf{k}}-h_{n\mathbf{k'}}$,
Class I contribution becomes:
\begin{equation}
M_{\alpha n,\alpha'm}^{(1)}=\delta_{\alpha\alpha'}\frac{\pi\lambda^{2}}{N}\sum_{\mathbf{kk'}}d_{n\mathbf{k}\mathbf{k'}}^{*}d_{m\mathbf{kk'}}f_{\mathbf{k},\mathbf{k'}}^{2}V_{\mathbf{k'-k}}\delta\left(\varepsilon_{\mathbf{k}}\right)\delta\left(\varepsilon_{\mathbf{k'}}\right)
\end{equation}
Here we made use of symmetry under $\mathbf{k\leftrightarrow-\mathbf{k'}}$
to derive the above expression. Similarly, class II contributions
can be shown to be of the form
\begin{equation}
\begin{split}M_{\alpha n,\alpha'm}^{(2)} & =\frac{\pi^{2}\lambda^{4}}{N^{2}}\frac{1-\left(-1\right)^{n}}{2}\frac{1-\left(-1\right)^{m}}{2}\sum_{\mathbf{kq}}\frac{V_{\mathbf{q}}}{\gamma_{\mathbf{q}}}f_{\mathbf{k},\mathbf{k}+\mathbf{q}}^{2}\delta\left(\varepsilon_{\mathbf{k}}\right)\delta\left(\varepsilon_{\mathbf{k+q}}\right)\\
 & \times\left[d_{n\mathbf{k}\mathbf{k'}}^{*}\sum_{\mathbf{k'}}h_{m\mathbf{k'}}f_{\mathbf{k}',\mathbf{k}'-\mathbf{q}}^{2}\delta\left(\varepsilon_{\mathbf{k'}}\right)\delta\left(\varepsilon_{\mathbf{k'-q}}\right)\right]
\end{split}
\end{equation}
For a convex Fermi surface and a given momentum transfer $\mathbf{q}$,
to place all four momentum states $\left\{ \mathbf{k},\mathbf{k+q},\mathbf{k'},\mathbf{k'-q}\right\} $
on the Fermi surface implies that $\mathbf{k'}=\mathbf{-k}$ or $\mathbf{k'}=\mathbf{k+q}$.
As a result, the bracket on the second line can be simplied to be
\[
\frac{\gamma_{\mathbf{q}}}{2\pi\lambda^{2}}d_{n\mathbf{k}\mathbf{k'}}^{*}\left(h_{m\mathbf{k+q}}+h_{m,-\mathbf{k}}\right).
\]

By change of variables: $\mathbf{k+q}\rightarrow\mathbf{k'}$, we
can combine the two class of diagrams to arrive at the final expression
for memory matrix in the harmonics basis:
\begin{equation}
\begin{split}M_{\alpha n,\alpha'm} & =\frac{\pi\lambda^{2}}{N}\sum_{\mathbf{kk'}}d_{n\mathbf{kk'}}^{*}d_{m\mathbf{\mathbf{kk'}}}V_{\mathbf{k'-k}}f_{\mathbf{k},\mathbf{k}'}^{2}\delta\left(\varepsilon_{\mathbf{k}}\right)\delta\left(\varepsilon_{\mathbf{k'}}\right)\\
 & \times\left[\delta_{\alpha\alpha'}-\frac{1}{N}\frac{1-\left(-1\right)^{n}}{2}\frac{1-\left(-1\right)^{m}}{2}\right]
\end{split}
\label{eq:harmonic_basis}
\end{equation}

Eq.~(\ref{eq:harmonic_basis}) shows a dichotomy between even/odd
parity modes. The second term in the bracket vanishes for even parity
modes, and as a result, these modes are fast-relaxing from critical
nematic fluctuations. In contrast, the relaxation of odd-parity modes
is strongly renormalized by scattering processes described by class
II diagrams. For a given pair of odd values of $\left\{ n,m\right\} $,
$M_{\alpha n,\alpha'm}$ has a simple $N\times N$ matrix structure
in the flavor basis, given by $1-\frac{1}{N}$ along the diagonal
and $-\frac{1}{N}$ otherwise:
\[
\left(\begin{array}{ccc}
1-\frac{1}{N} & -\frac{1}{N} & \cdots\\
-\frac{1}{N} & 1-\frac{1}{N} & \cdots\\
\vdots & \vdots & \ddots
\end{array}\right)_{N\times N}
\]

There is one zero mode associated with this matrix structure, with
an eigenvector given by $\frac{1}{\sqrt{N}}\left(1,1,\cdots\right)$.
This corresponds to a conservation of the generalized current $J_{n}\equiv\sum_{\alpha\mathbf{k}}h_{n\mathbf{k}}c_{\alpha\mathbf{k}}^{\dagger}c_{\alpha\mathbf{k}}$,
whenever $n$ is odd. Note that the total electron momentum $\mathbf{P}$
can be represented as a linear combination of $J_{n}$, and therefore
is also conserved.

We point out that except for momentum, all other odd-parity currents
are only quasi-conserved. Scattering processes away from the Fermi
surface will lift these zero modes to a decay rate which is $\mathcal{O}\left(T^{2}/\varepsilon_{F}^{2}\right)$
smaller than the relaxation of even-parity modes \cite{maslov11,ledwidth17}.

\section{Generalization to systems with multiple Fermi sheets\label{sec:Generalization-to-systems}}

The generalization of the memory matrix to multiple electron bands
is straightforward. For simplicity we omit the flavor index, and consider
the case $N=1$. The generalization to the case of a general $N$
is straightforward. In addition, we assume that the critical nematic
fluctuations can only scatter electrons within each band, since the
fluctuations carry a small momentum. The Lagrangian is taken to be:
\begin{equation}
L_{\text{multi}}=L_{0,\text{multi}}+\lambda\sum_{i\mathbf{k}\mathbf{q}}\phi_{\mathbf{q}}f_{i,\mathbf{k,k+q}}c_{i\mathbf{k+q}}^{\dagger}c_{i\mathbf{k}}
\end{equation}
where
\begin{equation}
L_{0,\text{multi}}=\sum_{i,\mathbf{k}}c_{i\mathbf{k}}^{\dagger}(\partial_{\tau}+\varepsilon_{i\mathbf{k}})c_{i\mathbf{k}}+\frac{1}{2}\sum_{\mathbf{q}}D_{0,\mathbf{q}}^{-1}|\phi_{\mathbf{q}}|^{2}.
\end{equation}
Here $i=1\dots N_{\text{band}}$ is the band label, and $\varepsilon_{i\mathbf{k}}$
is the dispersion for the $i$-th electron band. The other terms in
The derivation of the memory matrix parallels that of the multi-flavor,
single-band scenario, with the only difference coming from different
dispersions and form factors for different electron bands. There are
two class of Feynman diagrams for the memory matrix, and at low temperatures
and in the dc limit, the expressions for the two class of diagrams
are:
\begin{equation}
\begin{split}M_{i\mathbf{k},j\mathbf{k'}}^{(1)} & =\delta_{ij}2\pi\lambda^{2}\sum_{\mathbf{q}}\left(\delta_{\mathbf{kk}'}-\delta_{\mathbf{k}'-\mathbf{k},\mathbf{q}}\right)f_{i,\mathbf{k},\mathbf{k+q}}^{2}V_{\mathbf{q}}\delta\left(\varepsilon_{i\mathbf{k}}\right)\delta\left(\varepsilon_{i\mathbf{k+q}}\right),\\
M_{i\mathbf{k},j\mathbf{k'}}^{(2,+)} & =-2\pi^{2}\lambda^{4}\sum_{\mathbf{q}}\frac{V_{\mathbf{q}}}{\gamma_{\mathbf{q}}}f_{i,\mathbf{k},\mathbf{k}+\mathbf{q}}^{2}f_{j,\mathbf{k}',\mathbf{k}'+\mathbf{q}}^{2}\delta\left(\varepsilon_{i\mathbf{k}}\right)\delta\left(\varepsilon_{i\mathbf{k+q}}\right)\delta\left(\varepsilon_{j\mathbf{k'}}\right)\delta\left(\varepsilon_{j\mathbf{k'+q}}\right),\\
M_{i\mathbf{k},j\mathbf{k'}}^{(2,-)} & =2\pi^{2}\lambda^{4}\sum_{\mathbf{q}}\frac{V_{\mathbf{q}}}{\gamma_{\mathbf{q}}}f_{i,\mathbf{k},\mathbf{k}+\mathbf{q}}^{2}f_{j,\mathbf{k}',\mathbf{k}'-\mathbf{q}}^{2}\delta\left(\varepsilon_{i\mathbf{k}}\right)\delta\left(\varepsilon_{i\mathbf{k+q}}\right)\delta\left(\varepsilon_{j\mathbf{k'}}\right)\delta\left(\varepsilon_{j\mathbf{k'-q}}\right).
\end{split}
\label{eq:mm_multiband}
\end{equation}

The Landau damping coefficient describes a sum of scattering processes
on each electron band:
\begin{equation}
\gamma_{\mathbf{q}}=\pi\lambda^{2}\sum_{i\mathbf{k}}f_{i,\mathbf{k},\mathbf{k+q}}^{2}\delta\left(\varepsilon_{i\mathbf{k}}\right)\delta\left(\varepsilon_{i\mathbf{k+q}}\right).
\end{equation}
Similar to the discussion in Sec. IV, we can rewrite the memory matrix
in the harmonics basis. Class I diagram can be shown to be:
\begin{equation}
M_{in,jm}^{(1)}=\delta_{ij}\pi\lambda^{2}\sum_{\mathbf{kk'}}d_{in,\mathbf{kk'}}^{*}d_{jm,\mathbf{kk'}}f_{i,\mathbf{k},\mathbf{k'}}^{2}V_{\mathbf{k'-k}}\delta\left(\varepsilon_{i\mathbf{k}}\right)\delta\left(\varepsilon_{i\mathbf{k'}}\right)
\end{equation}
where we have defined a short-hand notation: $d_{in,\mathbf{kk'}}=h_{in\mathbf{k}}-h_{in\mathbf{k'}}$.
Similarly, class II diagrams are expressed as:
\begin{equation}
\begin{split}M_{in,jm}^{(2)} & =\pi^{2}\lambda^{4}\frac{1-\left(-1\right)^{n}}{2}\frac{1-\left(-1\right)^{m}}{2}\sum_{\mathbf{kq}}\frac{V_{\mathbf{q}}}{\gamma_{\mathbf{q}}}f_{i,\mathbf{k},\mathbf{k}+\mathbf{q}}^{2}\delta\left(\varepsilon_{i\mathbf{k}}\right)\delta\left(\varepsilon_{i\mathbf{k+q}}\right)\\
 & \times\left[d_{in,\mathbf{k}\mathbf{k'}}^{*}\sum_{\mathbf{k'}}h_{jm\mathbf{k'}}f_{j,\mathbf{k}',\mathbf{k}'-\mathbf{q}}^{2}\delta\left(\varepsilon_{j\mathbf{k'}}\right)\delta\left(\varepsilon_{j\mathbf{k'-q}}\right)\right].
\end{split}
\end{equation}
The term on the second line evaluates to:
\begin{equation}
\frac{\gamma_{j\mathbf{q}}}{2\pi\lambda^{2}}d_{in,\mathbf{k}\mathbf{k'}}^{*}\sum_{\mathbf{p}_{j}}h_{jm\mathbf{p}_{j}}\label{eq:di}
\end{equation}
where $\gamma_{j\mathbf{q}}$ is the Landau damping due to $j$-th
band, such that $\gamma_{\mathbf{q}}=\sum_{j}\gamma_{j\mathbf{q}}$.
In Eq. (\ref{eq:di}), the $\mathbf{p}_{j}$'s are solutions of the
two equations $\varepsilon_{j\mathbf{p}_{j}}=0$, $\varepsilon_{j\mathbf{p}_{j}-\mathbf{q}}=0$.
Since the momentum is two-dimensional, there is a discrete set of
such solutions. For a convex, inversion symmetric Fermi surface of
the $j$th band, there are two solutions: $\mathbf{p}_{j}$ and $-\left(\mathbf{p}_{j}-\mathbf{q}\right)$.
We perform change of variables $\mathbf{k'}=\mathbf{k+q}$, and the
terms corresponding to the two classes to obtain the final expression:
\begin{equation}
\begin{split}M_{in,jm} & =\pi\lambda^{2}\sum_{\mathbf{kk'}}V_{\mathbf{k'-k}}f_{i,\mathbf{k},\mathbf{k}'}^{2}\delta\left(\varepsilon_{i\mathbf{k}}\right)\delta\left(\varepsilon_{i\mathbf{k'}}\right)\\
 & \times\left[\delta_{ij}d_{in,\mathbf{kk'}}^{*}d_{jm,\mathbf{\mathbf{kk'}}}+\frac{\gamma_{j\mathbf{k'-k}}}{\gamma_{\mathbf{k'-k}}}\frac{1-\left(-1\right)^{n}}{2}\frac{1-\left(-1\right)^{m}}{2}d_{in,\mathbf{k}\mathbf{k'}}^{*}\sum_{\mathbf{p}_{j}}h_{jm\mathbf{p}_{j}}\right]
\end{split}
\label{eq:mm_multiband_harmonics}
\end{equation}
It is easy to see how this reduces to the one band expression in Eq.
(\ref{eq:harmonic_basis}). In that case, the solutions to the equations
$\varepsilon_{\mathbf{p}}=\varepsilon_{\mathbf{p-q}}=0$ are $\mathbf{p}=-\mathbf{k}$
or $\mathbf{k'}$. One for the second term in Eq. (\ref{eq:mm_multiband_harmonics}),
we get:
\[
\left[1-\frac{1-\left(-1\right)^{n}}{2}\frac{1-\left(-1\right)^{m}}{2}\right]d_{n,\mathbf{kk'}}^{*}d_{m,\mathbf{\mathbf{kk'}}}
\]
which is the same expression compared to Eq. (\ref{eq:harmonic_basis})
in the case $N=1$.

\section{Connection to the Boltzmann equation and the non-equilibrium distribution
function\label{sec:boltzmann}}

Here we make explicit connection between our memory matrix approach
with the familiar Boltzmann equation, and identify the non-equilibrium
distribution function in the presence of an applied electric field
within the memory function approach.

We begin with writing down the Boltzmann equation for the quasiparticle
distribution function $\left\{ n_{\alpha\mathbf{k}}(t)\right\} $:
\begin{equation}
\partial_{t}n_{\alpha\mathbf{k}}+eE_{x}\partial_{\mathbf{k}_{x}}n_{\alpha\mathbf{k}}=I_{\text{coll}}
\end{equation}

In the absence of an electric field, the steady state is given by
the equilibrium Fermi-Dirac distribution $n_{F,\alpha\mathbf{k}}$.
When a small electric field $\mathbf{E}=E_{x}\hat{x}$ is applied,
the distribution function deviates away from $n_{F,\alpha\mathbf{k}}$,
with the dominant response occuring near the Fermi surface. We write
$n_{\alpha\mathbf{k}}=n_{F,\alpha\mathbf{k}}+\left(\partial_{\varepsilon}n_{F,\alpha\mathbf{k}}\right)\tilde{\Phi}_{\alpha\mathbf{k}}$,
and the linearized Boltzman equation for the non-equilibrium distribuion
$\left\{ \Phi_{\alpha\mathbf{k}}(t)\right\} $ has the form
\begin{equation}
\left(eE_{x}v_{\alpha\mathbf{k},x}\right)\left(\partial_{\varepsilon}n_{F,\alpha\mathbf{k}}\right)=-\left(\partial_{\varepsilon}n_{F,\alpha\mathbf{k}}\right)\partial_{t}\tilde{\Phi}_{\alpha\mathbf{k}}+\sum_{\alpha'\mathbf{k'}}\int_{t'}W_{\alpha\mathbf{k},\alpha't'}\left(t-t'\right)\tilde{\Phi}_{\alpha'\mathbf{k'}}
\end{equation}

Fourier transformation gives:
\begin{equation}
\left(-i\Omega\chi+W\right)\tilde{\Phi}=-e\chi V_{x}E_{x}
\end{equation}
where
\begin{equation}
\chi_{\mathbf{\alpha k},\alpha'\mathbf{k'}}\equiv-\frac{\partial n_{F,\alpha\mathbf{k}}}{\partial\varepsilon_{\alpha\mathbf{k}}}\delta_{\alpha\alpha'}\delta_{\mathbf{kk'}}
\end{equation}
is a thermodynamic susceptibility. $\tilde{\Phi}\equiv\left(\tilde{\Phi}_{\alpha_{1}\mathbf{k}_{1}},\tilde{\Phi}_{\alpha_{2}\mathbf{k}_{2}}\cdots\right)^{T}$
and $V_{x}\equiv\left(v_{\alpha_{1}\mathbf{k}_{1}x},v_{\alpha_{2}\mathbf{k}_{2}x}\cdots\right)^{T}$
are vectors. The electrical conductivity is given by
\begin{equation}
\sigma\left(\Omega\right)=\sum_{\alpha\mathbf{k}}ev_{\alpha\mathbf{k},x}\left(n_{\alpha\mathbf{k}}-n_{F,\alpha\mathbf{k}}\right)=e^{2}V_{x}^{T}\chi\Phi
\end{equation}
where $\Phi=\tilde{\Phi}/E_{x}$. In matrix form, we get
\begin{equation}
\sigma\left(\Omega\right)=e^{2}V_{x}^{T}\chi\frac{1}{W-i\Omega\chi}\chi V_{x}\label{eq:boltzmann}
\end{equation}
Comparison between Eq. (\ref{eq:boltzmann}) and Eq. (4) in the main
text shows that our memory matrix $M$ corresponds to the collision
kernel $W$ of the Boltzmann equation.

Due to the structure of the collision kernel, the non-equilibrium
distribution function due to an applied electric field can be very
different from the quasiparticle velocity vector $V_{x}$ . In particular
we have shown that
\begin{equation}
\Phi=-\frac{1}{W-i\Omega\chi}e\chi V_{x}\label{eq:Phi}
\end{equation}

In the main text, we address the functional form of $\Phi$ due to
critical nematic fluctuations and how they lead to a temperature dependent
resistivity very different than previously calculated.

\section{Connection to standard feynman diagrams for optical conductivity\label{sec:Connection-to-diagrams}}

\begin{figure}
\includegraphics[width=0.6\linewidth]{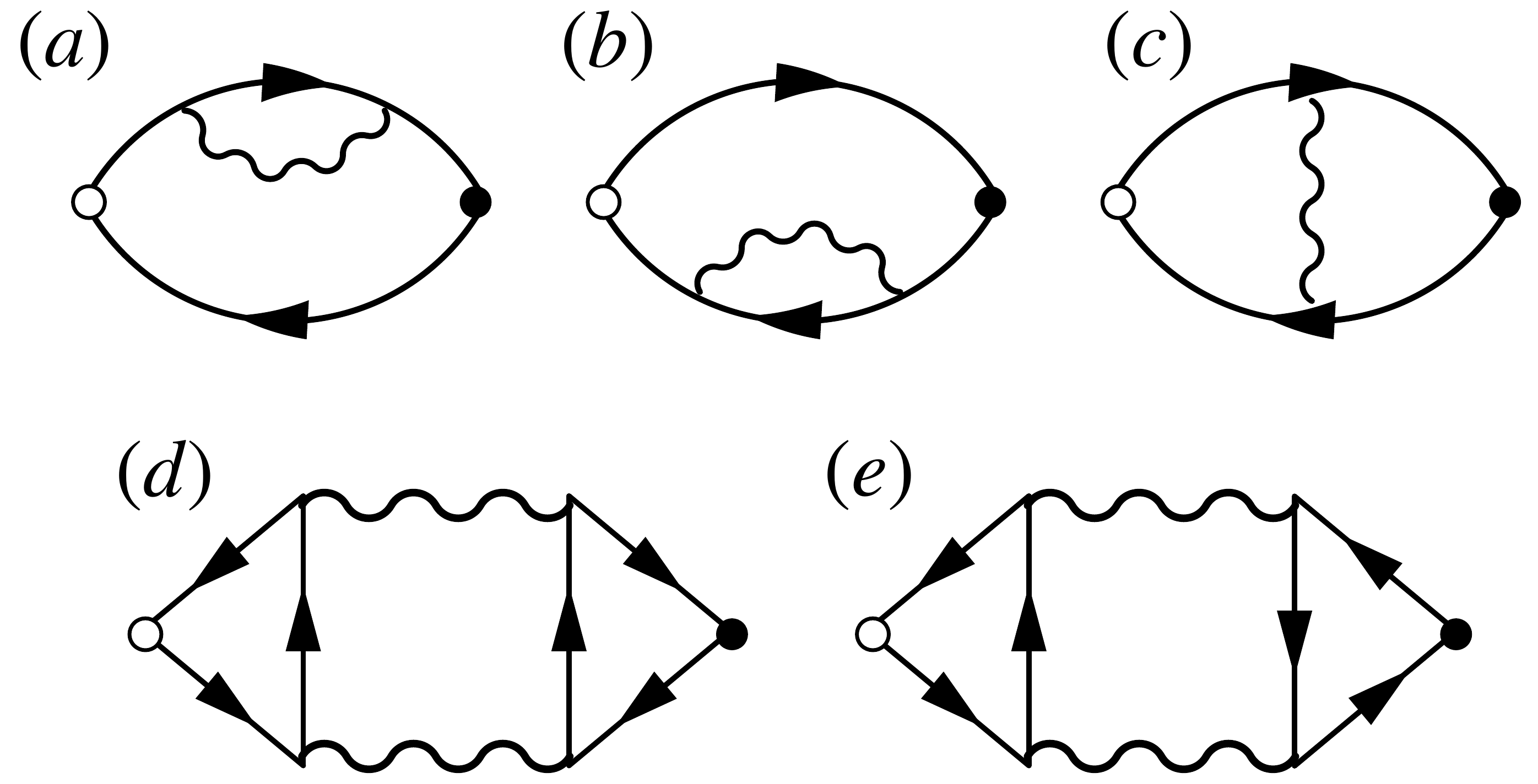}

\caption{\label{fig:fluctuation_diagram} Interaction contributions to optical
conductivity, to leading order in $1/N$. (a,b) are DOS diagrams,
(c) is MT, and (d,e) are AL diagrams. The current vertices are labeled
by circles. Open/solid circles represent the inflow/outflow of the
external frequency $\Omega$.}
\end{figure}
In this section we show that at high external frequencies, our derivation
of optical conductivity gives the same results as the standard perturbative
calculation. From Kubo linear response theory, the dissipative part
of optical conductivity is calculated from the imaginary part of the
retarded current-current response function:
\begin{equation}
\sigma'\left(\Omega\right)=\frac{1}{\Omega}\text{Im}\mathcal{G}_{JJ}\left(\Omega\right)
\end{equation}
To leading order in $1/N$, there are three types of diagrams that
contribute to $\sigma(\Omega)$, depicted in Figure \ref{fig:fluctuation_diagram}(a-e).
These are the so-called density of states (DOS), Maki-Thompson (MT),
and Aslamazov-Larkin (AL) diagrams. We discuss these contributions
in Matsubara frequency, and comment on the analytic continuation $i\Omega_{n}\rightarrow\Omega+i\delta$.

The DOS diagrams contribute two terms (a,b), related to each other
by a change of $i\Omega_{n}\rightarrow-i\Omega_{n}$. They can be
expressed as:
\begin{equation}
\begin{split}\mathcal{G}_{JJ}^{\text{DOS}}(i\Omega_{n}) & =-\lambda^{2}T^{2}\sum_{\nu_{n}\omega_{n}}\sum_{\mathbf{k}\mathbf{q}}v_{\mathbf{k},x}^{2}f_{\mathbf{k,k+q}}^{2}D_{\mathbf{q},\nu_{n}}G_{\mathbf{k}+\mathbf{q},\omega_{n}+\nu_{n}}\\
 & \times G_{\mathbf{k},\omega_{n}}^{2}\left(G_{\mathbf{k},\omega_{n}+\Omega_{n}}+G_{\mathbf{k},\omega_{n}-\Omega_{n}}\right).
\end{split}
\end{equation}
Here $G_{\mathbf{k},\omega_{n}}$ is short hand notation for the non-interacting
fermionic Green's function $G_{0}\left(\mathbf{k},i\omega_{n}\right)=\left(i\omega_{n}-\varepsilon_{\mathbf{k}}\right)^{-1}$.
Making use of the identity $G_{\mathbf{k},\omega_{n}}G_{\mathbf{k},\omega_{n}+\Omega_{n}}=\frac{1}{i\Omega_{n}}\left(G_{\mathbf{k},\omega_{n}}-G_{\mathbf{k},\omega_{n}+\Omega_{n}}\right)$,
it is straightforward to show:

\begin{equation}
\begin{split}\mathcal{G}_{JJ}^{\text{DOS}}(i\Omega_{n}) & =\frac{\lambda^{2}T^{2}}{\Omega_{n}^{2}}\sum_{\nu_{n}\omega_{n}}\sum_{\mathbf{k}\mathbf{q}}v_{\mathbf{k},x}^{2}f_{\mathbf{k,k+q}}^{2}D_{\mathbf{q},\nu_{n}}G_{\mathbf{k}+\mathbf{q},\omega_{n}+\nu_{n}}\left(G_{\mathbf{k},\omega_{n}+\Omega_{n}}+G_{\mathbf{k},\omega_{n}-\Omega_{n}}\right)\\
 & -\frac{\lambda^{2}T^{2}}{\Omega_{n}^{2}}\sum_{\nu_{n}\omega_{n}}\sum_{\mathbf{k}\mathbf{q}}v_{\mathbf{k},x}^{2}f_{\mathbf{k,k+q}}^{2}D_{\mathbf{q},\nu_{n}}G_{\mathbf{k}+\mathbf{q},\omega_{n}+\nu_{n}}G_{\mathbf{k},\omega_{n}}.
\end{split}
\end{equation}
The term on the second line does not contribute to $\sigma'\left(\Omega\right)$,
since following analytic continuation, it is the real part of the
retarded response function. Focusing on the first term, by a suitable
change of variables, the dependence on the external frequency can
be shifted entirely into the nematic propagator:
\begin{equation}
\begin{split}\mathcal{G}_{JJ}^{\text{DOS}}(i\Omega_{n}) & =-\frac{\lambda^{2}T}{\Omega_{n}^{2}}\sum_{\nu_{n}\omega_{n}}\sum_{\mathbf{k}\mathbf{q}}v_{\mathbf{k},x}^{2}\sum_{\zeta=\pm1}f_{\mathbf{k,k+\zeta q}}^{2}D_{\mathbf{q},\nu_{n}+\Omega_{n}}R\left(\mathbf{k},\mathbf{k+\zeta q},i\zeta\nu_{n}\right)\end{split}
\label{eq:dos}
\end{equation}
Here, as before $R\left(\mathbf{k},\mathbf{k+q},i\nu_{n}\right)\equiv-T\sum_{\omega_{n}}G_{\mathbf{k}+\mathbf{q},\omega_{n}+\nu_{n}}G_{\mathbf{k},\omega_{n}}$.

The MT term (c) can be expressed as:
\begin{equation}
\begin{split}\mathcal{G}_{JJ}^{\text{MT}}(i\Omega_{n}) & =-\lambda^{2}T^{2}\sum_{\nu_{n}\omega_{n}}\sum_{\mathbf{k}\mathbf{k'}\mathbf{q}}v_{\mathbf{k},x}v_{\mathbf{k'},x}f_{\mathbf{k,k'}}^{2}D_{\mathbf{q},\nu_{n}}\\
 & \times G_{\mathbf{k},\omega_{n}}G_{\mathbf{k}+\mathbf{q},\omega_{n}+\Omega_{n}}G_{\mathbf{k}',\omega_{n}+\nu_{n}+\Omega_{n}}G_{\mathbf{k}',\omega_{n}+\nu_{n}}
\end{split}
\end{equation}
Following similar procedure as above, the MT contribution can be reduced
to:
\begin{equation}
\mathcal{G}_{JJ}^{\text{MT}}(i\Omega_{n})=\frac{\lambda^{2}T}{\Omega_{n}^{2}}\sum_{\mathbf{kk'}}v_{\mathbf{k},x}v_{\mathbf{k'},x}\sum_{\mathbf{q}\nu_{n}}D_{\mathbf{q},\nu_{n}+\Omega_{n}}\sum_{\zeta=\pm1}\delta_{\mathbf{k'-k,\zeta q}}f_{\mathbf{k,k+\zeta q}}^{2}R\left(\mathbf{k},\mathbf{k+\zeta q},i\zeta\nu_{n}\right)\label{eq:mt}
\end{equation}

Eqs. (\ref{eq:dos}) and (\ref{eq:mt}) share a similar structure.
The combination gives:

\begin{equation}
\begin{split}\mathcal{G}_{JJ}^{\text{DOS+MT}}\left(i\Omega_{n}\right) & =\frac{\lambda^{2}T}{\Omega_{n}^{2}}\sum_{\mathbf{kk'}}v_{\mathbf{k},x}v_{\mathbf{k'},x}\sum_{\mathbf{q}\nu_{n}}D_{\mathbf{q},\nu_{n}+\Omega_{n}}\\
 & \times\sum_{\zeta=\pm1}\left(\delta_{\mathbf{k'-k,\zeta q}}-\delta_{\mathbf{kk'}}\right)f_{\mathbf{k,k+\zeta q}}^{2}R\left(\mathbf{k},\mathbf{k+\zeta q},i\zeta\nu_{n}\right).
\end{split}
\label{eq:dos+mt}
\end{equation}

The AL contribution (d,e) can be expressed as:
\begin{equation}
\begin{split}\mathcal{G}_{JJ}^{\text{AL}}(i\Omega_{n}) & =\lambda^{4}T\sum_{\mathbf{q},\nu_{n}}D_{\mathbf{q},\nu_{n}}D_{\mathbf{q},\nu_{n}+\Omega_{n}}\\
 & \times T\sum_{\mathbf{k}\omega_{n}}v_{\mathbf{k},x}G_{\mathbf{k},\omega_{n}-\Omega_{n}}G_{\mathbf{k},\omega_{n}}G_{\mathbf{k+q},\omega_{n}+\nu_{n}}\\
 & \times T\sum_{\mathbf{k'}\omega'_{n}}v_{\mathbf{k'},x}\left(G_{\mathbf{k'},\omega'_{n}-\Omega_{n}}G_{\mathbf{k'},\omega'_{n}}G_{\mathbf{k'+q},\omega'_{n}+\nu_{n}}+G_{\mathbf{k'},\omega'_{n}+\Omega_{n}}G_{\mathbf{k'},\omega'_{n}}G_{\mathbf{k'-q},\omega'_{n}-\nu_{n}}\right).
\end{split}
\end{equation}

Here the terms on the second and third lines represent the fermionic
triangle. Similarly, it can be reduced to:
\begin{equation}
\begin{split}\mathcal{G}_{JJ}^{\text{AL}}(i\Omega_{n}) & =-\frac{\lambda^{4}T}{\Omega_{n}^{2}}\sum_{\mathbf{k}\mathbf{k'}}v_{\mathbf{k},x}v_{\mathbf{k'},x}\sum_{\mathbf{q},\nu_{n}}D_{\mathbf{q},\nu_{n}}D_{\mathbf{q},\nu_{n}+\Omega_{n}}\\
 & \times f_{\mathbf{k,k+q}}^{2}\left[R\left(\mathbf{k},\mathbf{k+q},i\nu_{n}\right)-R\left(\mathbf{k},\mathbf{k+q},i\nu_{n}+i\Omega_{n}\right)\right]\\
 & \times\sum_{\zeta'=\pm1}f_{\mathbf{k'},\mathbf{k'+\zeta'q}}^{2}\left[R\left(\mathbf{k'},\mathbf{k'}+\zeta'\mathbf{q},i\zeta'\nu_{n}\right)-R\left(\mathbf{k'},\mathbf{k'}+\zeta'\mathbf{q},i\zeta'\nu_{n}+i\zeta'\Omega_{n}\right)\right].
\end{split}
\label{eq:al}
\end{equation}

Comparing Eqs. (\ref{eq:dos+mt}) and (\ref{eq:al}) with our class
I and class II diagrams for the memory matrix in Eqs (\ref{eq:mm_expressions}),
we see that DOS+MT correspond to class I diagram, and AL correspond
to class II diagram. Specifically,
\begin{equation}
\begin{split}\mathcal{G}_{JJ}^{\text{DOS+MT}}\left(i\Omega_{n}\right) & =\frac{1}{\Omega_{n}^{2}}\sum_{\mathbf{kk'}}v_{\mathbf{k},x}M_{\mathbf{kk'}}^{(1)}\left(i\Omega_{n}\right)v_{\mathbf{k'},x}\\
\mathcal{G}_{JJ}^{\text{AL}}\left(i\Omega_{n}\right) & =\frac{1}{\Omega_{n}^{2}}\sum_{\mathbf{kk'}}v_{\mathbf{k},x}M_{\mathbf{kk'}}^{(2)}\left(i\Omega_{n}\right)v_{\mathbf{k'},x}
\end{split}
\end{equation}

We note, however, that $\sigma(i\Omega_{n})$ computed to lowest order
in perturbation theory is equivalent to our memory matrix approach
only at the high frequency limit. This is seen by expanding Eq. (\ref{eq:mm_conductivity-1})
to leading order in the memory matrix:
\begin{equation}
\begin{split}\sigma\left(i\Omega_{n}\right) & \approx\frac{1}{\Omega_{n}}\left(\chi_{J_{x}\mathbf{k}}\chi_{\mathbf{kk'}}^{-1}\chi_{\mathbf{k}'J_{x}}\right)-\frac{1}{\Omega_{n}^{2}}\chi_{J_{x}\mathbf{k}}\chi_{\mathbf{k}\mathbf{k}_{1}}^{-1}M_{\mathbf{k}_{1}\mathbf{k}_{2}}\chi_{\mathbf{k}_{1}\mathbf{k}'}^{-1}\chi_{\mathbf{k}'J_{x}}\\
 & \approx\frac{1}{\Omega_{n}}\sum_{\mathbf{k}}v_{\mathbf{k},x}^{2}\frac{\partial n_{F}}{\partial\varepsilon_{\mathbf{k}}}-\frac{1}{\Omega_{n}^{2}}\sum_{\mathbf{kk'}}v_{\mathbf{k},x}M_{\mathbf{kk'}}(i\Omega_{n})v_{\mathbf{k}',x}
\end{split}
\end{equation}
Repeated indices are summed over. In the second step we made use of
$\mathbf{J}=\sum_{\mathbf{k}}\mathbf{v}_{\mathbf{k}}c_{\mathbf{k}}^{\dagger}c_{\mathbf{k}}$,
and $\chi_{\mathbf{kk'}}=\delta_{\mathbf{kk'}}\left(-\partial_{\varepsilon_{\mathbf{k}}}n_{F}\right)$.

In the dc limit, $\sigma'(\Omega\rightarrow0)\propto M^{-1}$, which
cannot be obtained perturbatively without resumming an infinite set
of diagrams.

\section{Totally compensated metal\label{sec:Totally-compensated-metal}}

In a general band structure, the momentum and the current operators
are not identical, and therefore electron-electron collisions can
change the total current even if they conserve momentum. However,
since the total electrical current has a finite overlap with the total
momentum, meaning that the thermodynamic susceptibility $\chi_{\mathbf{JP}}\neq0$,
the dc resistivity is still zero as long as momentum is conserved,
because then the current cannot relax to zero.

\begin{figure}
\includegraphics[width=0.3\linewidth]{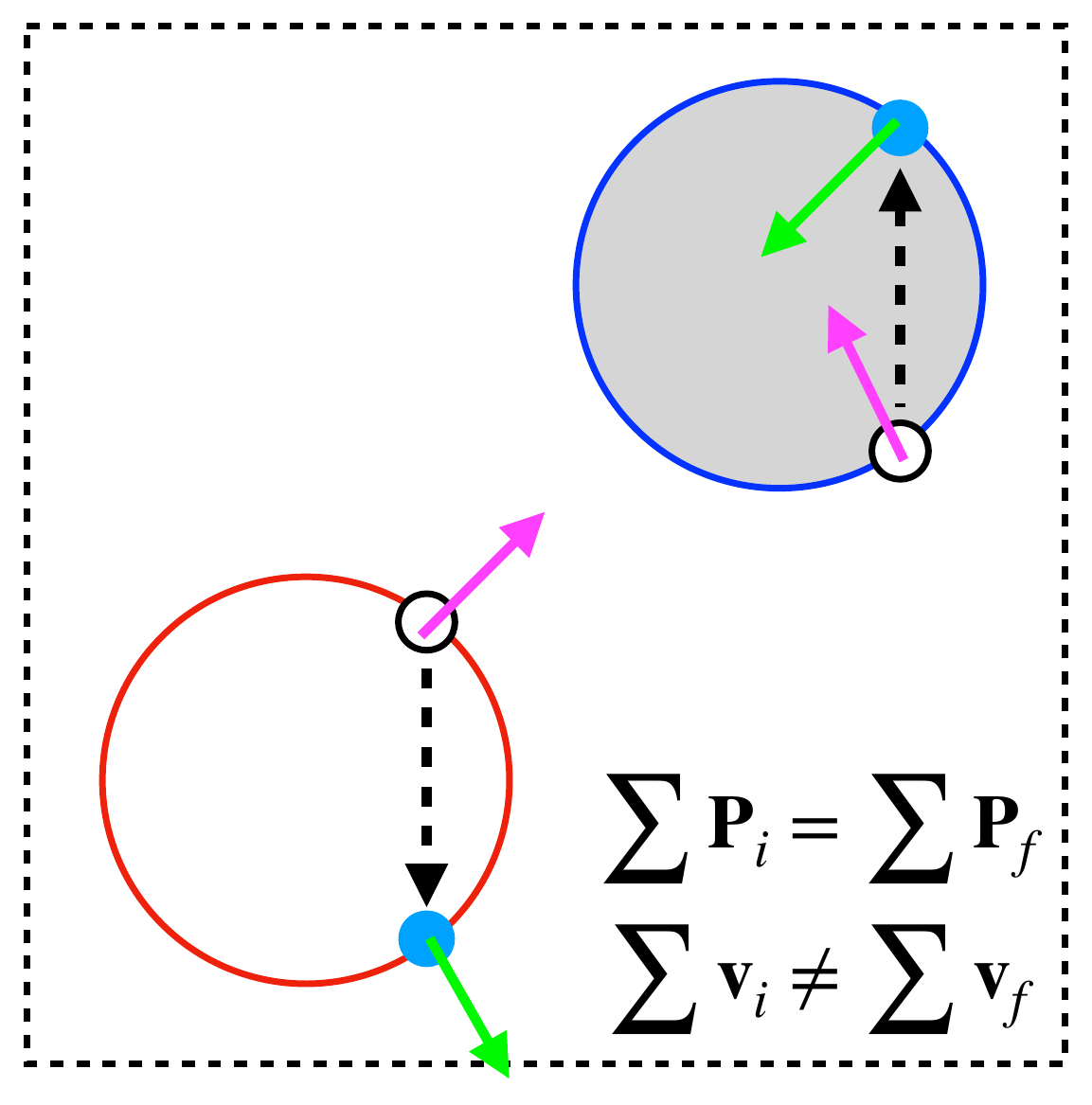}\caption{\label{fig:compensated_metal}Sketch of a two-electron scattering
process for compensated metal. Blue Fermi pockets are electron-like
and red Fermi pocket is hole-like. The purple (green) arrows label
the Fermi velocities in the initial (final) states. Such a process
flips the direction of electrical current from upward to downward.}
\end{figure}
It is well-known, however, that this is not the case for a totally
compensated metal, where the densities of electron and hole-like charge
carriers are equal. This is because in a compensated metal, $\chi_{\mathbf{JP}}=0$.
Below, we derive this result for completeness, first for non-interacting
electrons and then a general interacting model. , As a result, in
a compensated metal, the dc resistivity is non-zero even if momentum
is perfectly conserved.

\subsection{Non-interacting electrons}

We consider a system with $i=1,\dots,l$ Fermi sheets. The current
operator is given by $\mathbf{J}=-e\sum_{i\mathbf{k}}\mathbf{v}_{i\mathbf{k}}c_{i\mathbf{k}}^{\dagger}c_{i\mathbf{k}}$,
where $\mathbf{v}_{ik}=\nabla_{\mathbf{k}}$$\varepsilon_{i\mathbf{k}}$,
and $\varepsilon_{i\mathbf{k}}$ is the dispersion relation of the
$i$th sheet, while the momentum operator is given by $\mathbf{P}=\sum_{i\mathbf{k}}\mathbf{k}c_{i\mathbf{k}}^{\dagger}c_{i\mathbf{k}}$.
In the free electron limit, the thermodynamic susceptibility: $\chi_{\mathbf{JP}}=\int_{0}^{\beta}\mathrm{d}\tau\langle\mathbf{J}\left(\tau\right)\cdot\mathbf{P}(0)\rangle$
is given by:
\begin{equation}
\begin{split}\chi_{\mathbf{JP}} & =e\sum_{i\mathbf{k}}\mathbf{k}\cdot\mathbf{v}_{i\mathbf{k}}T\sum_{i\omega_{n}}G_{0i}^{2}\left(i\omega_{n},\varepsilon_{i\mathbf{k}}\right)\end{split}
\label{eq:chi_JP}
\end{equation}
Summing over Matsubara frequency gives $\frac{\partial n_{F}\left(\varepsilon_{i\mathbf{k}}\right)}{\partial\varepsilon_{i\mathbf{k}}}$.
At temperatures much smaller than the Fermi energy of any of the Fermi
sheets, we may approximate $\frac{\partial n_{F}\left(\varepsilon_{i\mathbf{k}}\right)}{\partial\varepsilon_{i\mathbf{k}}}\approx-\delta(\varepsilon_{\mathbf{k}})$.
Then we have
\begin{align}
\chi_{\mathbf{JP}} & =-eL^{2}\sum_{i}\int\frac{d^{2}k}{\left(2\pi\right)^{2}}\mathbf{k}\cdot\mathbf{v}_{i\mathbf{k}}\delta(\varepsilon_{i\mathbf{k}})=-\frac{eL^{2}}{\left(2\pi\right)^{2}}\sum_{i}\oint_{FS}dk\mathbf{k}\cdot\hat{\mathbf{v}}_{i\mathbf{k}}.\label{eq:chi_JP1}
\end{align}
Here, $L^{2}$ is the total area of the system. Note that $\oint_{FS}dk\mathbf{k}\cdot\hat{\mathbf{v}}_{i\mathbf{k}}$
is the area of the $i$th sheet in momentum space, with a positive
sign if $\mathbf{v}_{i\mathbf{k}}$ is pointing outward (electron-like),
and a negative sign if $\mathbf{v}_{i\mathbf{k}}$ is pointing inward
(hole-like). Thus, by Eq. (\ref{eq:chi_JP1}), $\chi_{\mathbf{JP}}$
is proportional to the total area of all the Fermi pockets, each weighted
with a sign that corresponds to the type of carriers. In a compensated
metal, this is zero by definition, and hence $\chi_{\mathbf{JP}}=0$.

\subsection{Interacting compensated metal}

We now derive this result in the presence of arbitrary interactions.
We consider a general first-quantized Hamiltonian of the form
\begin{equation}
H=\sum_{j}\varepsilon_{j}\left(-i\nabla_{j}+q_{j}\mathbf{A}\right)+V(\{\mathbf{r}_{j}\})
\end{equation}
where $\varepsilon_{j}(\mathbf{k})$ is the dispersion relation of
the $j$th particle, and $V\left(\{\mathbf{r}_{j}\}\right)$ is an
arbitrary interaction potential, assumed to be translationally invariant.
Here, we have allowed for different types of carriers with different
charges $q_{j}$. We would like to compute the overlap between the
momentum and current operators:
\begin{equation}
\chi_{\mathbf{JP}}=\int_{0}^{\beta}\mathrm{d}\tau\left\langle \mathbf{P}(\tau)\cdot\left(\frac{\partial H}{\partial\mathbf{A}}\right)_{\mathbf{A}=0}\right\rangle
\end{equation}
where $\mathbf{A}$ is spatially uniform. Since momentum is conserved,
we can replace $\mathbf{P}(\tau)=\mathbf{P}(0)=\mathbf{P}.$ We write
$\chi_{\mathbf{JP}}$ as
\begin{equation}
\begin{split}\chi_{\mathbf{JP}} & =\frac{1}{T}\frac{1}{Z}\text{Tr}\left[e^{-\beta H(\mathbf{A})}\left(\frac{\partial H}{\partial\mathbf{A}}\right)_{\mathbf{A}=0}\cdot\mathbf{P}\right]\\
 & =-\frac{1}{T}\frac{1}{Z}\left\{ \frac{\partial}{\partial\mathbf{A}}\cdot\text{Tr}\left[e^{-\beta H(\mathbf{A})}\mathbf{P}\right]\right\} _{\mathbf{A}=0}\\
 & =-\frac{1}{T}\frac{1}{Z}\left\{ \frac{\partial}{\partial\mathbf{A}}\cdot\text{Tr}\left[e^{-\beta H(\mathbf{A}=0)}(\mathbf{P}-\sum_{j}q_{j}\mathbf{A})\right]\right\} =\sum_{j}q_{j}
\end{split}
\end{equation}
where in the last line, we have performed a gauge transformation that
shifts $-i\nabla_{j}\rightarrow-i\nabla_{j}-q_{j}\mathbf{A}$, and
hence removes the $\mathbf{A}$ dependence from $H$ \footnote{There is a subtlety at this step: if we use periodic boundary conditions,
then a gauge transformation can only shift the vector potential by
integer multiples of $2\pi/L$, where $L$ is the system's spatial
dimention. This can be remedied by approximating, e.g., $[\nabla_{A_{x}}e^{-\beta H(A_{x})}]_{A_{x}=0}\approx(e^{-\beta H(A_{x})}-e^{-\beta H(0)})/A_{x}$
with $A_{x}=2\pi/L$, and taking the limit $L\rightarrow\infty$.}. In particular, if $\sum_{j}q_{j}=0$, i.e., in a totally compensated
metal, then $\chi_{\mathbf{JP}}=0$. .

\section{Effect of the form factor and cold spots in compensated metal\label{sec:form-factor}}

In this Appendix, we derive the we derive the asymptotic behavior
of $\rho(T)$ for a clean, compensated metal discussed in Sec. \ref{subsec:Compensated-metal}.
As indicated in Fig. 7, the results depend qualitatively on the configuration
of ``cold spots'' on the different Fermi sheets. Using a variational
ansatz for the non-distribution function, we show that one can readily
recover the different low-temperature exponents obtained numerically.

We use Eqs.~(\ref{eq:boltzmann},\ref{eq:Phi}) to write the dc
resistivity as:
\begin{equation}
\rho(T)=\frac{\Phi^{T}M\Phi}{e^{2}(\Phi^{T}\chi V_{x})^{2}},\label{eq:rho_func}
\end{equation}
where, as before, we treat $\Phi$ as a vector in momentum, Fermi
sheet, and flavor space. The computation of the resistivity can be
treated as a minimization problem: $\Phi$ that satisfies Eq. (\ref{eq:boltzmann})
also minimizes the functional on the right hand side of Eq. (\ref{eq:rho_func})
\citep{ziman}. Hence, we can use Eq. (\ref{eq:rho_func}) to bound
the resistivity from above, by inserting a variational ansatz for
$\Phi$. Crucially, for the ansatz we use below, the denominator of
Eq. (\ref{eq:rho_func}) will be essentially temperature independent.
Hence, to find the temperature dependence of $\rho(T)$, we need to
compute the numerator of (\ref{eq:rho_func}) with our variational
ansatz for $\Phi$.

From Eq. (\ref{eq:mm_multiband_harmonics}), we observe that
\begin{equation}
\begin{split}\rho\left(T\right)\propto & \lambda^{2}\sum_{ij;\mathbf{kk'}}V_{\mathbf{k'-k}}f_{i,\mathbf{k},\mathbf{k}'}^{2}\delta\left(\varepsilon_{i\mathbf{k}}\right)\delta\left(\varepsilon_{i\mathbf{k'}}\right)\\
 & \times\left[\delta_{ij}\left(\Phi_{i\mathbf{k}}-\Phi_{i\mathbf{k'}}\right)^{2}-\frac{\gamma_{j\mathbf{k'-k}}}{\gamma_{\mathbf{k'-k}}}\left(\Phi_{i\mathbf{k}}-\Phi_{i\mathbf{k}'}\right)\left(\Phi_{j\mathbf{k}}-\Phi_{j\mathbf{k}'}\right)\right]
\end{split}
\label{eq:multiband_resistivity}
\end{equation}

where $i,j$ are band labels, and we have replaced $h_{in\mathbf{k}}$
with $\Phi_{i\theta_{\mathbf{k}}}$, and replaced $d_{in,\mathbf{kk'}}$
with $\Phi_{i\mathbf{k}}-\Phi_{i\mathbf{k}'}$. By symmetry, $\Phi_{i\mathbf{k}}=-\Phi_{i,-\mathbf{k}}$.
For concreteness, it is useful to consider a simple model with two
circular Fermi pockets with opposite Fermi velocities; we will comment
about the generalization to more generic situations below. In the
two-pocket model, since the Fermi velocities of the two pockets are
opposite, $\Phi_{1\mathbf{k}}=-\Phi_{2\mathbf{k}}$. Carrying out
the summation over band as well as performing momentum integration
along the direction perpendicular to the Fermi surface, we get:
\begin{equation}
\begin{split}\rho\left(T\right)\propto & \frac{k_{F}^{2}\lambda^{2}}{v_{F}^{2}}\oiint_{\theta\theta'}\left(\Phi_{\theta}-\Phi_{\theta'}\right)^{2}V_{\mathbf{k'-k}}(T)\frac{f_{1\mathbf{k},\mathbf{k'}}^{2}f_{2\mathbf{k},\mathbf{k'}}^{2}}{f_{1\mathbf{k},\mathbf{k'}}^{2}+f_{2\mathbf{k},\mathbf{k'}}^{2}}\end{split}
\label{eq:compensated_resistivity}
\end{equation}
Here, as before, $V_{\mathbf{q}}(T)=F(T/\omega_{\mathbf{q}})/\gamma_{\mathbf{q}}$,
where $\omega_{\mathbf{q}}=r_{\mathbf{q}}/\gamma_{\mathbf{q}}$, $|\mathbf{k-k'}|=2k_{F}\sin|\frac{\theta-\theta'}{2}|$,
and $F\left(x\right)=\frac{1}{x}\int_{-\infty}^{+\infty}\frac{\mathrm{d}u}{\pi}\frac{u^{2}}{1+u^{2}}\sinh^{-2}\left(\frac{u}{2x}\right)$.
The asymptotic behavior of $F(x)$ is: for $x\ll1$, $F\left(x\right)\propto x^{2}$;
whereas for $x\gg1$, $F\left(x\right)\propto x$.

Equation (\ref{eq:compensated_resistivity}) forms the basis for carrying
out scaling analysis. We analyze cases where either there are no cold
spots on either pocket, a case with cold spots on both pockets, and
a case where one pocket has cold spots and the other does not.

\subsection{Absence of cold spots on either Fermi surface}

In this case, we assume for simplicity that $f_{1}=f_{2}=1$. Since
our the problem is rotationally invariant, the angular distribution
function for an electric field in the $x$ direction is $\Phi_{\theta}=\cos\theta$
The dominant contribution to resistivity is due to small angle scattering.
We define relative angle $\alpha\equiv|\theta-\theta'|$, and the
integrand of Eq. (\ref{eq:compensated_resistivity}) for $\alpha\ll1$:
\begin{equation}
\rho\left(T\right)\propto\int_{0}^{\theta_{0}}\mathrm{d}\alpha\alpha^{3}F\left(\frac{\theta_{T}}{\alpha{}^{3}}\right)
\end{equation}
Here $\theta_{0}\sim1$ is an upper cutoff. We defined $\theta_{T}\equiv T/\Omega_{L}$
to be a dimensionless temperature.

The integral is convergent both for $\alpha\rightarrow0$ and $\alpha\rightarrow\infty$,
due to the asymptotic behavior of the scaling function $F$ discussed
above. Therefore, we can rescale $\tilde{\alpha}\equiv\alpha/\theta_{T}^{1/3}$,
and extend the upper cutoff to infinity, resulting in $\rho(T)\propto\theta_{T}^{4/3}\sim T^{4/3}$.

\subsection{Cold spots on both Fermi surfaces}

As shown in Figs. 8(b) and (c), when cold spots are present, $\Phi_{\theta}$
deviates strongly from $\cos\theta$. In particular, $\Phi_{\theta}$
becomes nearly constant between each pair of cold spots. This motivates
us to consider the following variational ansatz for $\Phi_{\theta}$:

\begin{equation}
\Phi_{\theta}\approx\begin{cases}
1 & \theta\in\left(-\frac{\pi}{4},\frac{\pi}{4}\right),\\
-1 & \theta\in\left(\frac{3\pi}{4},\frac{5\pi}{4}\right),\\
0 & \text{elsewhere}.
\end{cases}\label{eq:var}
\end{equation}
The contribution to Eq. (\ref{eq:compensated_resistivity}) then comes
purely from scattering processes relating different angular regions.
Let us consider the vicinity of the cold spot at $\theta=\frac{\pi}{4}$.
Near the cold spot, we get:
\begin{equation}
\rho\left(T\right)\propto\int_{0}^{\theta_{0}}\alpha\mathrm{d}\alpha\int_{0}^{\alpha}\mathrm{d}\varphi F\left(\frac{\theta_{T}\varphi^{2}}{\alpha^{3}}\right).
\end{equation}
Here we defined $\varphi\equiv\left|\frac{\theta+\theta'}{2}-\frac{\pi}{4}\right|$
and $\alpha\equiv|\theta-\theta'|$. Compared to the case when cold
spots are absent, the scaling properties are not only determined the
relative angle $\alpha$ between the momentum states, but also the
average angle $\varphi$ with respect to the location of the cold
spot. To extract the scaling of the resisitivity with $T$, we split
the integration domain into two regions:
\begin{enumerate}
\item $\theta_{T}<\alpha<\theta_{0}$: Since in this regime the argument
of $F$ in the integrand is smaller than unity, we estimate the integral
by replacing $F(x)$ by its small $x$ behavior: $F\sim x^{2}$. The
contribution of this regime is then estimated as $\Delta\rho_{2}\sim\int_{0}^{\theta_{0}}\alpha\mathrm{d}\alpha\int_{0}^{\alpha}\mathrm{d}\varphi\left(\theta_{T}\varphi^{2}\alpha^{-3}\right)^{2}\sim\theta_{0}\theta_{T}^{2}$.
Notice that this contribution arises from large angle scattering,
and hence it is proportional to $\theta_{0}$. This justifies considering
$\alpha$ as much larger than $\theta_{T}$.
\item $0<\alpha<\theta_{T}$: in this regime, we perform a change of variables
as follows: $\alpha=\tilde{\alpha}/\theta_{T}$, $\varphi=\tilde{\varphi}/\theta_{T}$.
The integral becomes $\Delta\rho_{2}\sim\theta_{T}^{3}\int_{0}^{1}\tilde{\alpha}\mathrm{d}\tilde{\alpha}\int_{0}^{\tilde{\alpha}}\mathrm{d}\tilde{\varphi}F\left(\tilde{\varphi}^{2}\tilde{\alpha}^{-3}\right)\sim\theta_{T}^{3}$.
This contribution is subleading compared to $\Delta\rho_{1}$.
\end{enumerate}
As a result, at low temperatures and when the nematic cold spots are
present on both Fermi surfaces, $\rho(T)\propto T^{2}$ due to scattering
off non-critical nematic fluctuations carrying large momenta.

\subsection{Cold spots on part of the Fermi surfaces}

We consider a system with cold spots on the first Fermi sheet but
not on the second. In this case, we may replace $f_{2}=1.$ For $\mathbf{q}$
pointing near the direction of the cold spots, the Landau damping
coefficient $\gamma_{\mathbf{q}}\approx\gamma_{2,\mathbf{q}}$, and
we can neglect its angular dependence. We use the same variational
ansatz as for case B {[}Eq. (\ref{eq:var}){]}. Again expanding near
$\theta\left(\theta'\right)\approx\frac{\pi}{4}$, we find that
\begin{equation}
\rho(T)\propto\int_{0}^{\theta_{0}}\alpha\mathrm{d}\alpha\int_{0}^{\alpha}\mathrm{d}\varphi\varphi^{2}F\left(\frac{\theta_{T}}{\alpha^{3}}\right),\label{eq:cold_spot_on_one_band}
\end{equation}
where the factor of $\varphi^{2}$ comes from $f_{1}$ in Eq. (\ref{eq:compensated_resistivity}).
As a result, $\rho(T)\sim\int_{0}^{\theta_{0}}\alpha\mathrm{d}\alpha\int_{0}^{\alpha}\mathrm{d}\varphi\varphi^{2}F\left(\frac{\theta_{T}}{\alpha^{3}}\right)\sim\theta_{T}^{5/3}$.
In this case, the resistivity comes mostly from small-angle scattering
in the vicinity of $\theta\left(\theta'\right)\approx\frac{\pi}{4}$.

We comment on the generalization for the more generic case with more
two pockets, and where the shape of the pockets is non-circular. In
this case, we use a variational ansatz where $\Phi$ is of the form
(\ref{eq:var}) on the pockets with cold spots, and zero on the pockets
where cold spots are absent. Evalutaing Eq. (\ref{eq:multiband_resistivity})
along the same lines as above gives again $\rho(T)\sim\theta_{T}^{5/3}\sim T^{5/3}$,
as in our two-pocket toy model. It is worth noting that this is an
upper bound on the resistivity, and it only provides a lower bound
on the resistivity exponent at low temperature. The numerical results
for the two-pocket case {[}Fig. \ref{fig:rho_compensated}(b){]} suggest
the exponent is indeed $5/3$.

\section{Numerical construction of the memory matrix\label{sec:Numerical-construction-of}}

\begin{figure}
\includegraphics[width=0.4\linewidth]{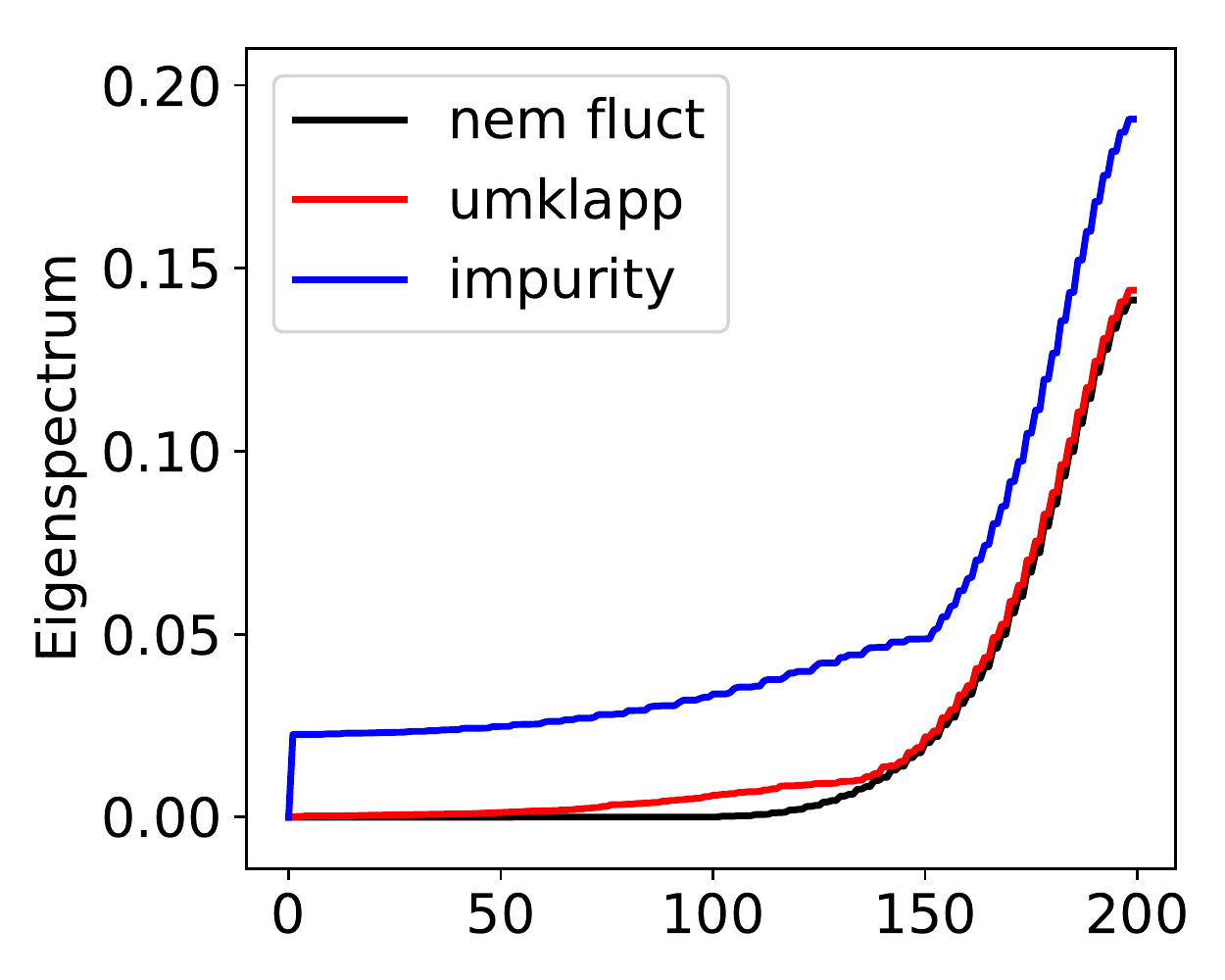}\caption{\label{fig:eigenspectrum}Eigenspectrum of the memory matrix $M_{\theta,\theta'}$
for a single Fermi surface when only quantum critical scattering is
considered (black), when umklapp scattering is included (red), and
when quenched random impurity is included (blue). In this calculation,
we used the dispersion $\varepsilon_{\mathbf{k}}=-2t\left(\cos k_{x}+\cos k_{y}\right)-4t'\cos k_{x}\cos k_{y}-\mu$
, with $t=1$, $t'=-0.3$ and $\mu=-1$. The nematic coupling strength
is $\lambda^{2}\approx2.62\varepsilon_{F}$, and impurity strength
is $g_{\text{imp}}\approx0.02\varepsilon_{F}$. The spectrum is obtained
at temperature $T\approx0.04\varepsilon_{F}$. We used 200 points
to uniformly discretize the Fermi surface.}
\end{figure}
In this section we provide details on the construction of the memory
matrix in the dc limit, as well as further numerical results.

We always work in the limit that temperature is much smaller than
the Fermi energy. This allows us to project all scattering processes
onto the Fermi surface, and perform integration over the momentum
direction perpendicular to the Fermi surface. In practice, we use
expressions in Eqs. (\ref{eq:mm_classI-1}) and (\ref{eq:mm_classII-1})
to construct the radially-integrated memory matrix. For a generic
multi-Fermi surface system,
\begin{equation}
M_{i\theta,j\theta'}=\int\frac{\mathrm{d}k_{i,\perp}\mathrm{d}k_{j,\perp}'}{\left(2\pi\right)^{4}}M_{i\mathbf{k},j\mathbf{k'}}
\end{equation}
and the dc conductivity is calculated as
\begin{equation}
\sigma=\frac{e^{2}}{\hbar}\int\frac{\mathrm{d}k_{i,\parallel}\mathrm{d}k'_{j,\parallel}}{\left(2\pi\right)^{4}}v_{i\mathbf{k}}M_{i\mathbf{\theta},j\theta'}^{-1}v_{j\mathbf{k}'}
\end{equation}

Here $v_{i\mathbf{k}}$ is the Fermi velocity of the $i$th band,
and $\theta$ is the angle of $\mathbf{k}$ with respect to the $x$
direction. $\mathrm{d}k_{i,\parallel}=k_{iF}(\theta)\mathrm{d}\theta$.

The off-diagonal elements ($\mathbf{\theta\neq\theta'}$) of the memory
matrix are constructed from Eqs. (\ref{eq:mm_multiband}) by solving
for the condition that all participating momentum states are on the
Fermi surface. For class I diagram, the following expression is used:
\begin{equation}
\begin{split}M_{i\theta,j\theta'}^{(1)} & =-\delta_{ij}\frac{\lambda^{2}}{8\pi^{3}}\frac{f_{i,\mathbf{k},\mathbf{k'}}^{2}V_{\mathbf{k'-k}}}{v_{i\mathbf{k}}v_{i\mathbf{k}'}}\end{split}
\end{equation}
where on the right hand side both $\mathbf{k}$ and $\mathbf{k'}$
are taken to be on the Fermi surface. Class II diagrams are constructed
using:
\begin{equation}
\begin{split}M_{i\theta,j\theta'}^{(2,+)} & =-\frac{\lambda^{4}}{32\pi^{4}}\sum_{\mathbf{q}=\mathbf{q}_{c}}\frac{V_{\mathbf{q}}}{\gamma_{\mathbf{q}}}f_{i,\mathbf{k},\mathbf{k}+\mathbf{q}}^{2}f_{j,\mathbf{k}',\mathbf{k}'+\mathbf{q}}^{2}\frac{1}{v_{i\mathbf{k}}v_{j\mathbf{k'}}}\frac{1}{\left|\mathbf{v}_{i\mathbf{k+q}}\times\mathbf{v}_{j\mathbf{k'+q}}\right|},\\
M_{i\theta,j\theta'}^{(2,-)} & =\frac{\lambda^{4}}{32\pi^{4}}\sum_{\mathbf{q}=\mathbf{q}_{c}}\frac{V_{\mathbf{q}}}{\gamma_{\mathbf{q}}}f_{i,\mathbf{k},\mathbf{k}+\mathbf{q}}^{2}f_{j,\mathbf{k}',\mathbf{k}'-\mathbf{q}}^{2}\frac{1}{v_{i\mathbf{k}}v_{j\mathbf{k'}}}\frac{1}{\left|\mathbf{v}_{i\mathbf{k+q}}\times\mathbf{v}_{j\mathbf{k'-q}}\right|}
\end{split}
\end{equation}

Here all four momentum states $\mathbf{k},\mathbf{k'},\mathbf{k+q}_{c}$
and $\mathbf{k'\pm q}_{c}$ are on the Fermi surface. Note that the
head-on collisions (namely when $\mathbf{k'}=-\mathbf{k}$) need to
be treated differently, as in this case, there is a one-dimensional
manifold of $\mathbf{q}_{c}$ satisfying the momentum-energy conservation
constraints. Specifically, we incoporate these processes using the
following expression:
\[
M_{i\theta,i\left(\theta+\pi\right)}^{(2)}=\int\mathrm{d}k'_{i,\parallel}M_{i\theta,i\theta'}^{(2,+)}
\]

Finally, the diagonal elements of the memory matrix are constructed
from number conservation, namely,
\[
M_{i\theta,i\theta}=\sum_{j}\int\mathrm{d}k'_{j,\parallel}M_{i\theta,j\theta'}
\]

In numerical calculations, we discretize the angle along the Fermi
surface. The number of points is usually taken to be 500. However,
to extract proper scaling behavior at the lowest temperatures, we
checked the convergence of the results for up to 2000 points.

In Figure \ref{fig:eigenspectrum}, we illustrate the properties of
the eigenvalues of the memory matrix for a single Fermi surface. We
present the case when only quantum critical scattering is present,
as well as when current-relaxing mechanisms such as impurity and umklapp
processes are incorporated.

In the absence of impurity or umklapp, there is a large number of
zero modes, corresponding to the conservation of total electron number
as well as the conservation of all odd-parity modes. This is expected
since we are projecting scattering processes onto the Fermi surface.
When either impurity or umklapp scattering is added, all odd-parity
modes (including momentum) are lifted from zero, and only total number
is conserved. However, the way these zero modes get lifted is different
depending on the mechanism. In the case of impurity scattering, all
odd-parity modes gain a finite decay rate $\sim g_{\text{imp}}^{2}\nu_{F}$.
However when umklapp scattering is dominant, the odd-parity modes
are smoothly lifted from $0$, and form a continuous spectrum.

\twocolumngrid

\bibliography{references}

\end{document}